\documentclass[preprint,12pt]{elsarticle}

\usepackage{amssymb}
\usepackage{amsmath}
\usepackage{setspace}

\usepackage{lineno}
\usepackage{multirow}
\usepackage{xcolor}
\usepackage{textcomp}
\usepackage{soul}
\usepackage{graphicx}
\usepackage{enumitem}
\usepackage{mhchem}
\usepackage{tikz,graphics,color,fullpage,float,epsf,caption,subcaption}

\setstcolor{blue}

\journal{International Journal of Heat and Mass Transfer}
\usepackage{graphicx}
\usepackage{epstopdf}
\usepackage{hyperref}
\hypersetup{
    colorlinks=false,
    linkcolor=black,
    filecolor=black,  
    citecolor=black,
    }

\usepackage{geometry}
\begin{document}
\graphicspath{{bilder/}}
\begin{frontmatter}


\title{
Wettability-dependent dissolution dynamics of oxygen bubbles on Ti64
substrates}


\author[Dai,Dai2]{Hongfei Dai\corref{cor1}}
\ead{hongfei.dai1@tu-dresden.de}
\author[Dai,Dai2]{Xuegeng Yang}
\author[Dai,Dai2]{Karin Schwarzenberger}
\author[Dai,Dai2]{Julian Heinrich}
\author[Dai,Dai2]{Kerstin Eckert\corref{cor1}}
\ead{kerstin.eckert@tu-dresden.de}

\address[Dai]{Institute of Process Engineering and Environmental Technology, Technische Universität Dresden, Helmholtzstr. 14, 01069 Dresden, Germany}
\address[Dai2]{Institute of Fluid Dynamics, Helmholtz-Zentrum Dresden-Rossendorf, Bautzner Landstr. 400, Dresden 01328, Germany}
\cortext[cor1]{Corresponding authors.}

\onehalfspacing
\begin{abstract}

In this study, the dissolution of a single oxygen bubble on a solid surface, here Titianium alloy Ti64, in ultrapure water with different oxygen undersaturation 
levels is investigated. For that purpose, a combination of shadowgraph technique and 
planar laser-induced fluorescence is used to measure simultaneously the changes in bubble geometry and in the dissolved oxygen concentration around the bubble. Two different wettabilities of the Ti64 surface are adjusted by using plasma-enhanced chemical vapour deposition.
The dissolution process on the solid surface involves two distinct phases, namely bouncing of the oxygen bubble at the Ti64 surface and the subsequent dissolution of  the bubble, primarily by diffusion. By investigating the features of oxygen bubbles bouncing, it was found that the boundary layer of dissolved oxygen surrounding the bubble surface is redistributed by the vortices emerging during bouncing. This establishes the initial conditions for the subsequent second dissolution phase of the oxygen bubbles on the Ti64 surfaces. In this phase, the mass transfer of $O_2$ proceeds non-homogenously across the bubble surface, leading to an oxygen accumulation close to the Ti64 surface. We further show that the main factor influencing the differences in the dynamics of $O_2$ bubble dissolution is the variation in the surface area of the bubbles available for mass transfer, which is determined by the substrate wettability. As a result, dissolution proceeds faster at the hydrophilic Ti64 surface due to the smaller contact angle, which provokes a larger surface area.

\end{abstract}

\begin{graphicalabstract}
\begin{figure}[h!]
	\centering
	\includegraphics[height=10cm]{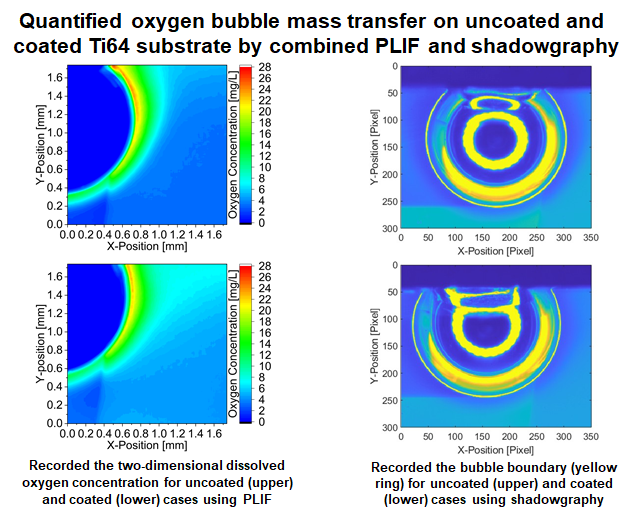} 
\end{figure} 
\end{graphicalabstract}

\begin{highlights}

\item Mass transfer around an oxygen bubble interacting with solid surfaces was quantified
\item Dissolved oxygen concentration and bubble shape could be measured simultaneously by combined PLIF and shadowgraphy
\item A bouncing and a dissolution phase were identified during bubble-surface interaction
\item Stronger redistributed bubble boundary layer after bouncing in more hydrophobic case





\end{highlights}


\end{frontmatter}


\clearpage
\onehalfspacing
\section{Introduction}
\label{sec:intro}

\textcolor{black}{The growth and dissolution of oxygen bubbles play a crucial role in water electrolysis, both for PEM and alkaline electrolysis \cite{angulo2020influence}. These mass transfer
processes are also significant in the aeration of activated sludge tanks in wastewater treatment plants \cite{reinecke4704929population, herrmann2021experimental}. 
Bubbles often come into contact with surfaces such as the tubings of the electrolyzer stacks, or solid materials like fibers in case of sludges. 
This contact often has a negative impact on the systems mentioned above, e.g. reducing the heat transfer efficiency of the heat exchanger in electrolysis system \cite{khani2021novel}.
To mitigate the impact of bubble contact on the systems mentioned above, it is essential to understand the mechanisms that facilitate rapid detachment and dissolution of bubbles from surfaces.}

As early as the 1950s, the growth and shrinkage of non-bound bubbles far away from solid walls have been studied, mainly based on Epstein-Plesset's model \cite{epstein1950stability}, Scriven's model \cite{scriven1995dynamics} and Cable \& Evans' model \cite{cable1967spherically}. Compared to the non-bound case, the growth or shrinkage analysis of bound bubbles is more complex, because both the influence of the substrate on bubble mass transfer \cite{enriquez2014quasi, li2014growth} and the substrate surface wettability \cite{kovats2018characterizing, rubio2021superhydrophobic} need to be taken into account. The substrate surface wettability, i.e. the hydrophilicity and hydrophobicity, as well as surface roughness, affects the bubble shape in static \cite{chen2018contact, moraila2019wetting} and dynamic conditions \cite{xia2021influence, jo2016single, shen2018effect, kibar2017bubble, huynh2015plastron}. 
\textcolor{black}{Moreover, it has been shown recently  
that hydrophobic Ti64 samples significantly enhance the oxygen nucleation by generating a higher number of nucleation spots \cite{heinrich2024functionalization}.
} 

Mass transfer through the gas-liquid interface is accompanied by changes in the bubble shape \cite{Wang2016Investi, popov2005evaporative}. Currently, the description of the mass transfer of bound bubbles is mainly based on the shadowgraph method to determine bubble shape changes, combined with theoretical calculations \cite{paruya2021numerical, soto2017gas} or pure simulations \cite{vachaparambil2020modeling, lee2008growth, diddens2021competing}. 
\textcolor{black}{However
, measurements are lacking which couple the bubble shape changes to highly space- and time-resolved dissolved oxygen concentration distributions around oxygen bubbles interacting with solid surface. This is a missing element to better understand the influence of different surface properties on mass transfer.}


To visualize the mass transfer around bubbles, various optical methods are used, such as LIF (Laser Induced Fluorescence) \cite{babu2019experimental, xu2018mass}, colorimetric methods \cite{dietrich2018visualisation, felis2019using, kherbeche2020hydrodynamics}, UV-induced fluorescence \cite{zhang2020effective, qi2021modelling} and UV-LED induced fluorescence \cite{chen2013measurement, kim2016quantitative}. The LIF method can visualize concentration fields and mass transfer processes with high spatial and temporal resolution compared to other optical methods \cite{ruttinger2018laser, xu2020comparison}. In recent years, LIF was used in bubble wake studies under multidimensional situations \cite{jimenez2013mass, von20193d} and in investigations of the effect of surfactants on mass transfer rate \cite{lebrun2022effect, huang2015influence, kovats2020influence}. For the measurement of \textcolor{black}{the mass transfer at oxygen bubbles}, planar laser-induced fluorescence (PLIF) is often applied, which is an important method in LIF \cite{kuck2009analyse, butler2016modelling, lebrun2021gas, gerke2024planar}. 

\textcolor{black}{In this study, we employ a direct combination of PLIF and shadowgraphy to measure simultaneously the concentration field of dissolved oxygen around the bubble and its geometry during the dissolution. On this basis, 
a comprehensive understanding of the highly dynamic processes of mass transfer and flow phenomena during the interaction of a dissolving bubble with two different solid surfaces \textcolor{black}{has been achieved}. As a substrate, we use Ti64 as an important material for technological processes, focusing particularly on the effects of a hydrophobic coating deposited on the substrate surface.}


\section{Experimental methods}
\label{sec:Exp}
\subsection{Experimental Setup}
\label{sec:Experimental_setup}

\begin{figure}[h!]
	\centering
	\includegraphics[height=6.5cm]{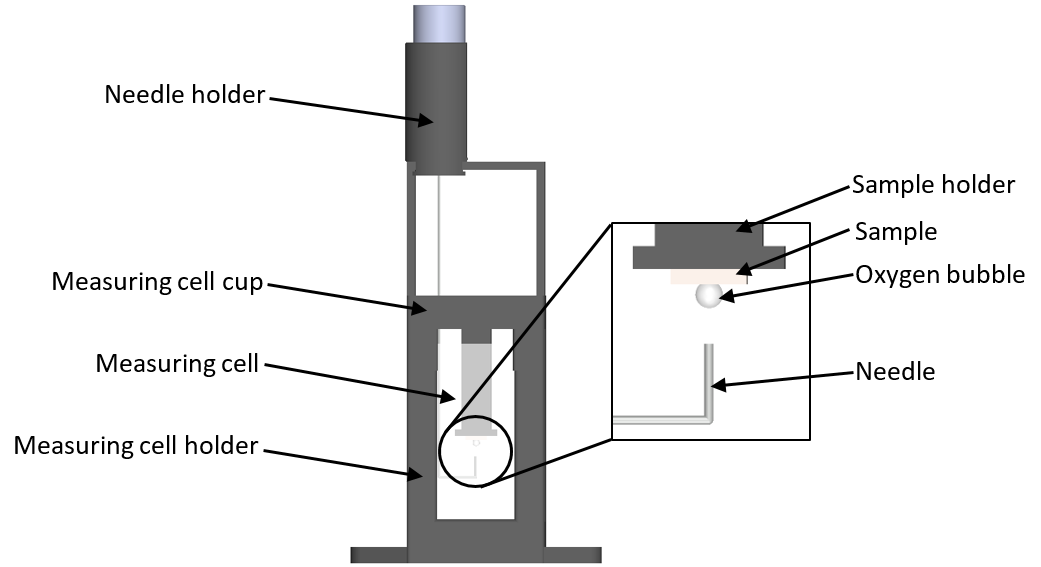} 
	\caption{Schematic of experimental setup used for holding the needle, measuring cell and samples as well as for observing single oxygen bubble on sample surfaces.}
	\label{fig:schema}		
\end{figure}

The experimental setup to study the dissolution of single oxygen bubbles is shown in Fig. \ref{fig:schema}. It is composed of three parts: the measuring cell, the cell holder and the injection needle with its holder. The holders were printed with a 3D printer (Ultimaker S5, USA) using PLA (polylactic acid, Ultimaker, USA) as printing filament. The needle to inject the oxygen bubble 
has an inner diameter of 160 
$ \mu$m, and can slide on the slider track of the measuring cell cap so that the bubbles can be accurately captured in the center of the sample after its detachment.
Titanium alloy Ti64 (Ti-6Al-4V, Goodfellow) was chosen as the sample material due to its high mechanical stability, high temperature resistance, and high workability, particularly its superior corrosion resistance in acidic conditions within the anodic compartment of PEM electrolyzers compared to conventional metals.
The sample size was 5 mm $\times$ 5 mm $\times$ 1 mm. 
The surface of the Ti64 samples was polished, yielding a roughness of $S_a$ = 0.0749 $\mu$m, $S_q$ = 0.1020 $\mu$m, and $S_z$ = 2.7600 $\mu$m, where $S_a$, $S_q$ and $S_z$ denote arithmetical mean height, root-mean-square height and maximum height, respectively. 
To obtain more hydrophobic surfaces for comparison, polished samples were coated by plasma-enhanced chemical vapour deposition in a low-pressure plasma device (Nano, Diener, Germany) with a C6 monomer (1H,1H,2H,2H-perfluorooctylacrylate, CAS: 17527-29-6). \textcolor{black}{After coating, the contact angle of a sessile $O_2$ bubble increases from $51^\circ \pm 2^\circ$ for the uncoated Ti64 substrate to $70.5^\circ \pm 1.5^\circ$ as determined through three measurements.}  
The sample was attached on the sample holder in the measuring cell with dimensions of 10 mm $\times$ 20 mm $\times$ 40 mm (101 - Macro cells, Hellma). 
The measuring cell was placed in its holder and fixed on a goniometer (GO 65S-W30, OWIS, Germany), which was adjusted to keep the sample horizontal. 

The injection needle was submerged in the measurement cell filled with the aqueous phase. By regulating the internal pressure via a micro pressure controller
, an oxygen bubble is generated at the tip of the needle. The bubble size at detachment depends on the inner diameter of the needle \cite{zawala2017bubble}. For the needle inner diameter of 160 $\mu$m, nearly mono-disperse oxygen bubbles with diameters of 1.70 - 1.92 mm were produced. After a short rise over a distance of 7 mm the bubble reaches the Ti64 samples.

\begin{figure}[h!]
	\centering
	\begin{subfigure}{0.8\textwidth}
		\centering
		\captionsetup{justification=centering}
		\includegraphics[height=5.5cm]{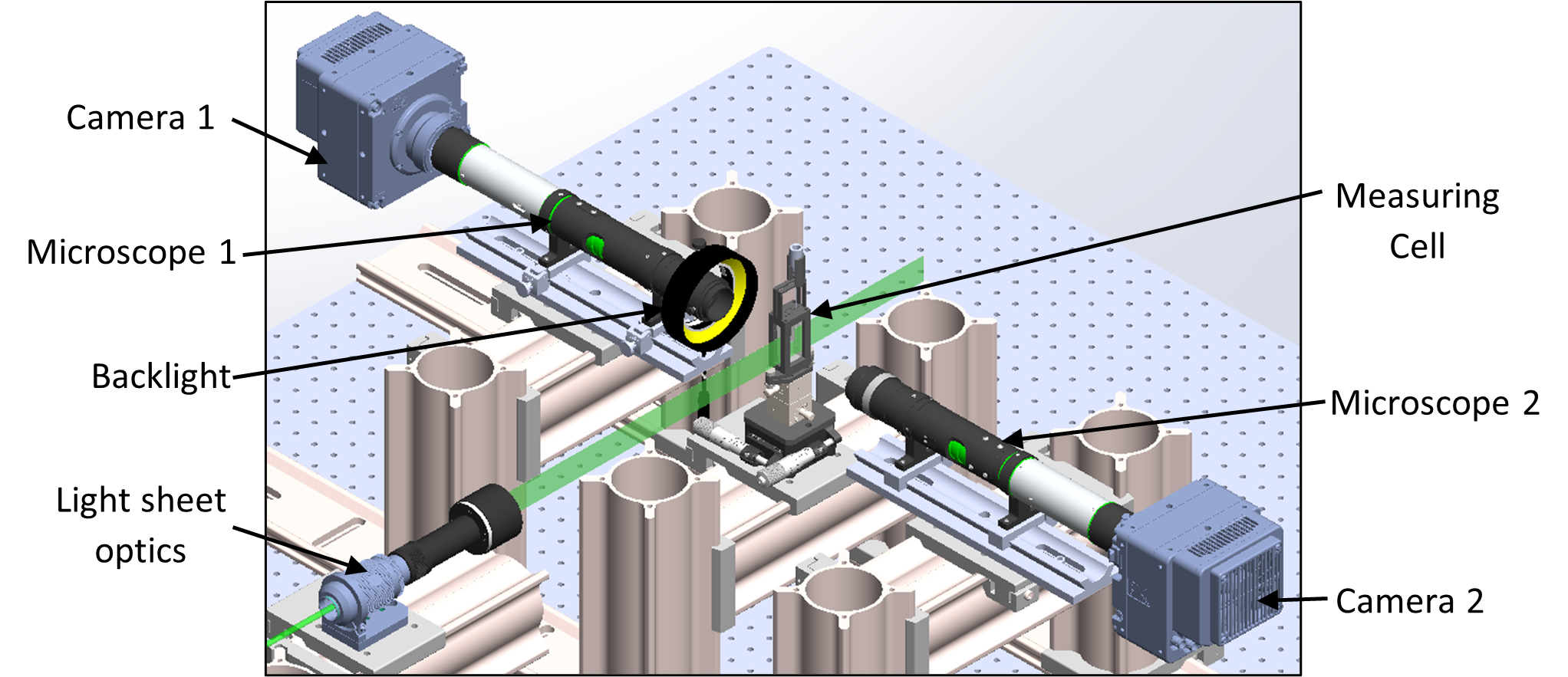}
		\caption{}
		\label{fig:system}
	\end{subfigure} \\
		\begin{subfigure}{0.45\textwidth}
		\centering
		\captionsetup{justification=centering}
		\includegraphics[height=5.1cm]{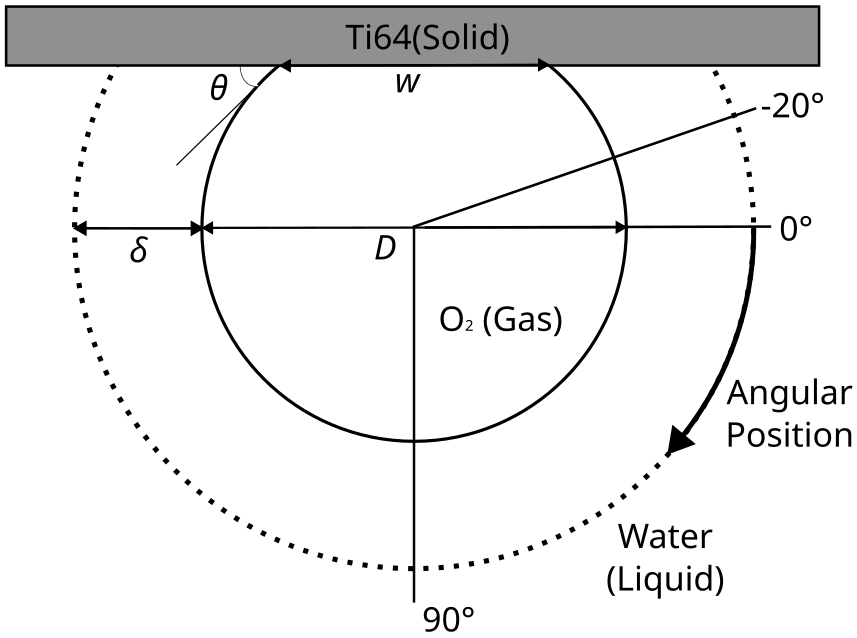}
		\caption{}
	\label{fig:single_bubble}
	\end{subfigure} 	\ \ \ \
		\begin{subfigure}{0.45\textwidth}
		\centering
		\captionsetup{justification=centering}
		\includegraphics[height=5.5cm]{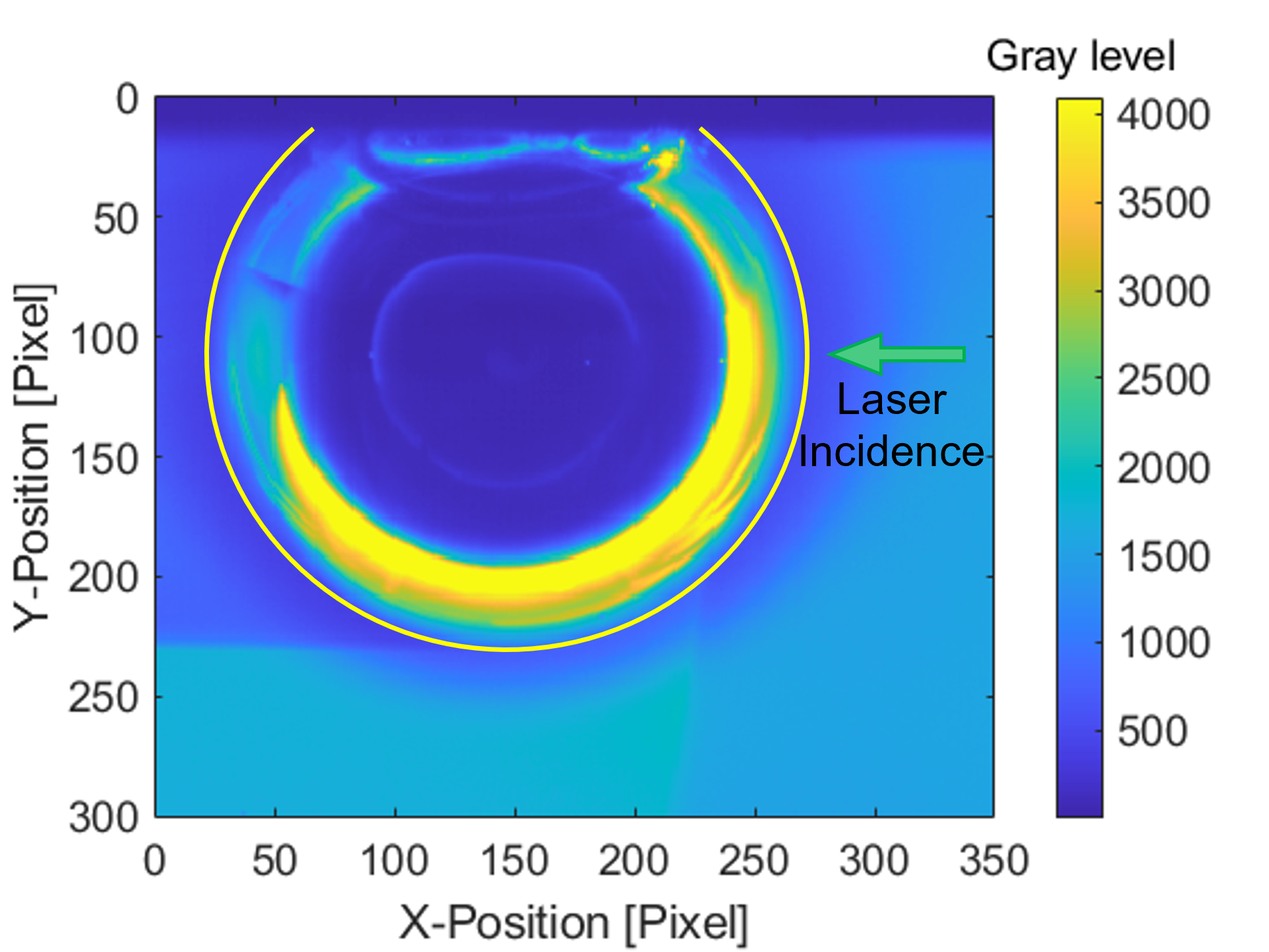}
		\caption{}
		\label{fig:shadow}
	\end{subfigure}  
   \\
	\caption{(a) Schematics of the optical measurement system: Combination of camera 1 and microscope 1 for PLIF measurement and combination of camera 2 and microscope 2 for shadowgraph measurement; (b)  Sketch of single oxygen bubble on sample surface, where $D$, $w$, $\theta$ and $\delta$ are bubble diameter, diameter of the three-phase contact line, contact angle and the thickness of the concentration boundary layer, respectively. The angular position (used in Figs. \ref{fig:v_p_3_a} and \ref{fig:concentration_profile}) increases clockwise from -20\textdegree{} to 90\textdegree{}; (c) Recorded PLIF image: The thin yellow circle shows the bubble boundary. The green arrow indicates the laser incidence. The dark to light blue region around the bubble boundary shows the oxygen concentration distribution. \textcolor{black}{The light area on the lower left indicates laser refraction, while the dark area on the left shows the bubble shadow. These regions are not considered in the analysis.}}
	\label{fig:2}		
\end{figure}

The experimental cell described in Fig. \ref{fig:schema} is integrated into the optical setup depicted in Fig. \ref{fig:system}. It combines planar laser-induced flurorescence (PLIF) to measure the concentration field 
and shadowgraphy to monitor the bubble geometry.

In order to measure the concentration field of dissolved oxygen around the single oxygen bubble simultaneously with the resulting change of bubble shape, two high-speed cameras (Phantom VED-410L, USA), each providing a dynamic range of 12 bits and a resolution of $1280 \times 800 \, \text{pixels}^{2}$ with a corresponding viewing window size of $9.0 \times 5.7 \, \text{mm}^{2}$, were used, denoted by 1 and 2 in Fig. \ref{fig:system}. Each of them is equipped with a precision micro-imaging microscope with a magnification of 3 (Optem FUSION, USA), which were arranged opposite each other. A longpass filter (LP 565 HT, Schneider, Germany) with 565 nm cut-on wavelength was fitted in front of each microscope to avoid the laser's influence on the recording of fluorescence signals. The measuring cell was located in the middle of the two opposite microscopes with a working distance of 100 mm.
The shadowgraph method to analyze the bubble shape \cite{merzkirch2012flow, karnbach2016interplay} employed 
microscope 2 in Fig. \ref{fig:system} together a white LED ring light (35-07-41-000, Optem FUSION, USA) as light source. The LED ring light was nested in the lower lens part of the microscope 1. The PLIF measurements are explained next section. 

Before the experiment starts, the aqueous solution 
was stirred for 1h using a magnetic stirrer in order to fully dissolve the fluorescent dye 
in water. During stirring a vacuum pump (N 840 G LABOPORT, KNF) was used to degas this solution, 
which is quantified through the supersaturation $\zeta$ in our study. This means that the supersaturation value in this paper is always negative, indicating undersaturation. It is defined as:
\begin{equation}
   \zeta=\frac{(c_{\infty}-c_{0})}{c_{0}}.
   \label{eq:zeta}
\end{equation}
where $c_{\infty}$ and $c_{0}$ show the gas concentration in the bulk solution and the gas-saturation concentration in the bulk solution, respectively. 
Different degassing times have been used to achieve various levels of (negative) oxygen supersaturation, ranging from -0.76 to -0.24.

At the beginning of the experiment, undersaturated water mixed with the fluorescent dye 
was cautiously 
filled in the measurement cell using a syringe.
A single $O_2$ bubble was generated through the single-bubble generator. The $O_2$ bubble rose and bounced freely until it was eventually captured on the Ti64 sample surface facing downwards. After that, the $O_2$ bubble rested on the sample surface, thereby undergoing dissolution.



The bubble geometry in terms of diameter ($D$), diameter of three-phase contact line ($w$), and contact angle ($\theta$), as shown in Fig. \ref{fig:single_bubble}, is obtained from binarized shadowgraphy images.
Considering the bubble as axisymmetric, the maximum horizontal extension of the
bubble is taken as the bubble diameter.
The diameter of the three-phase contact line is determined by the distance from where the bubble boundary intersects the sample surface.
\textcolor{black}{
The bubble boundary close to the Ti64 substrate, see Fig. \ref{fig:shadow}, is fitted by a circle 
and the contact angle is calculated using this circle's geometric relations.}

\subsection{Planar laser-induced flurorescence (PLIF) measurements}
\label{sec:PLIF}

The PLIF measurements employ microscope 1, see Fig. \ref{fig:system}, and use a quenching dye, which is a ruthenium complex (dichlorotris(1,10-phenanthroline)ruthenium(II) hydrate, Sigma-Aldrich, CAS: 207802-45-7).
Fluorescence of the dye is excited by a Nd:YAG-Laser (Photonics, USA) with a wavelength of 527 nm.
To obtain a clear fluorescence signal (sufficient signal-to-noise ratio), a dye concentration of 175 mg/L was required instead of 30 mg/L \cite{kursula2022unsteady} and 75 mg/L \cite{xu2020comparison} as reported earlier. 

This ruthenium complex dye is very suitable for two reasons: (i) It does not react chemically with oxygen and (ii) it is not surface active. This was verified by measuring the surface tension of the water-dye mixture at 175 mg/L concentration. The value obtained (71.30 mN/m) is close to the value of DI water (72.80 mN/m  at 20 \textdegree{}C). Thus, the dye neither affects the bubble shape nor does it create a Marangoni convection  \cite{babich2023situ}.
Different fluorescence intensities, corresponding to different dissolved oxygen concentration  are represented  by different gray levels in the recorded images, as exemplarily shown in Fig. \ref{fig:shadow}.

\begin{figure}[h!]
	\centering
	\includegraphics[height=8.5cm]{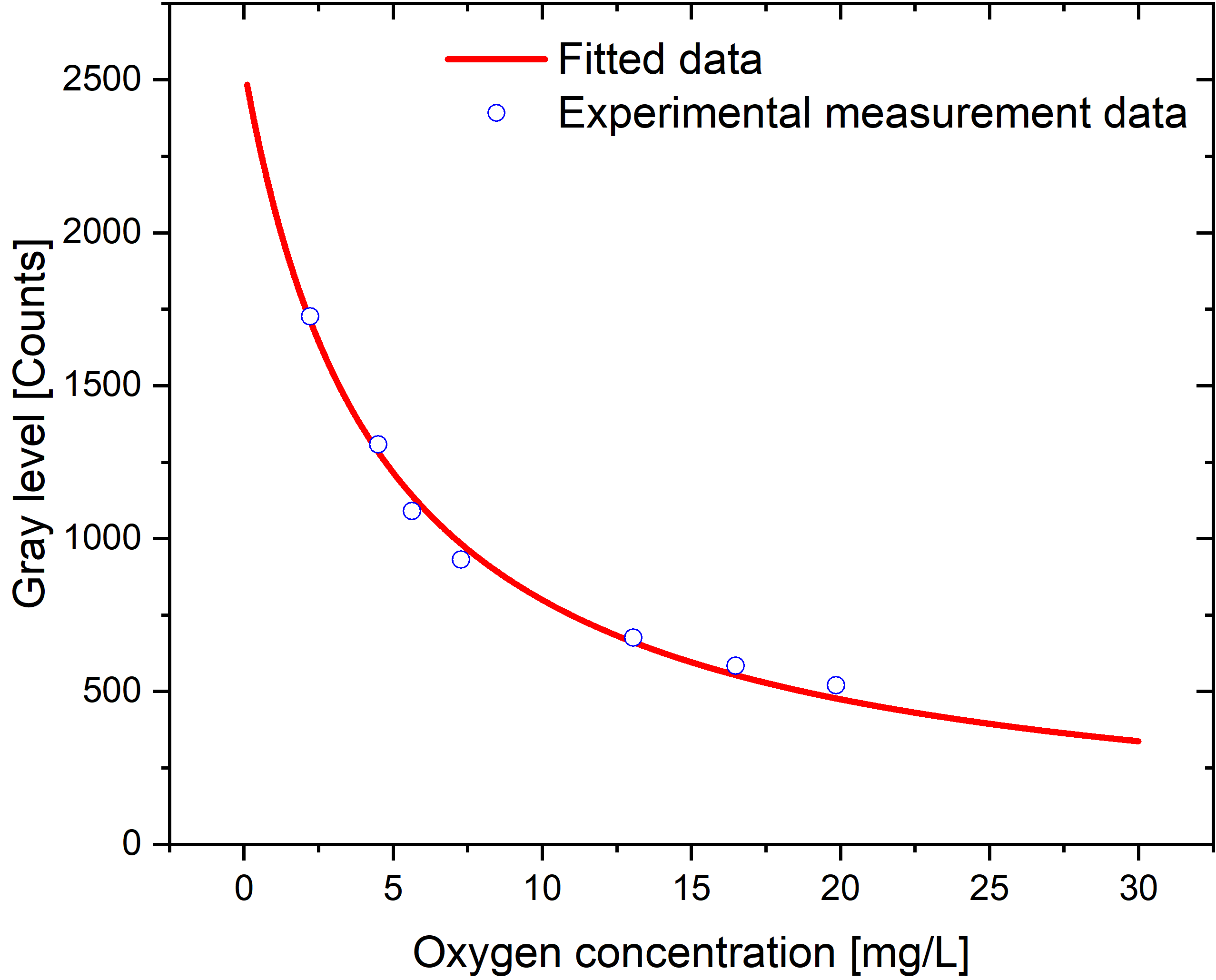} 
	\caption{Calibration curve: gray level vs. oxygen concentration at seven different oxygen concentrations (open circles) at a concentration of 175 mg/L of the ruthenium complex dye excited by a laser with wavelength of 527 nm.
	}
	\label{fig:calibration}		
\end{figure}

\textcolor{black}{To obtain real oxygen concentration values, a calibration relationship between oxygen concentration and gray level is required. For that purpose, an aqueous solution with a concentration of 175 mg/L ruthenium complex was prepared. 
After 1 h of stirring and simultaneous degassing, the oxygen concentration in the aqueous solution is minimized. The minimum oxygen concentration for the calibration is 2.225 mg/L, which was measured by an oxygen sensor (OXYBase WRBlue, PreSens). PLIF was used to obtain the gray level of the image at this oxygen concentration. After that, oxygen was passed into the solution to obtain other six different oxygen concentrations ranging from 4.510 mg/L to 19.850 mg/L and the corresponding gray levels of images. The measured values were processed and fitted using Stern-Volmer relationships, as shown in Fig. \ref{fig:calibration}, to obtain the equation for the gray level (fluorescence intensity $I$), as a function of oxygen concentration, as follows:}

\begin{equation}
   I=\frac{1}{(0.00008565\,c_{\infty}/(\text{mg/L})+0.0003938)}.
   \label{eq:Fluorescence}
\end{equation}

\section{Mass transport during bubble dissolution}
\label{sec:Theoretical}

For simplicity we consider a free oxygen bubble sufficiently away from the titanium substrate, which is exposed to water undersaturated with dissolved oxygen. 
We refer to \cite{yang2020analytical} or \cite{popov2005evaporative} for more elaborate 
models for growth of bubbles or evaporation of droplets on substrates which consider the details of the three-phase contact line explicitly.
As a result of the undersaturation, mass transport of oxygen from the free oxygen bubble into the water phase takes place. The transport of dissolved oxygen in the water is governed by \cite{Gro2017diffusion}:
\begin{equation}
\frac{\partial c}{\partial t}+\nabla \cdot (\boldsymbol{U} c)-{D_{O_2} {\nabla}^2 c}=0 ,
\label{eq:cDistri}
\end{equation}
where $\boldsymbol{U}$ is the velocity field in the water phase and $D_{O_2}$ \textcolor{black}{($2.10 \times 10^{-9} \ \text{m}^2/\text{s}$ at 25\textdegree{}C \cite{cussler2009diffusion})} is the diffusion coefficient of dissolved oxygen. At the oxygen-liquid interface ($r=r_b$, where $r_b$ refers to the bubble radius), the boundary condition is given by $c(r_b)=c_{sat}$. The saturation concentration $c_{sat}$ follows from \cite{lohse2015pinning}
\begin{equation}
    c_{sat}=H^{cp}(P_{ext}+\frac{2\gamma}{r_b}),
    \label{eqn:Kh}
\end{equation} 
where $H^{cp}$, $P_{ext}$ and $\gamma$ are the Henry's constant, the pressure above the bulk solution and interfacial tension of the gas-liquid interface, respectively.

By solving Eq. \ref{eq:cDistri}, the distribution of the dissolved oxygen concentration around the bubble can be obtained.
The bubble's dissolution is driven by the molar flux of gaseous oxygen, $\Vec{j}$, across the oxygen-water interface which can be described by the Fick’s first law or by introducing a mass transfer coefficient $k$
\cite{wang2016investigations}:
\begin{equation}
\left| \Vec{j}\right|=j=D_{O_2} \left| \nabla c \right|=k (c_{sat}-c_\infty).
\label{eqn:j}
\end{equation}
The rate by which the bubble loses mass, $m$, due to dissolution can be evaluated 
by integrating $\Vec{j}$ over the gas-liquid interface:
\begin{equation}
   \frac{dm}{dt}=
   M_g \frac{dn}{dt}=
   M_g\dot{n}=
M_g\int \Vec{j}  d{\Vec{A}},
   \label{eqn:dmdt1}
\end{equation}
with $M_g$, $\dot{n}$ and $d\Vec{A}$ representing the molar mass of the gas (g/mol)
, the amount of gas leaving the bubble in unit time (mol/s)
and the increment of the surface area ($ \text{m}^2$), respectively.  




The presence of the Ti64 substrate modifies the mass transfer as analyzed in the next section. To compare with  the ideal case of a spherical, non-bound bubble, the corresponding concentration profiles were obtained by  solving the 1D version of Eq. \ref{eq:cDistri} without the advective term using the “pdepe” function in MATLAB R2018b.  The advective term is neglected here based on the small value of the  Peclet number \cite{huysmans2005review} given by: $Pe = {L \cdot u}/{D_{O_2}} = 2.63 \times 10^{-3}$, in which $L$ is reference length (the bubble diameter $D$) and $u$ is the velocity due to the bubble size change by dissolution.

\section{Results and discussion}
\label{sec:Results}

The dissolution process of oxygen bubbles in the present study consists of two phases. In phase I, termed bouncing, the oxygen bubble released in the bulk, cf. Section \ref{sec:Experimental_setup}, rises a short distance, approx. 7 mm, towards the Ti64 substrate. Unexpectedly, it turned out that this bouncing phase displays a rich dynamic, which influences the oxygen transfer, thereby setting the initial conditions for the subsequent dissolution phase II. Thus this section is divided into two parts. Section \ref{sec:bouncing} investigates the features of the bouncing phase I at both uncoated and coated substrates. The second part is devoted to the pure dissolution of the oxygen bubbles after the bouncing oscillations have been damped, in terms of both the concentration field in Section \ref{sec:concentration_gradient} and the bubble geometry in Section \ref{sec:Geometry_bubble}.


\begin{figure}[h!]
	\centering
		\begin{subfigure}{0.2\textwidth}
		\centering
		\captionsetup{justification=centering}
		\includegraphics[height=3.5cm]{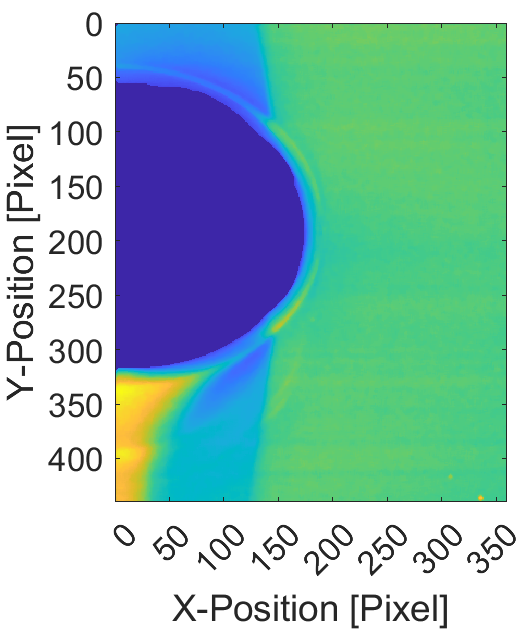}
		\caption{$t=0\, s$}
		\label{fig:2.410_1658}
	\end{subfigure}  \ \ \ \
	\begin{subfigure}{0.2\textwidth}
		\centering
		\captionsetup{justification=centering}
		\includegraphics[height=3.5cm]{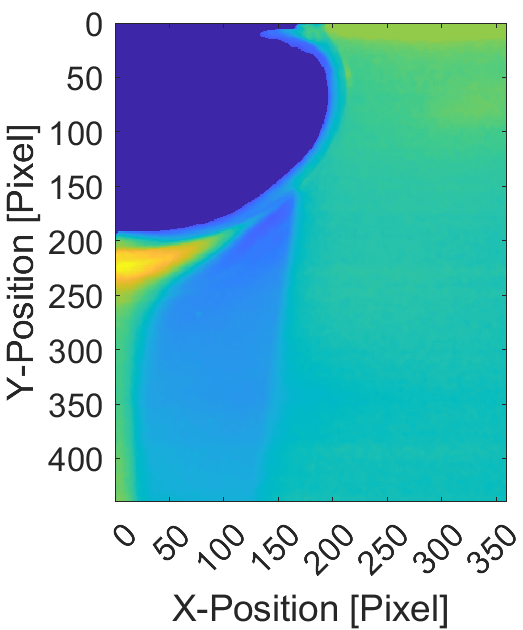}
		\caption{$t=0.006\, s$}
		\label{fig:2.410_1661}
	\end{subfigure}    \ \ \ \
	\begin{subfigure}{0.2\textwidth}
		\centering
		\captionsetup{justification=centering}
		\includegraphics[height=3.5cm]{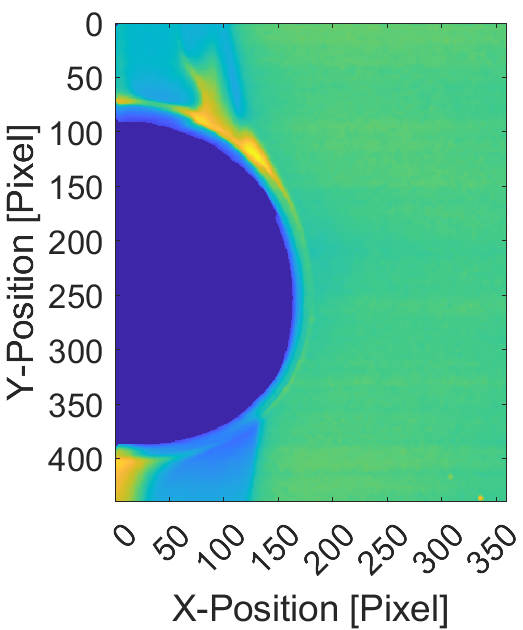}
		\caption{$t=0.020\, s$ }
		\label{fig:2.410_1668}
	\end{subfigure}    \ \ \ \
	\begin{subfigure}{0.2\textwidth}
		\centering
		\captionsetup{justification=centering}
		\includegraphics[height=3.5cm]{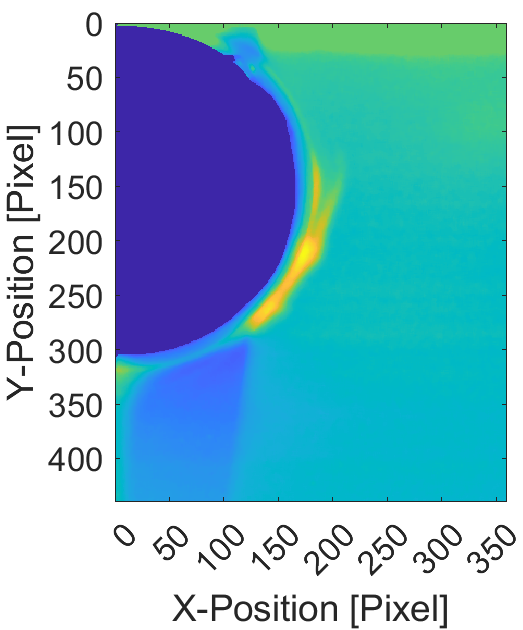}
		\caption{$t=0.054\, s$ }
		\label{fig:2.410_1685}
	\end{subfigure}  \\
	\begin{subfigure}{0.2\textwidth}
		\centering
		\captionsetup{justification=centering}
		\includegraphics[height=3.5cm]{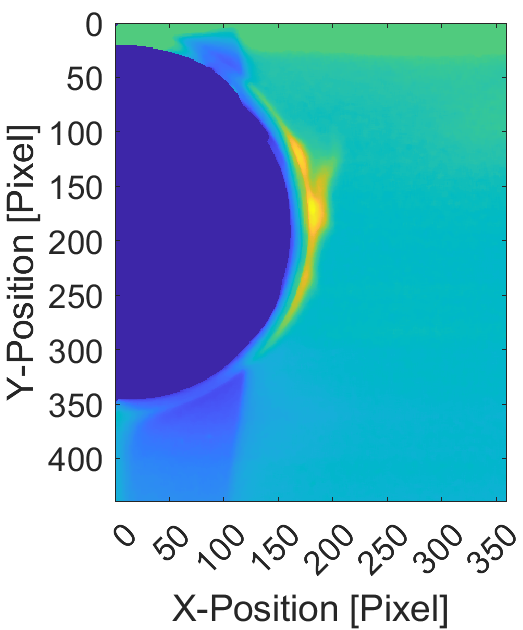}
		\caption{$t=0.062\, s$ }
		\label{fig:2.410_1689}
	\end{subfigure}  \ \ \ \
	\begin{subfigure}{0.2\textwidth}
		\centering
		\captionsetup{justification=centering}
		\includegraphics[height=3.5cm]{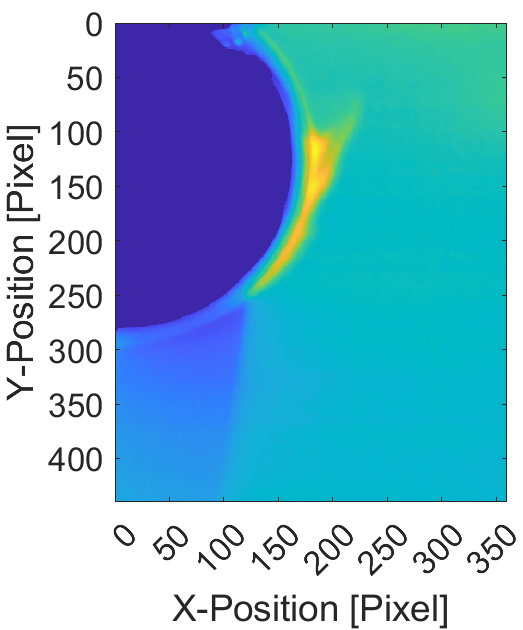}
		\caption{$t=0.098\, s$}
		\label{fig:2.410_1709}
	\end{subfigure}   \ \ \ \
	\begin{subfigure}{0.2\textwidth}
		\centering
		\captionsetup{justification=centering}
		\includegraphics[height=3.5cm]{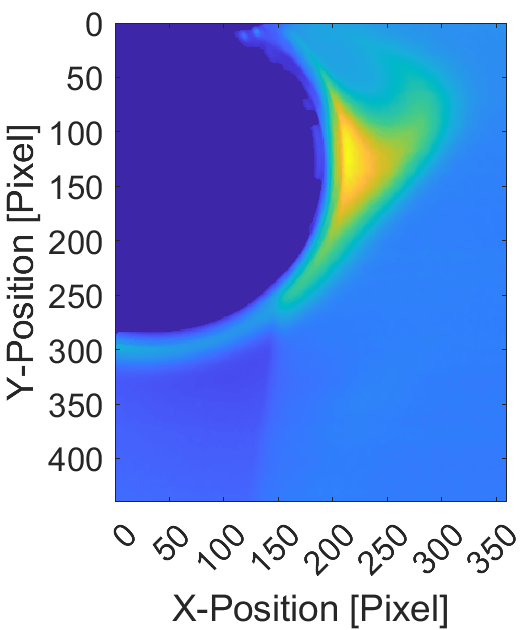}
		\caption{$t=5.392\, s$}
		\label{fig:2.410_4363}
		\end{subfigure}   \ \ \ \
	\begin{subfigure}{0.2\textwidth}
		\centering
		\captionsetup{justification=centering}
		\includegraphics[height=3.5cm]{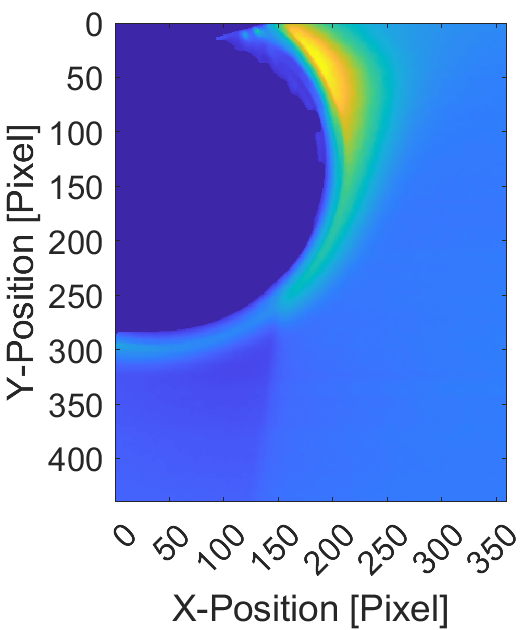}
		\caption{$t=16.512\, s$}
		\label{fig:2.410_9914}
	\end{subfigure}   \\	
	\vskip1cm
	\centering
	\begin{subfigure}{0.2\textwidth}
		\centering
		\captionsetup{justification=centering}
		\includegraphics[height=3.5cm]{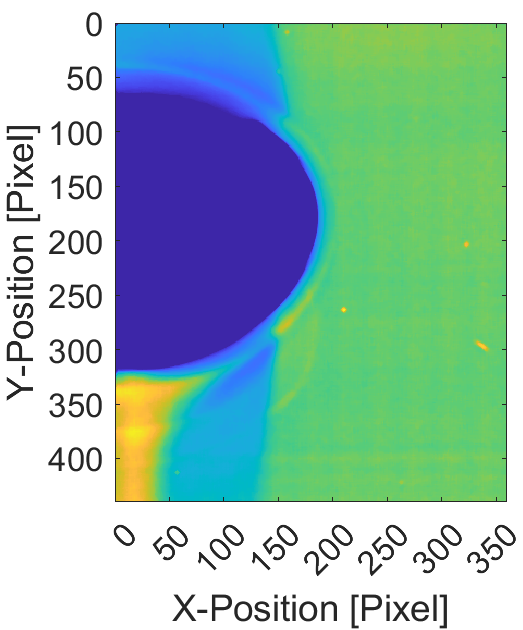}
		\caption{$t=0\, s$}
		\label{fig:3.915_1638}
	\end{subfigure}  \ \ \ \
	\begin{subfigure}{0.2\textwidth}
		\centering
		\captionsetup{justification=centering}
		\includegraphics[height=3.5cm]{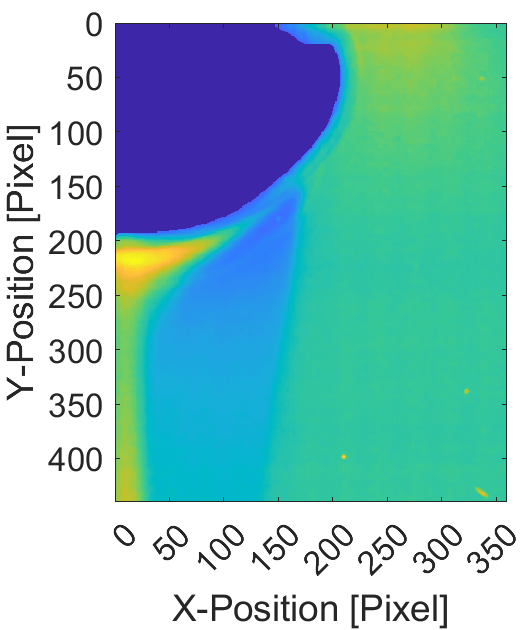}
		\caption{$t=0.006\, s$}
		\label{fig:3.915_1641}
	\end{subfigure}  \ \ \ \
	\begin{subfigure}{0.2\textwidth}
		\centering
		\captionsetup{justification=centering}
		\includegraphics[height=3.5cm]{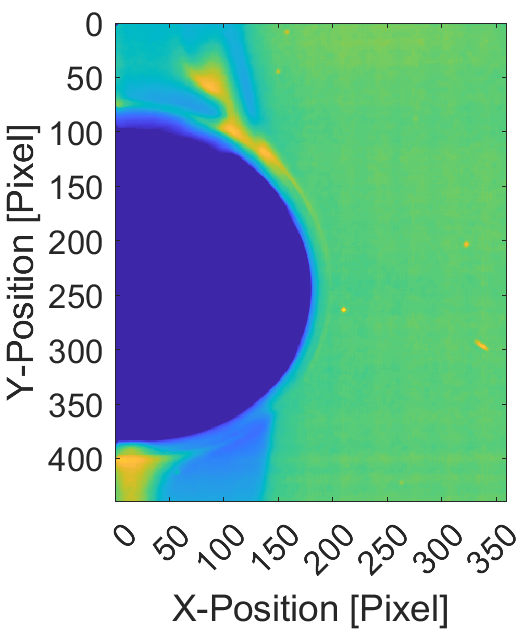}
		\caption{$t=0.020\, s$}
		\label{fig:3.915_1648}
	\end{subfigure}  \ \ \ \
	\begin{subfigure}{0.2\textwidth}
		\centering
		\captionsetup{justification=centering}
		\includegraphics[height=3.5cm]{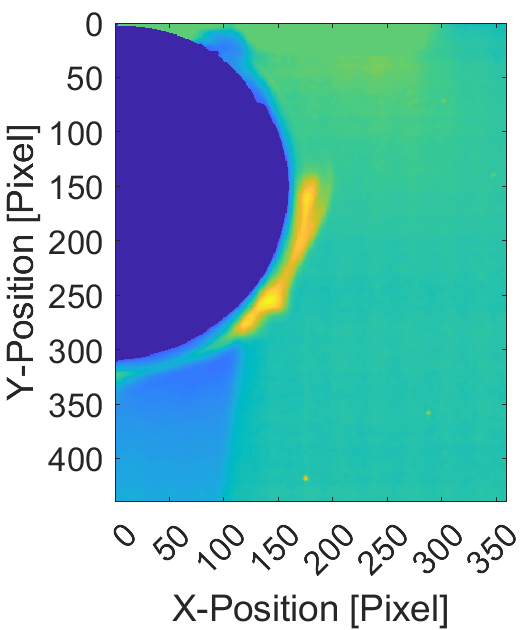}
		\caption{$t=0.054\, s$}
		\label{fig:3.915_1665}
	\end{subfigure}  \\
	\begin{subfigure}{0.2\textwidth}
		\centering
		\captionsetup{justification=centering}
		\includegraphics[height=3.5cm]{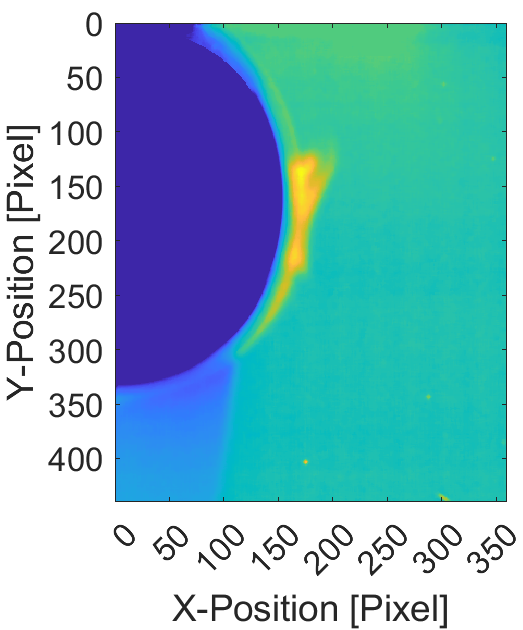}
		\caption{$t=0.062\, s$}
		\label{fig:3.915_1669}
	\end{subfigure}  \ \ \ \
	\begin{subfigure}{0.2\textwidth}
		\centering
		\captionsetup{justification=centering}
		\includegraphics[height=3.5cm]{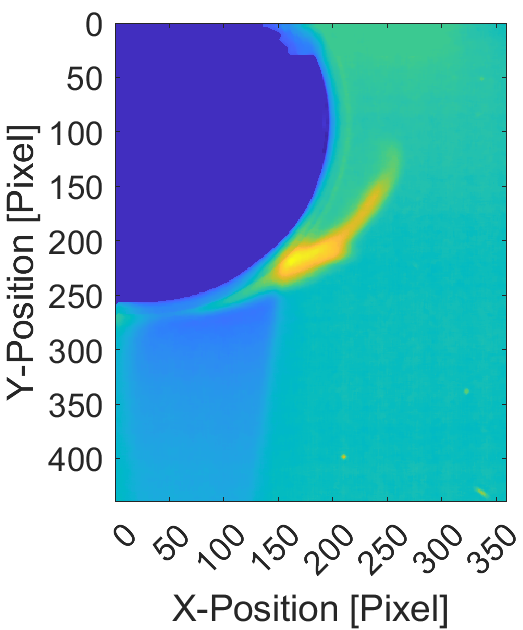}
		\caption{$t=0.098\, s$}
		\label{fig:3.915_1687}
	\end{subfigure} \ \ \ \
	\begin{subfigure}{0.2\textwidth}
		\centering
		\captionsetup{justification=centering}
		\includegraphics[height=3.5cm]{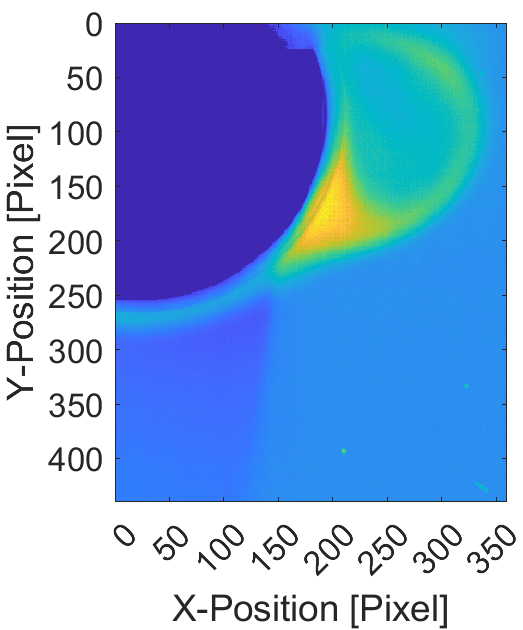}
		\caption{$t=5.392\, s$}
		\label{fig:3.915_4334}
		\end{subfigure}  \ \ \ \
	\begin{subfigure}{0.2\textwidth}
		\centering
		\captionsetup{justification=centering}
		\includegraphics[height=3.5cm]{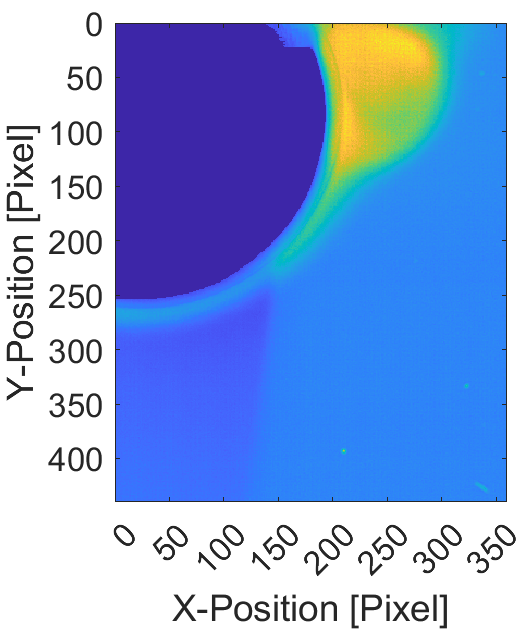}
		\caption{$t=16.512\, s$}
		\label{fig:3.915_9894}
	\end{subfigure}  \\
	\vskip1cm
\begin{subfigure}{0.26\textwidth}
		\centering
		\captionsetup{justification=centering}
		\includegraphics[height=2.8cm]{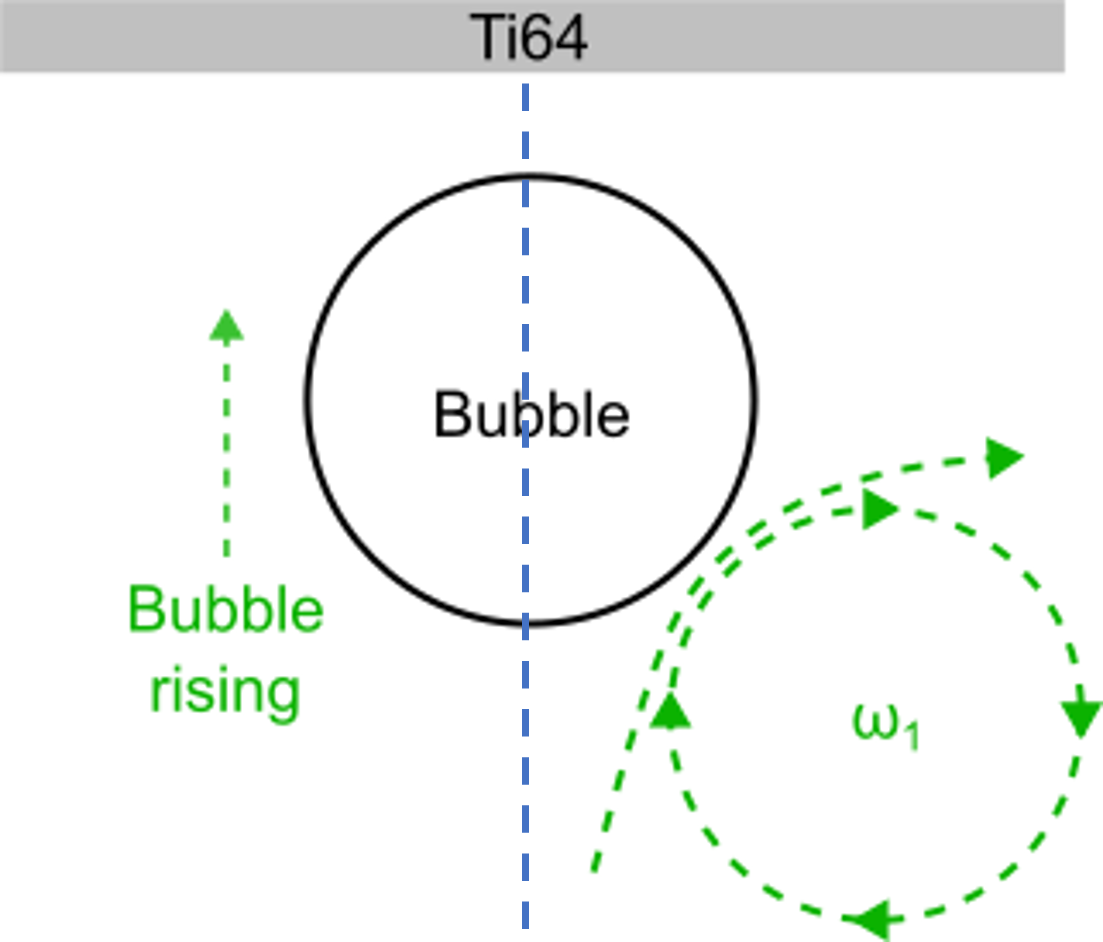}
		\caption{}
		\label{fig:rising}
	\end{subfigure}  \ \ \ \
	\begin{subfigure}{0.26\textwidth}
		\centering
		\captionsetup{justification=centering}
		\includegraphics[height=2.8cm]{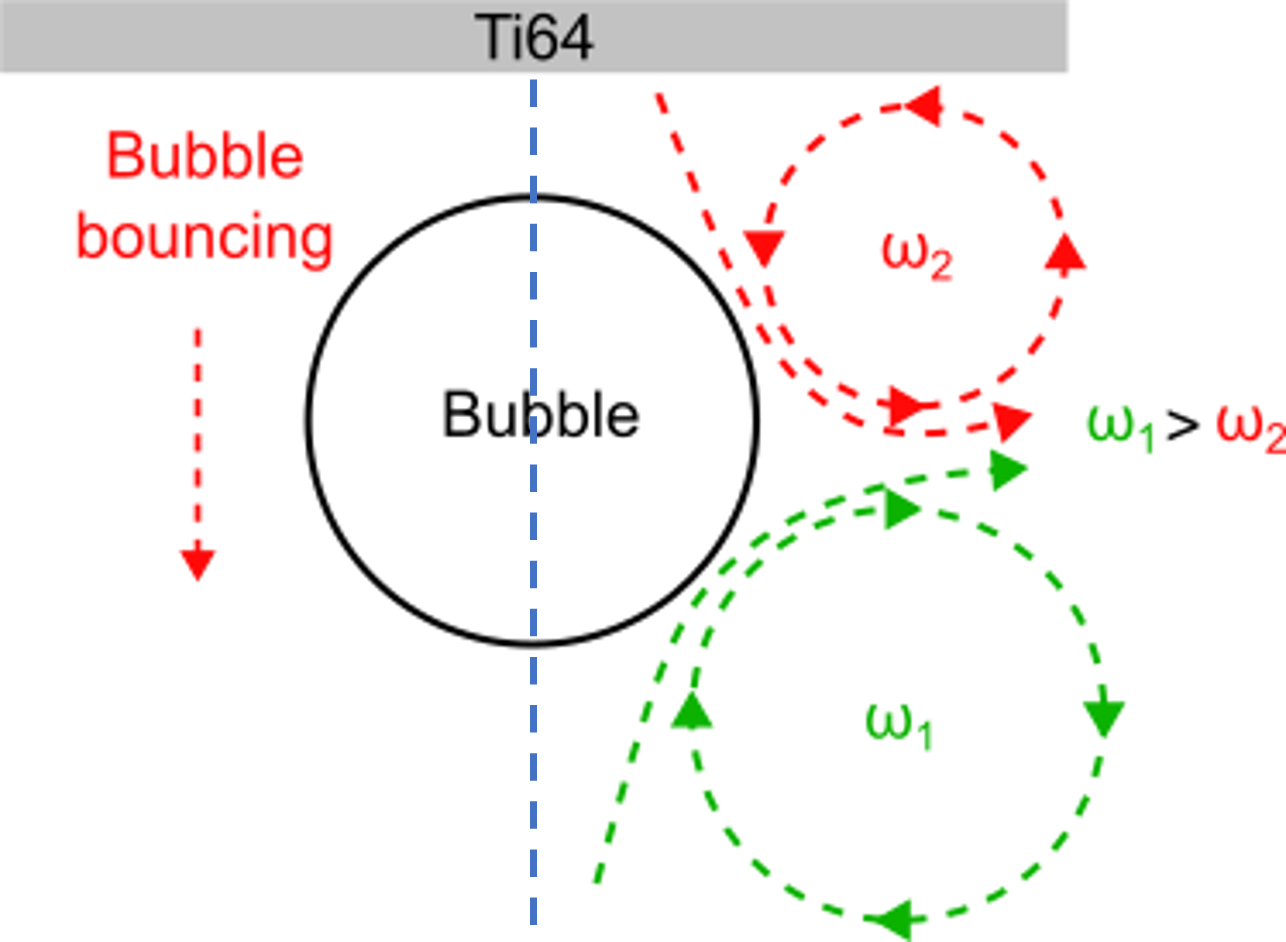}
		\caption{}
		\label{fig:rebouncing}
	\end{subfigure}   \ \ \ \
	\begin{subfigure}{0.26\textwidth}
		\centering
		\captionsetup{justification=centering}
		\includegraphics[height=2.8cm]{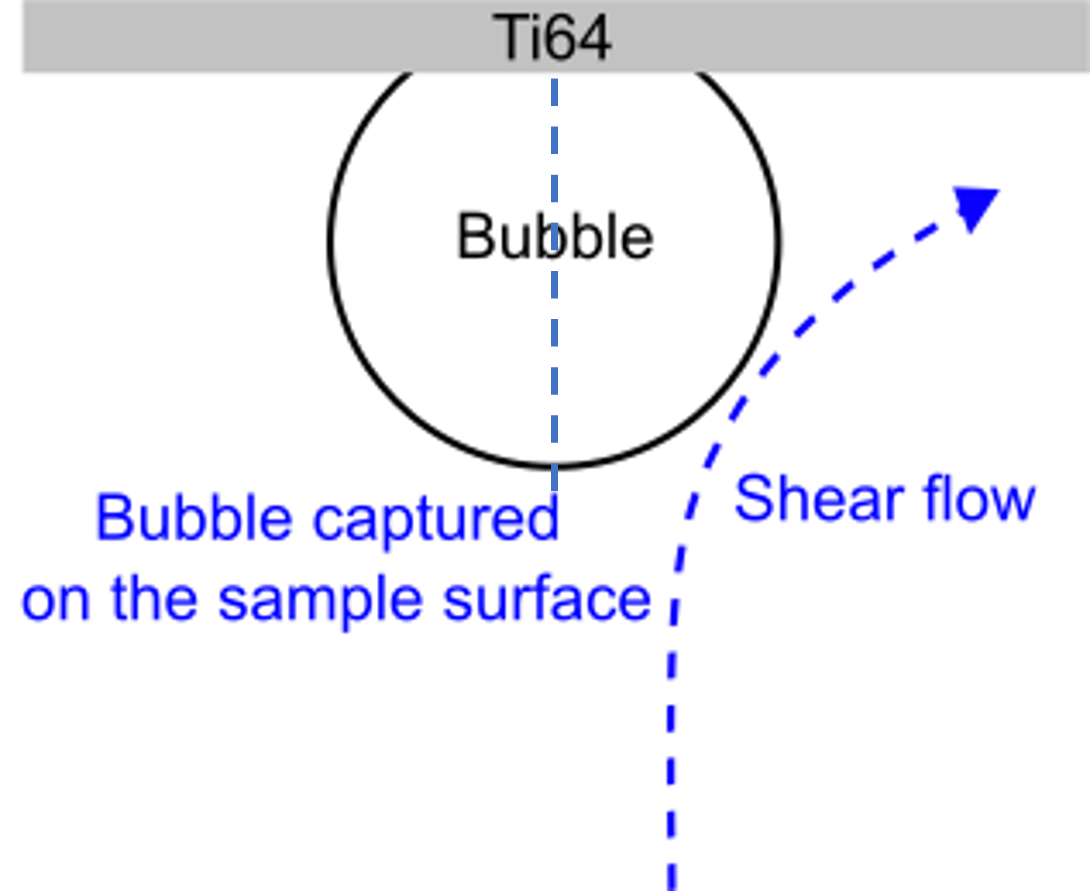}
		\caption{}
		\label{fig:shear_flow}
	\end{subfigure}   \\
	\caption{Screenshots of a single oxygen bubble during bouncing at  an uncoated (a-h) and a coated (i-p) Ti64 sample. Schematics of the vortices around the bubble during the bubble rise (q) and bouncing (r), as well as the shear flow after the bubble is captured on the sample surface (s). $\omega$ refers to the vorticity. 
	}
	\label{fig:bubble_bouncing_and_rebouncing}
			\end{figure}

\subsection{Phase I: 
Bouncing of the $O_2$ bubble}
\label{sec:bouncing}

\begin{figure}[h!]
	\begin{subfigure}{0.46\textwidth}
		\centering
		\captionsetup{justification=centering}
		\includegraphics[height=5.25cm]{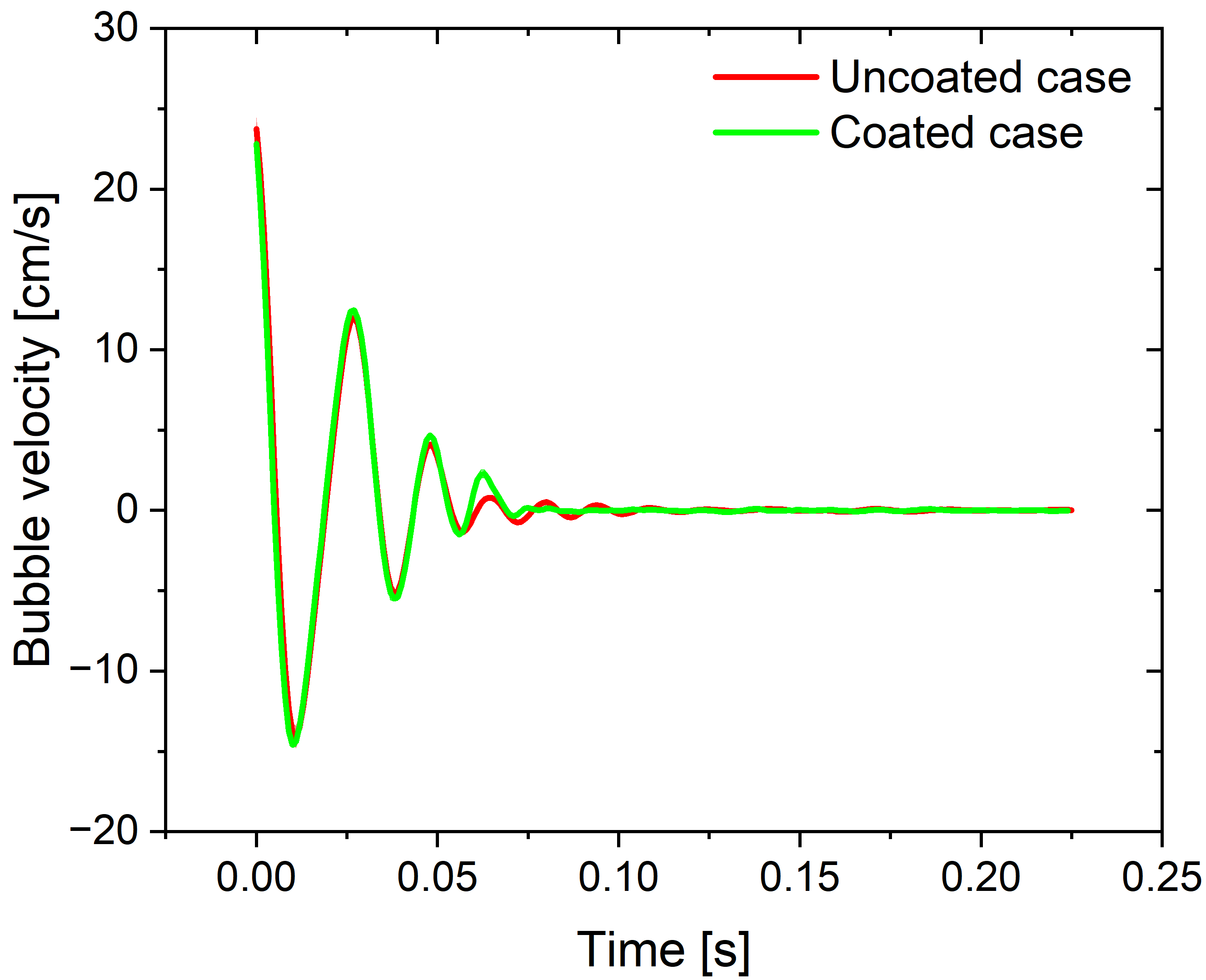}
		\caption{}
		\label{fig:velocity}
	\end{subfigure}   \ \ \ \
	\begin{subfigure}{0.46\textwidth}
		\centering
		\captionsetup{justification=centering}
		\includegraphics[height=5.25cm]{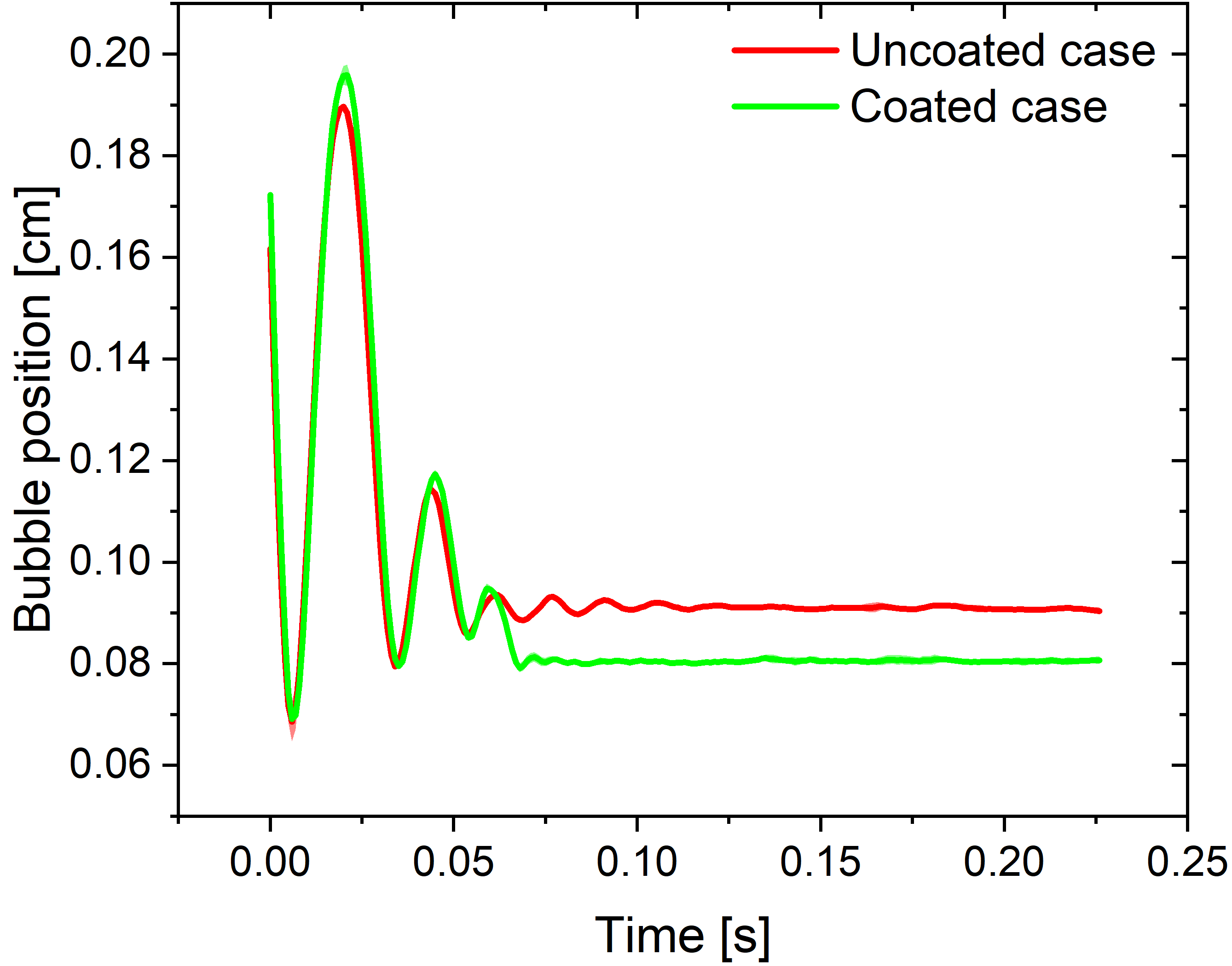}
		\caption{}
		\label{fig:position}
	\end{subfigure}   \\
	\begin{subfigure}{0.46\textwidth}
		\centering
		\captionsetup{justification=centering}
		\includegraphics[height=5.25cm]{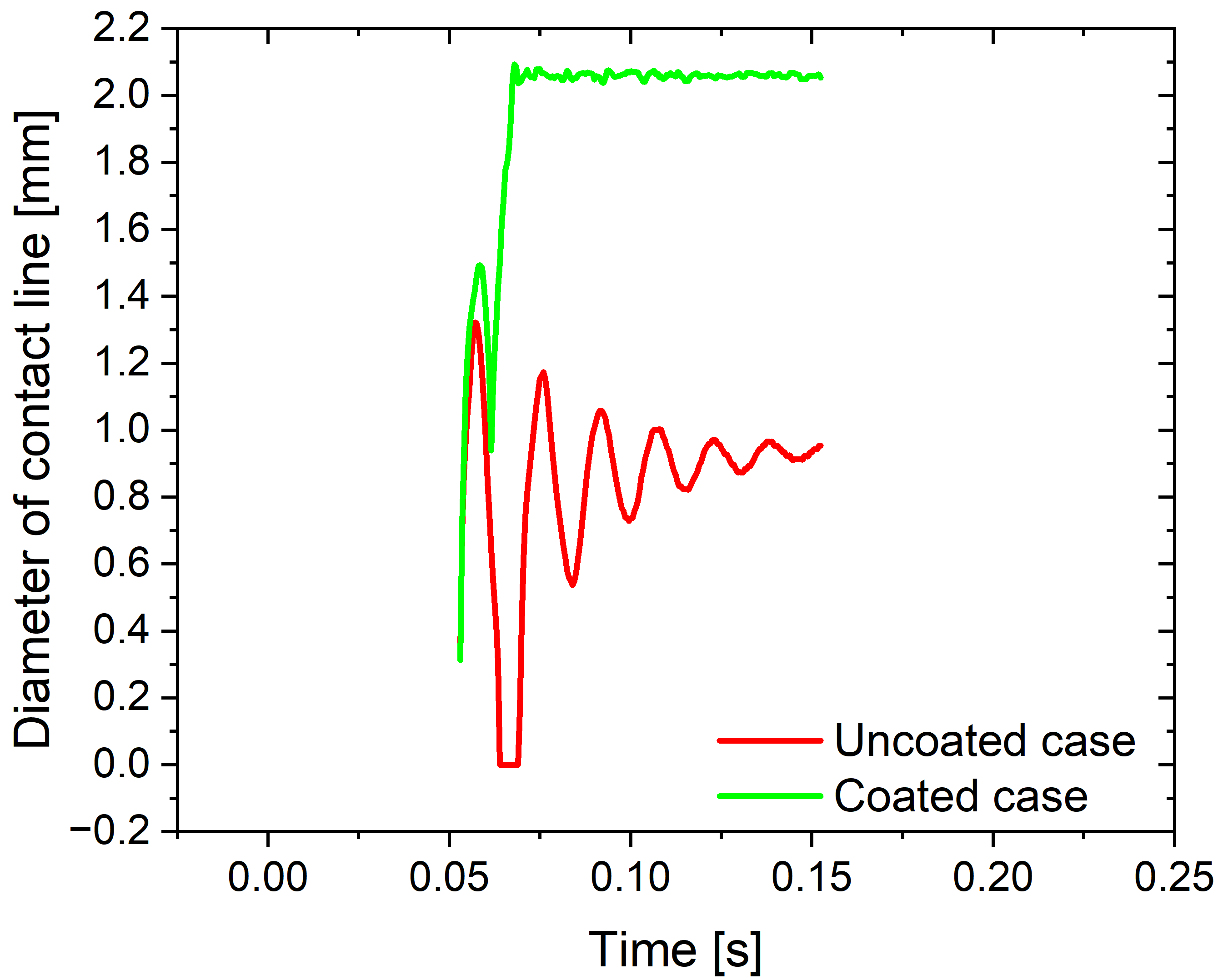}
		\caption{}
		\label{fig:bouncing_cl}
	\end{subfigure}   \ \ \ \
	\begin{subfigure}{0.46\textwidth}
		\centering
		\captionsetup{justification=centering}
		\includegraphics[height=5.25cm]{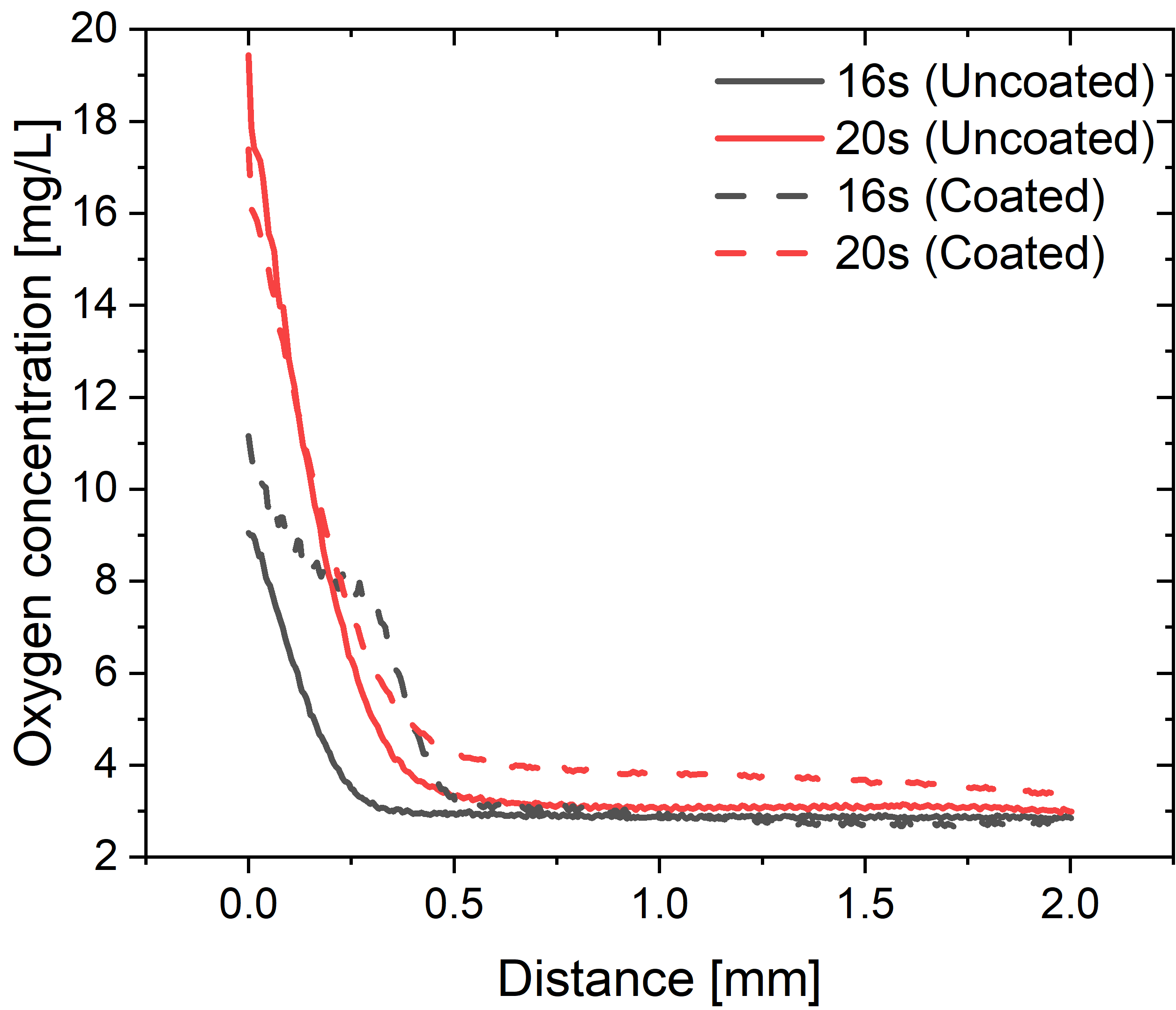}
		\caption{}
		\label{fig:16_20_145}
	\end{subfigure}   \\
	\caption{Bouncing phase I: Velocity (a), position (b) and extension of the diameter of the contact line (c) of a bouncing  $O_2$ bubble as a function of time at uncoated (red line) and coated (green line) Ti64 samples. In (b), the bubble position is defined as the absolute distance from the centerpoint of the bubble to the Ti64 surface. 
	In (c), the change in the diameter of the contact line between 0.053 s and 0.153 s is extracted to specifically investigate the time span with different bubble dynamics for the uncoated and coated cases.
	(d) Radial profiles of the dissolved oxygen concentration in the equatorial plane (angular position 0\textdegree{}) for the uncoated (solid lines) and the coated (dashed lines) Ti64 samples at similar oxygen supersaturation of $\zeta \sim -0.72 \pm 0.2$ at 16 s and 20 s (start of phase II). 
	}
	\label{fig:v_p_3_a}		
\end{figure}

The concentration field around the $O_2$ bubble  during the  bouncing is shown in Fig. (\ref{fig:2.410_1658} – \ref{fig:2.410_1709}) and Fig. (\ref{fig:3.915_1638} – \ref{fig:3.915_9894})  for an  uncoated and coated Ti64 sample, respectively, at oxygen supersaturation $\zeta \sim -0.72 \pm 0.2$.
For further shadowgraphy snapshots we refer to  the supplemental material (Section \ref{sec:Supplemental Material}).
Quantitative details of the bouncing periods in terms of velocity, center position and diameter of contact line extension of the bubble are depicted in Fig. \ref{fig:v_p_3_a}.
Inspecting Figs. \ref{fig:2.410_1658} and \ref{fig:3.915_1638} one can nicely recognize the wake of dissolved oxygen (in yellow) behind the rising bubbles. For a better visualization only the right half of the $O_2$ bubble directed toward the laser is plotted. The bubbles hit the Ti64 sample with a terminal velocity of about 24 cm/s. This forces a strong deformation of the bubble, which initiates the bouncing process. The latter lasts about 160 ms and includes approx. 6 periods, the amplitudes of which exponentially decay with time, cf. Fig. \ref{fig:velocity}.

Differences between the uncoated and coated cases become visible only beyond 60 ms, while the foregoing two bouncing periods proceed with nearly identical values of bouncing velocity, center position and diameter of contact line extension.
During the third bouncing period, the $O_2$ bubble is able to detach again from the uncoated Ti64 substrate but not from the coated one, cf. Fig. \ref{fig:position}.  Moreover, 
the bubble becomes arrested at the coated Ti64 for $t>60$ ms. This correlates with a fast extension of the diameter of the contact line in Fig.~\ref{fig:bouncing_cl} (green line) driven by the rupture of the thin water film between bubble and coated Ti64 substrate. 
\textcolor{black}{By contrast, the thin water film between the bubble and the uncoated Ti64 substrate remains intact for a while and gradually ruptures, reaching stable three-phase contact approximately at 2.56 s.} 




During the bouncing phase, an enhanced shear flow is caused by the rapid oscillation of the bubble. As schematically depicted in Fig. (\ref{fig:rising} - \ref{fig:shear_flow}), the flow field consists of the decaying wake of the rising phase of the bubble which is supplemented by a counter-rotating vortex resulting from bouncing. These vortices and their interaction are responsible for the advection of dissolved oxygen along the bubble surface and away from the boundary layer. The latter is clearly visible in the \textcolor{black}{Figs. \ref{fig:2.410_1709} and \ref{fig:3.915_1687}} corresponding to $t =$ 0.098 s. 
As a result, a significant  bulging of the boundary layer of the dissolved oxygen concentration in vicinity of the equatorial plane appears, see \textcolor{black}{Figs. \ref{fig:2.410_4363} and \ref{fig:3.915_4334}} at $t =$ 5.392 s,  after the bouncing phase.
The remaining weak shear flow around the resting bubble, sketched in Fig. \ref{fig:shear_flow}, further advects 
the bulged boundary layer. Nevertheless, a stronger fingerprint of the bouncing remains visible in the bulged boundary layer for the coated Ti64 in comparison with the uncoated one, see 
\textcolor{black}{Figs. \ref{fig:2.410_9914} and \ref{fig:3.915_9894}} at $t =$ 16.512 s.
However, the resulting shoulder in the concentration profile at the bubble equator at $t =$ 16 s, cf. black dashed line in Fig. \ref{fig:16_20_145}, is rapidly smeared out by the convective and diffusive transport at $t =$ 20 s, which we define as the start of phase II, the actual dissolution phase.

\subsection{Phase II - Dissolution of the $O_2$ bubble: Concentration field 
around the bubble}
\label{sec:concentration_gradient}

\begin{figure}[h!]
	\centering
		\begin{subfigure}{0.21\textwidth}
		\centering
		\captionsetup{justification=centering}
		\includegraphics[height=3.15cm]{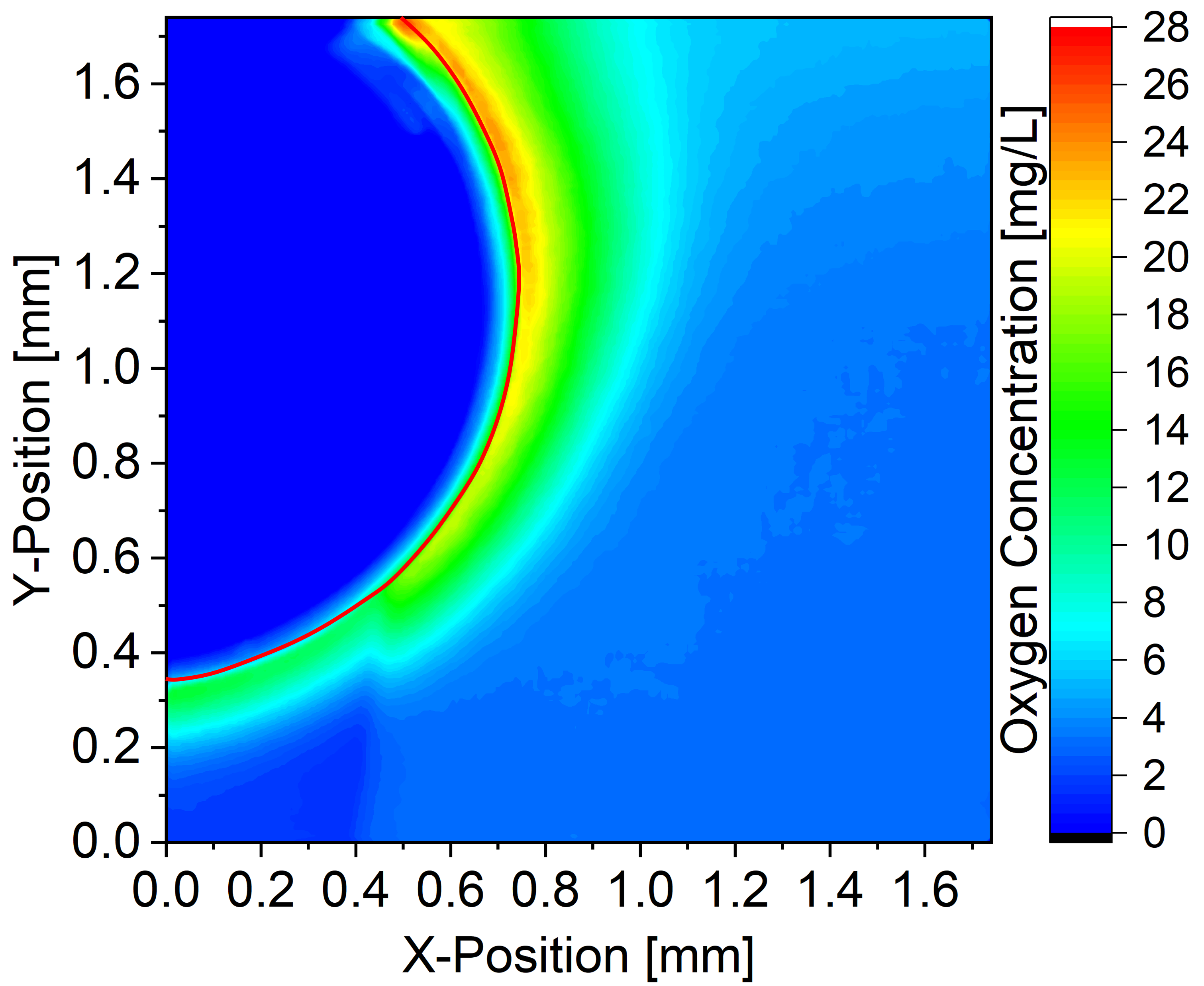}
		\caption{$t=20$ s}
		\label{fig:Q0}
	\end{subfigure}  \ \ \ \
	\begin{subfigure}{0.21\textwidth}
		\centering
		\captionsetup{justification=centering}
		\includegraphics[height=3.15cm]{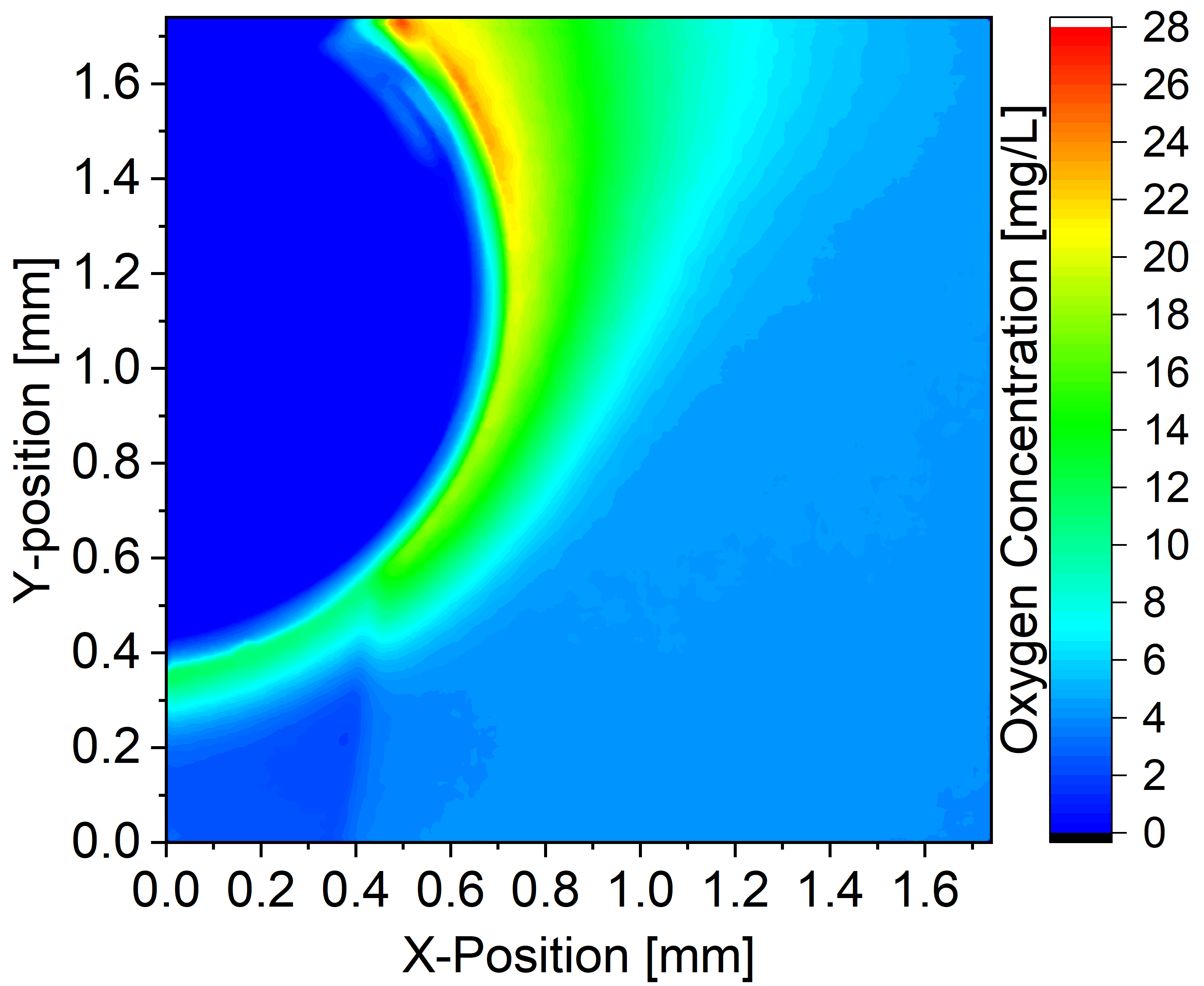}
		\caption{$t=145.01$ s}
		\label{fig:Q125.01}
	\end{subfigure}  \ \ \ \
	\begin{subfigure}{0.21\textwidth}
		\centering
		\captionsetup{justification=centering}
		\includegraphics[height=3.15cm]{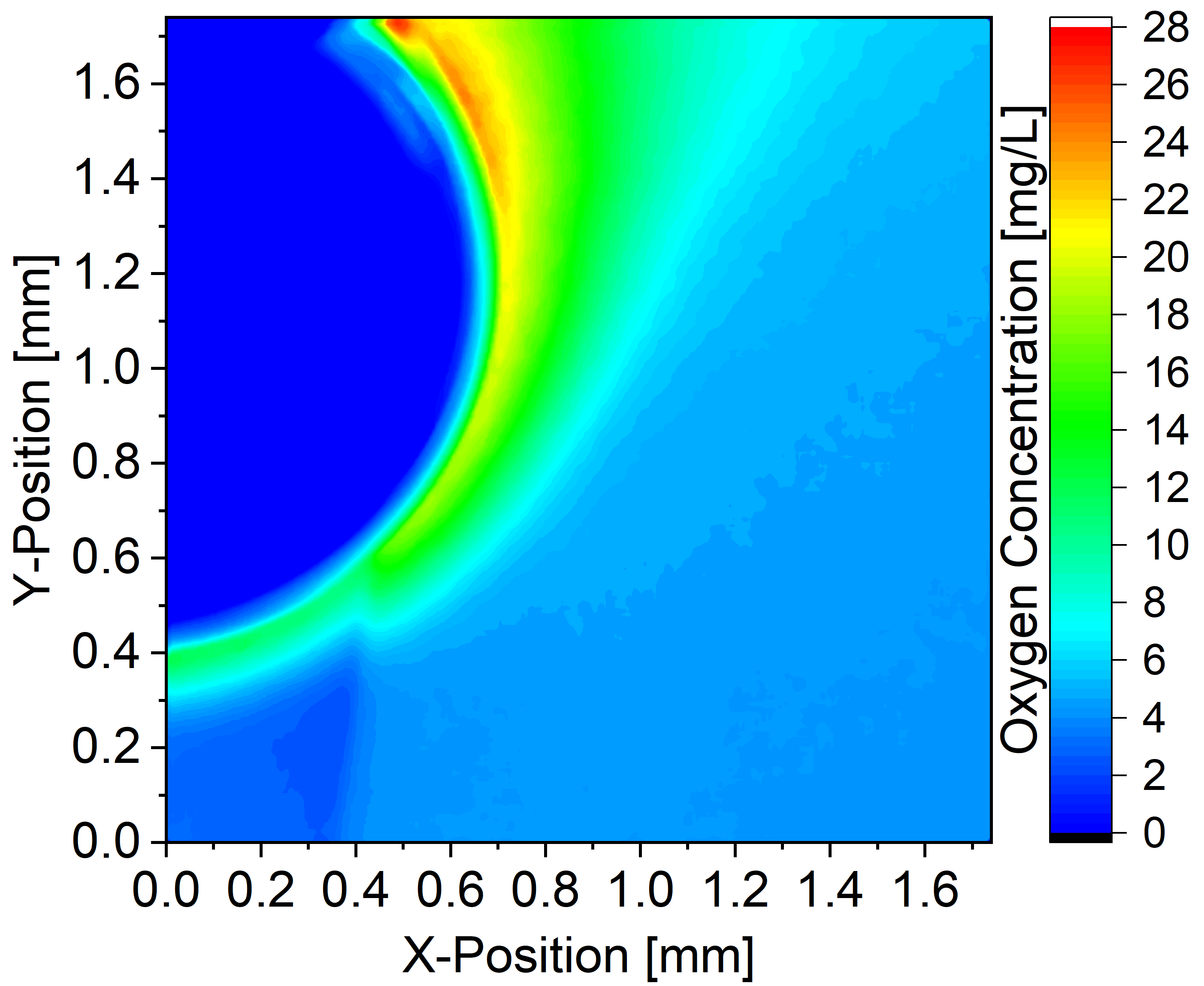}
		\caption{$t=270.02$ s}
		\label{fig:Q250.02}
	\end{subfigure}  \ \ \ \
	\begin{subfigure}{0.21\textwidth}
		\centering
		\captionsetup{justification=centering}
		\includegraphics[height=3.15cm]{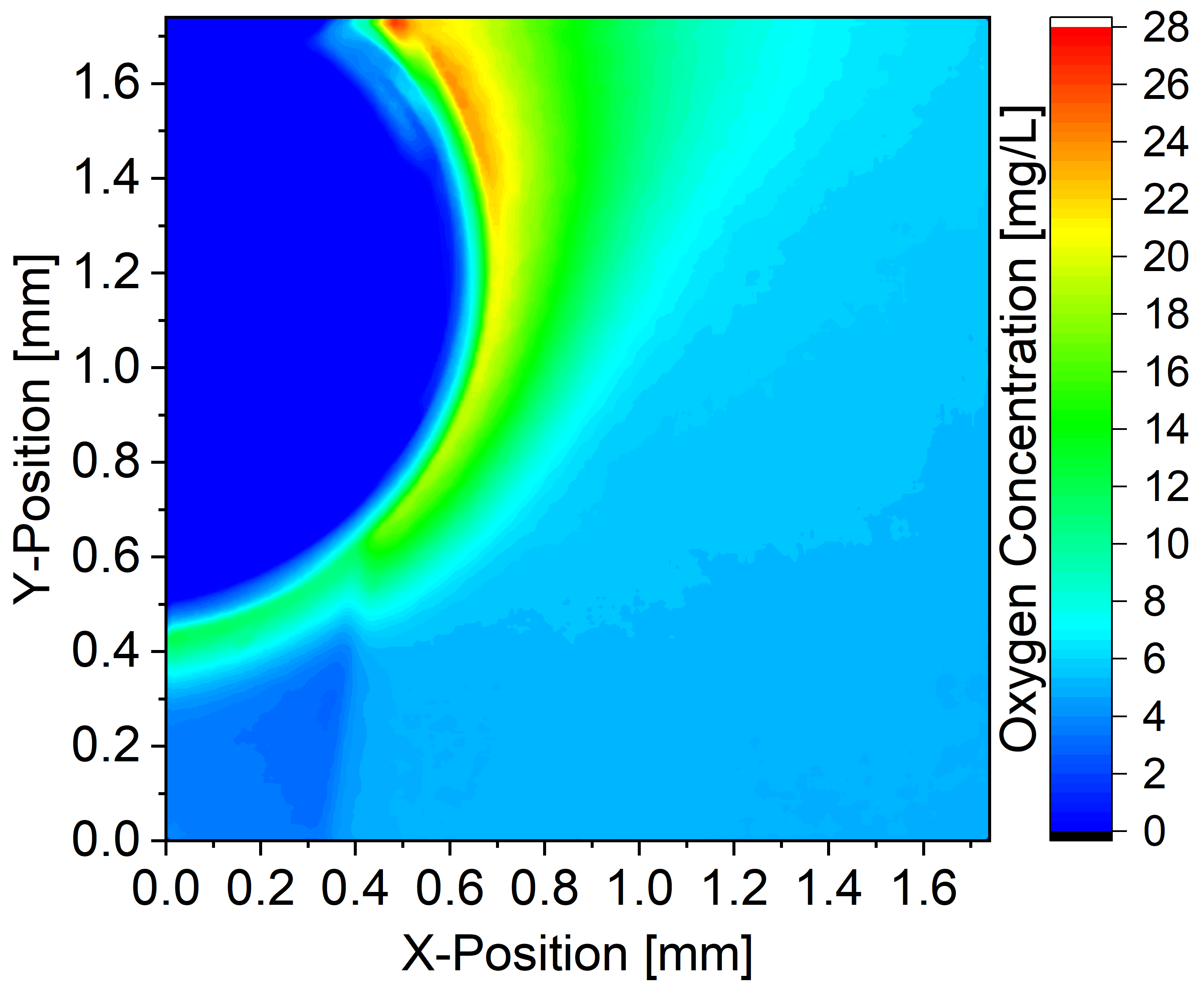}
		\caption{$t=395.03$ s}
		\label{fig:Q375.03}
	\end{subfigure}  
	\vskip0.5cm
	\centering
		\begin{subfigure}{0.21\textwidth}
		\centering
		\captionsetup{justification=centering}
		\includegraphics[height=3.15cm]{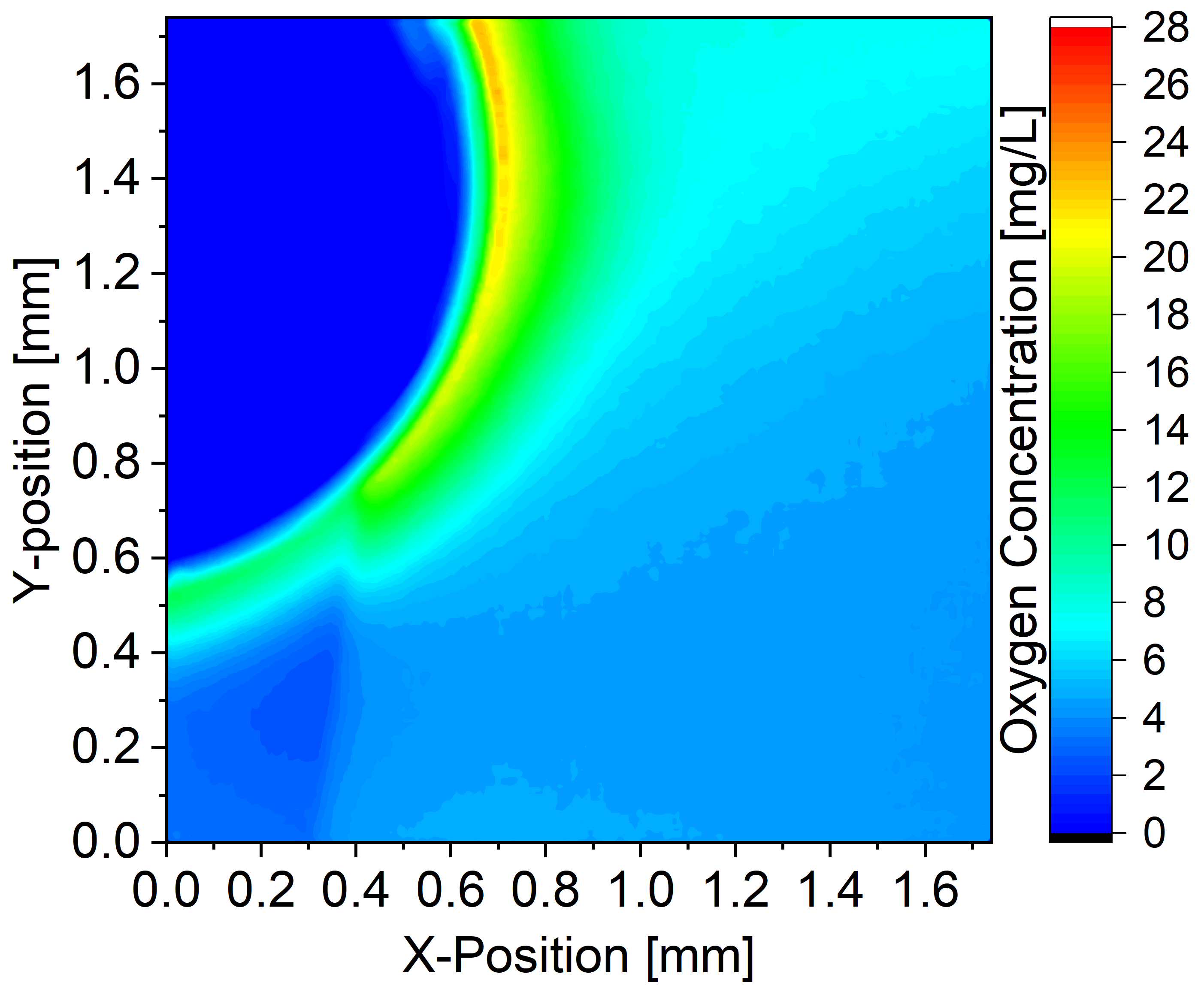}
		\caption{$t=20$ s}
		\label{fig:S0}
	\end{subfigure}  \ \ \ \
	\begin{subfigure}{0.21\textwidth}
		\centering
		\captionsetup{justification=centering}
		\includegraphics[height=3.15cm]{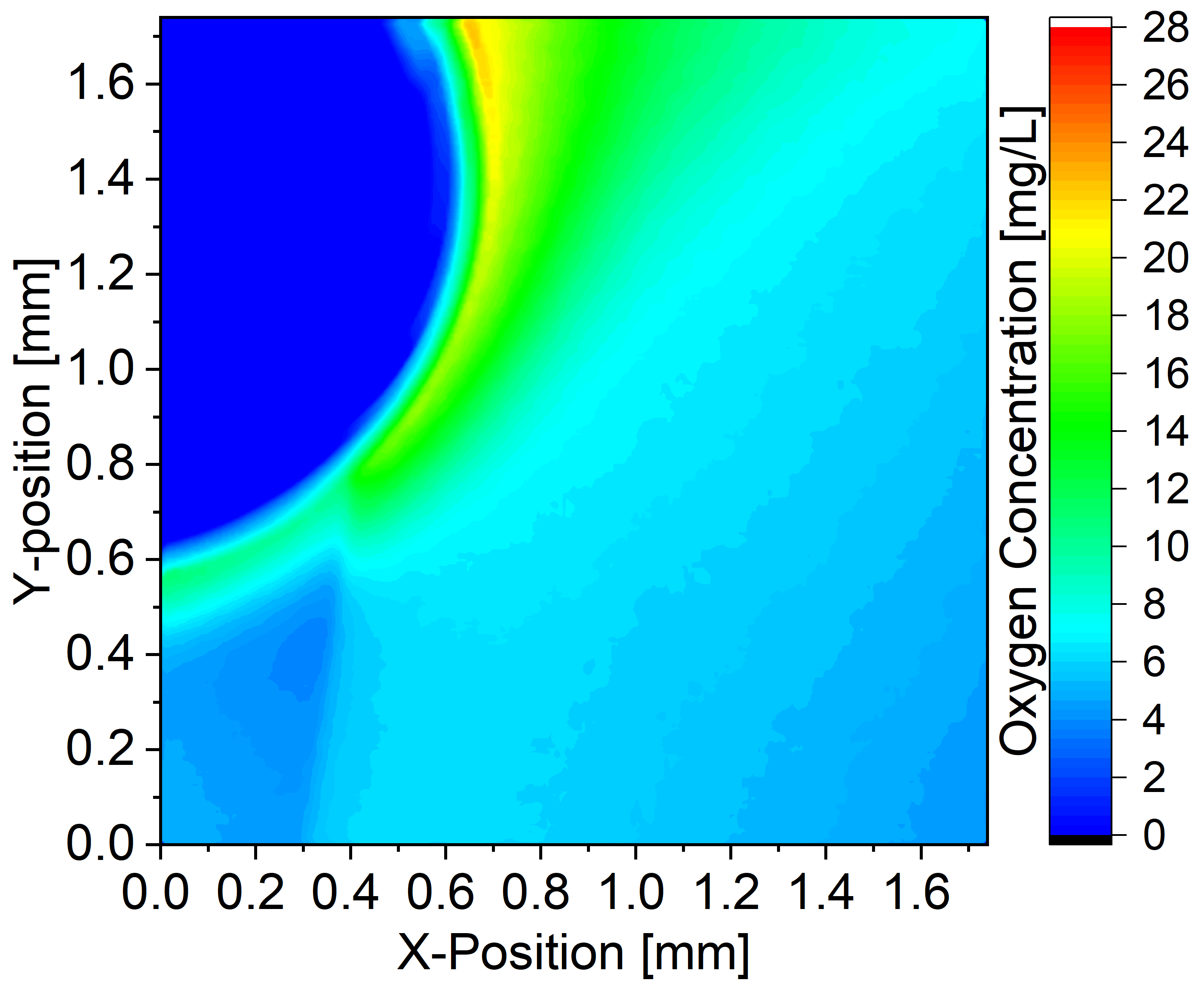}
		\caption{$t=145.01$ s}
		\label{fig:S125.01}
	\end{subfigure}  \ \ \ \
	\begin{subfigure}{0.21\textwidth}
		\centering
		\captionsetup{justification=centering}
		\includegraphics[height=3.15cm]{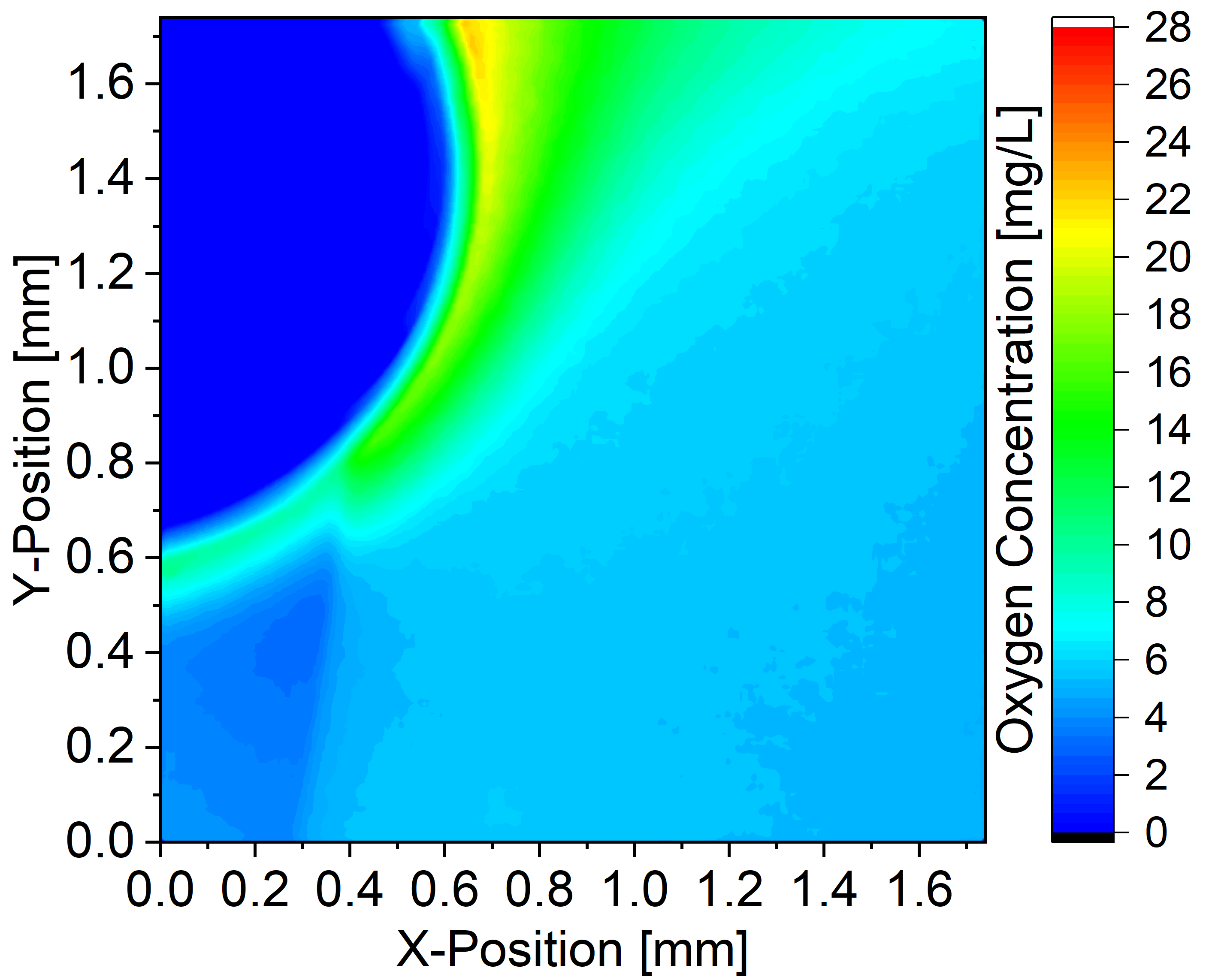}
		\caption{$t=270.02$ s}
		\label{fig:S250.02}
	\end{subfigure}  \ \ \ \
	\begin{subfigure}{0.21\textwidth}
		\centering
		\captionsetup{justification=centering}
		\includegraphics[height=3.15cm]{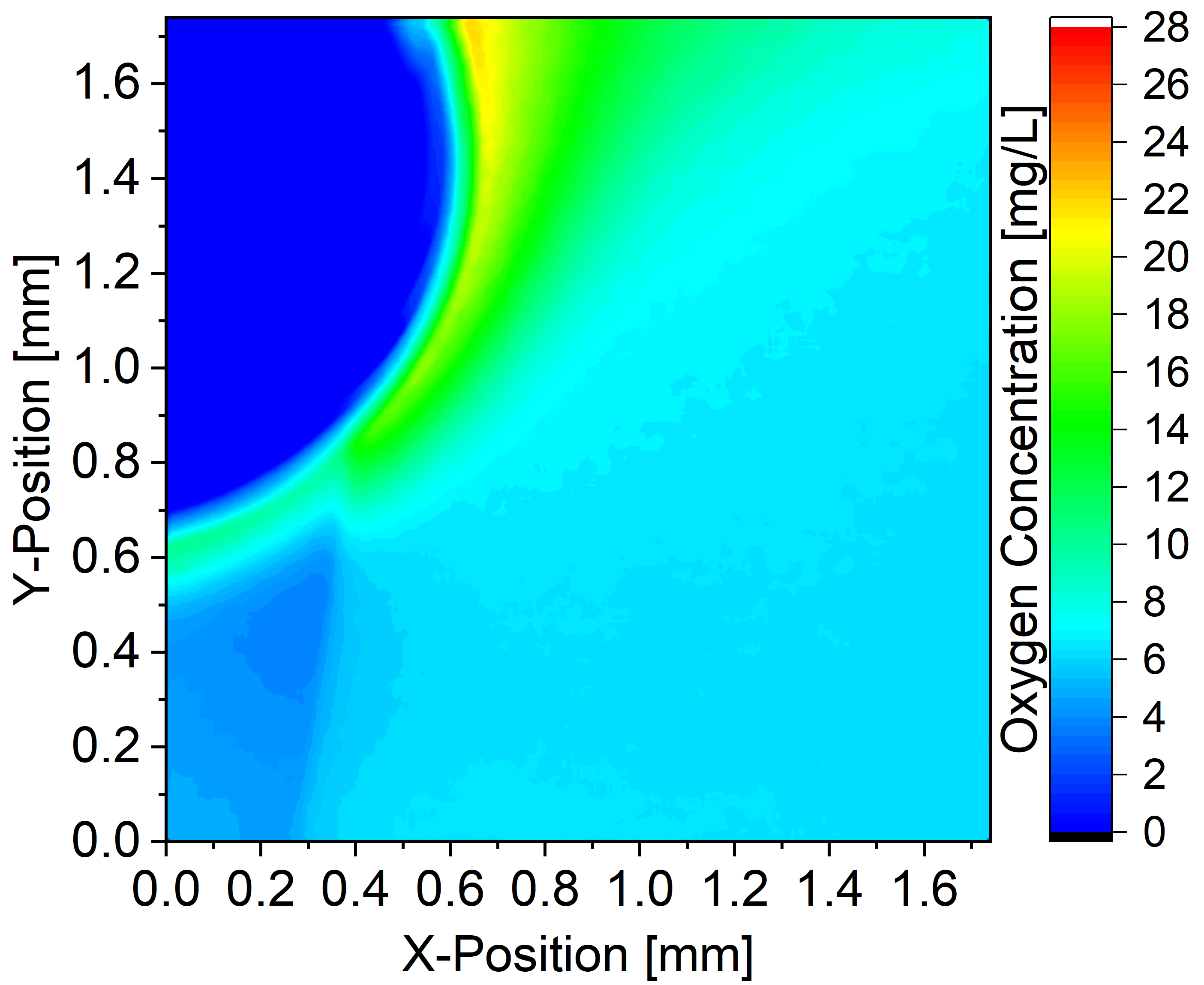}
		\caption{$t=395.03$ s}
		\label{fig:S375.03}
	\end{subfigure}  \\
	\caption{Dissolution phase II of a single oxygen bubble on an uncoated surface at an oxygen supersaturation $\zeta$ = -0.73 (a - d)
	and on a coated surface ($\zeta$ = -0.76)  at different times (e - g).
	The red arc in (a) shows the bubble boundary. \textcolor{black}{A green margin inside the bubble is due to the integral effect of fluorescence  signal, caused by the thickness of the laser sheet, in the view direction.} 
	The oxygen concentration is given by the color bar in mg/L.
	}
	\label{fig:Q}
			\end{figure}

The dissolution process of an $O_2$ bubble after bubble bouncing
is documented in Fig. \ref{fig:Q} in terms of the two-dimensional concentration of dissolved oxygen at the uncoated Ti64 sample (contact angle = $51^\circ \pm 2^\circ$)  in Fig. (\ref{fig:Q0} - \ref{fig:Q375.03}) and at the coated Ti64 sample (contact angle = $70.5^\circ \pm 1.5^\circ$)  in Fig. (\ref{fig:S0} - \ref{fig:S375.03}).
The figures show how the boundary layer of dissolved oxygen propagates from the $O_2$ bubble towards the bulk. Over time, there is a visible increase in the accumulation of dissolved oxygen along the Ti64 surface, which is caused by hindered diffusion due to the presence of the Ti64 sample.
This accumulation is more pronounced for the coated than the uncoated sample.
\begin{figure}[h!]
	\centering
	\begin{subfigure}{0.46\textwidth}
		\centering
		\captionsetup{justification=centering}
		\includegraphics[height=5.25cm]{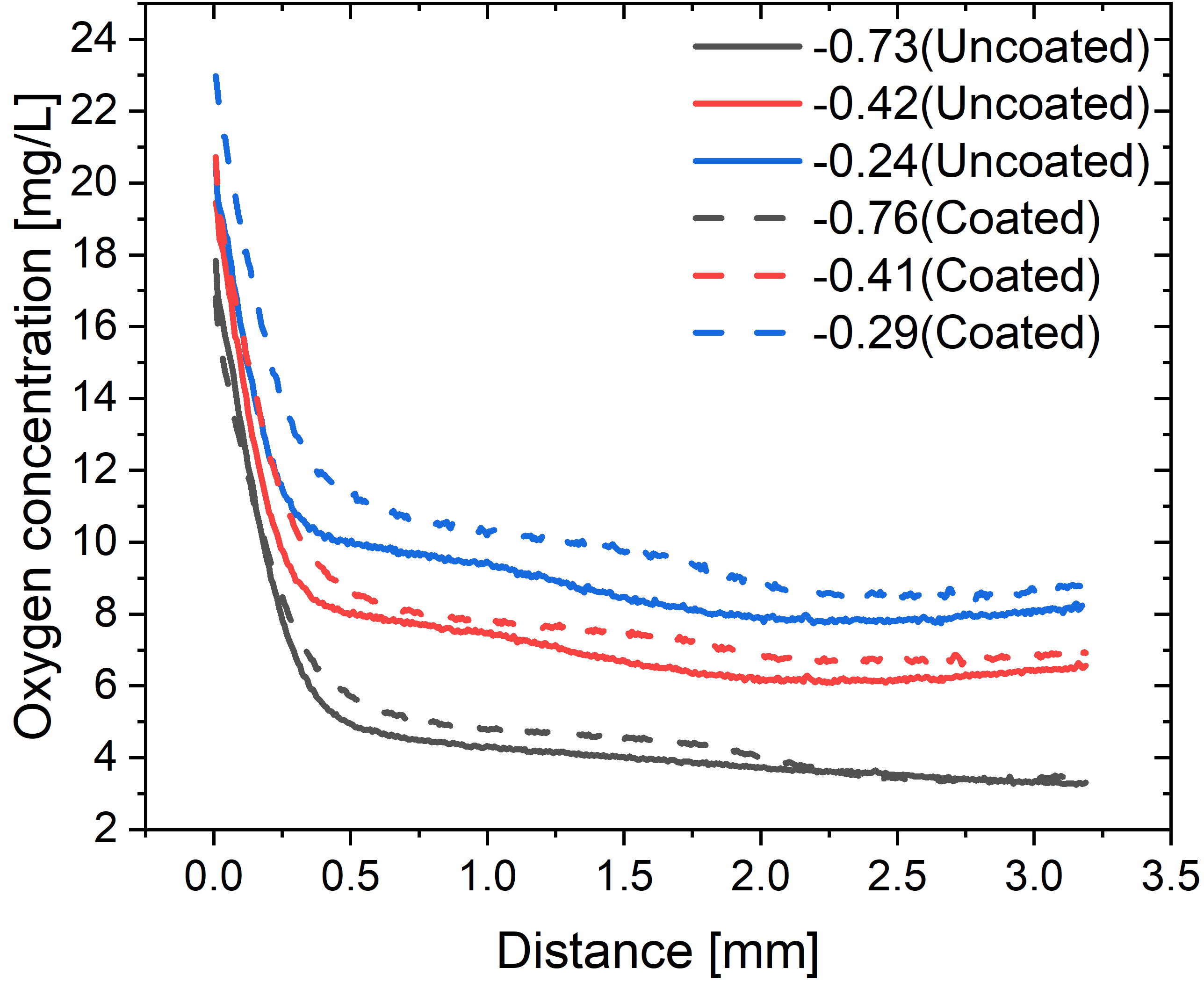}
		\caption{}
		\label{fig:concentration_qs}
	\end{subfigure} 
	\begin{subfigure}{0.46\textwidth}
		\centering
		\captionsetup{justification=centering}
		\includegraphics[height=5.25cm]{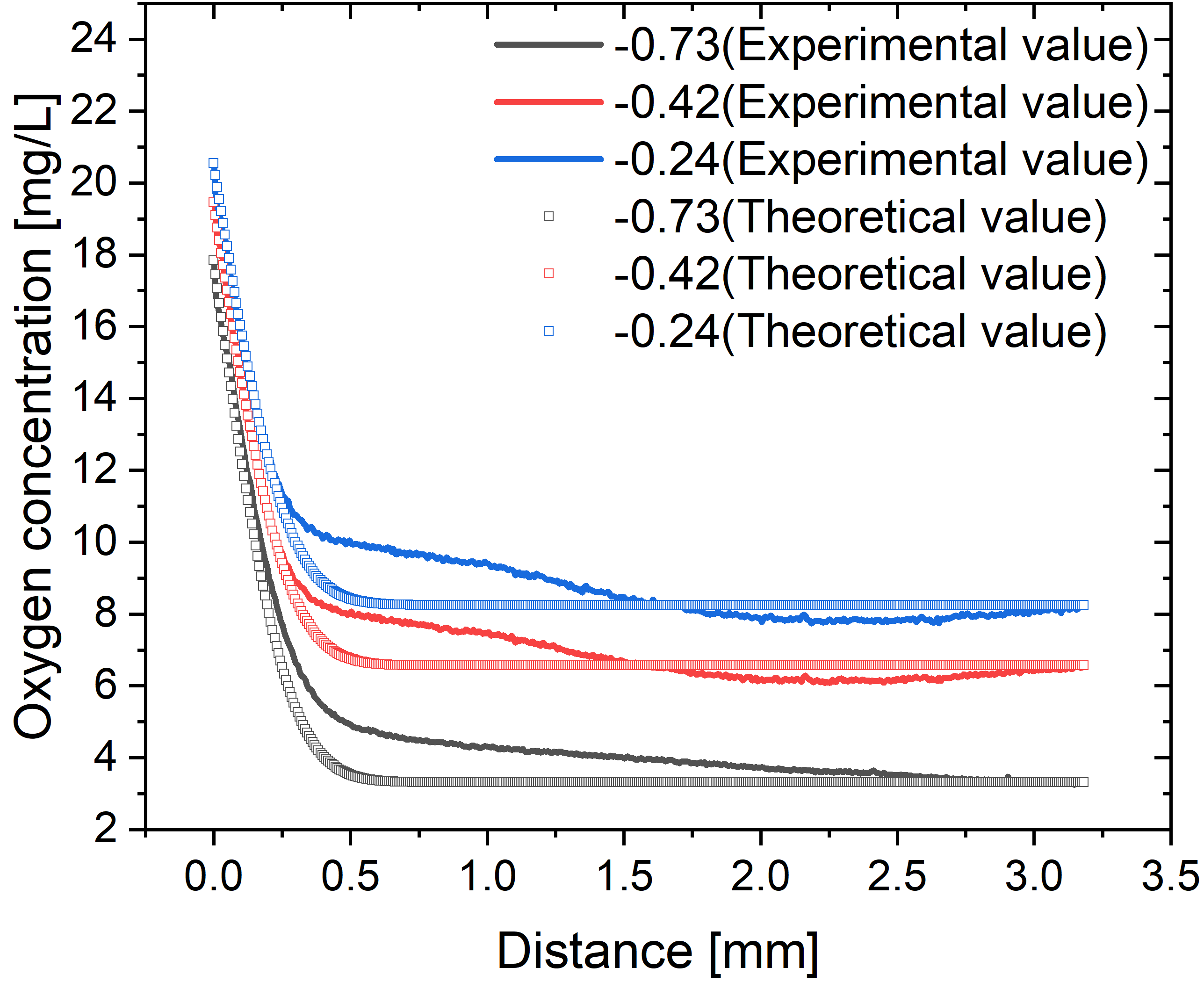}
		\caption{}
		\label{fig:fit_concentration}
	\end{subfigure}   
	\begin{subfigure}{0.46\textwidth}
		\centering
		\captionsetup{justification=centering}
		\includegraphics[height=5.25cm]{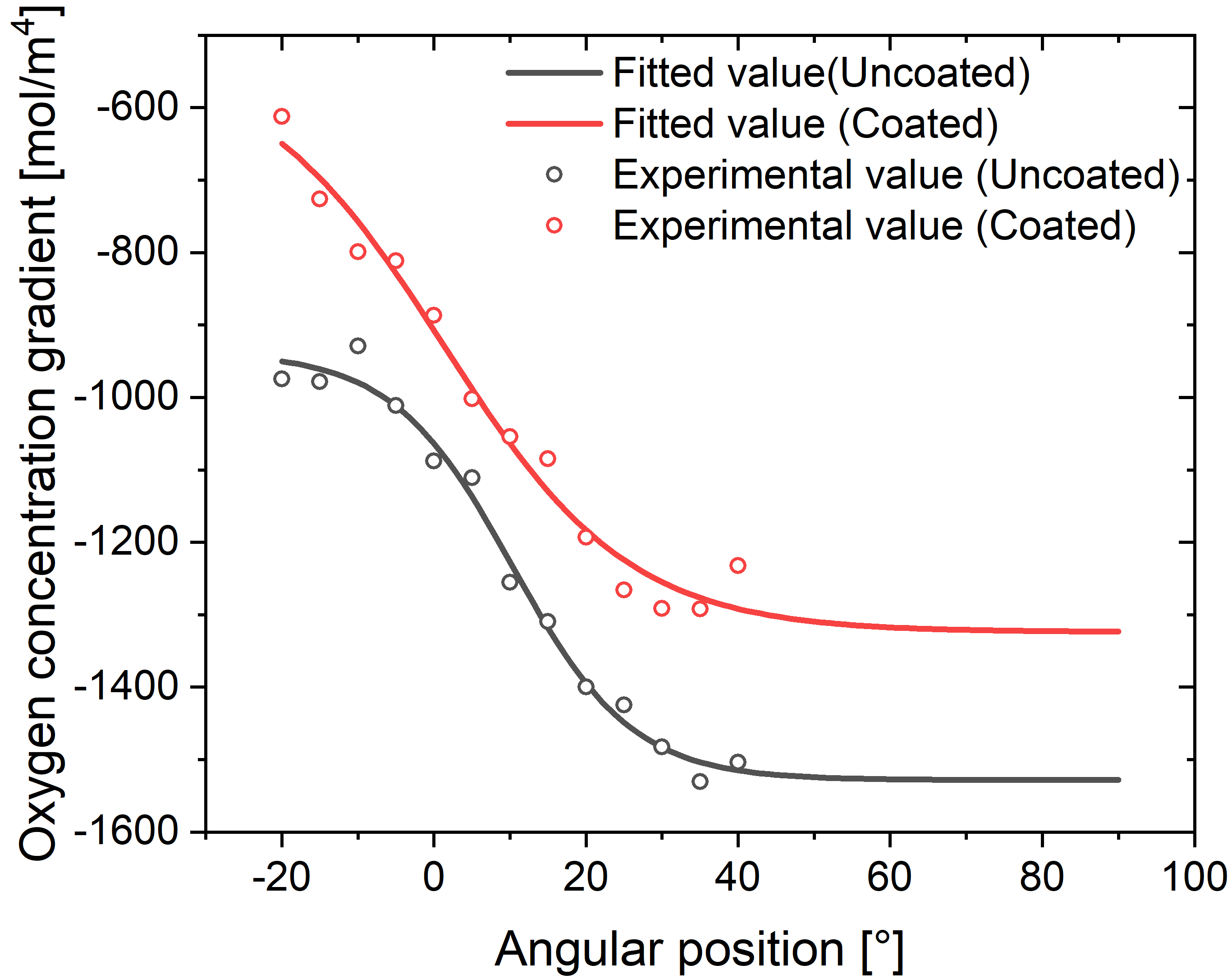}
		\caption{}
		\label{fig:con_gra_ang_pos}
	\end{subfigure} 
	\begin{subfigure}{0.46\textwidth}
		\centering
		\captionsetup{justification=centering}
		\includegraphics[height=5.25cm]{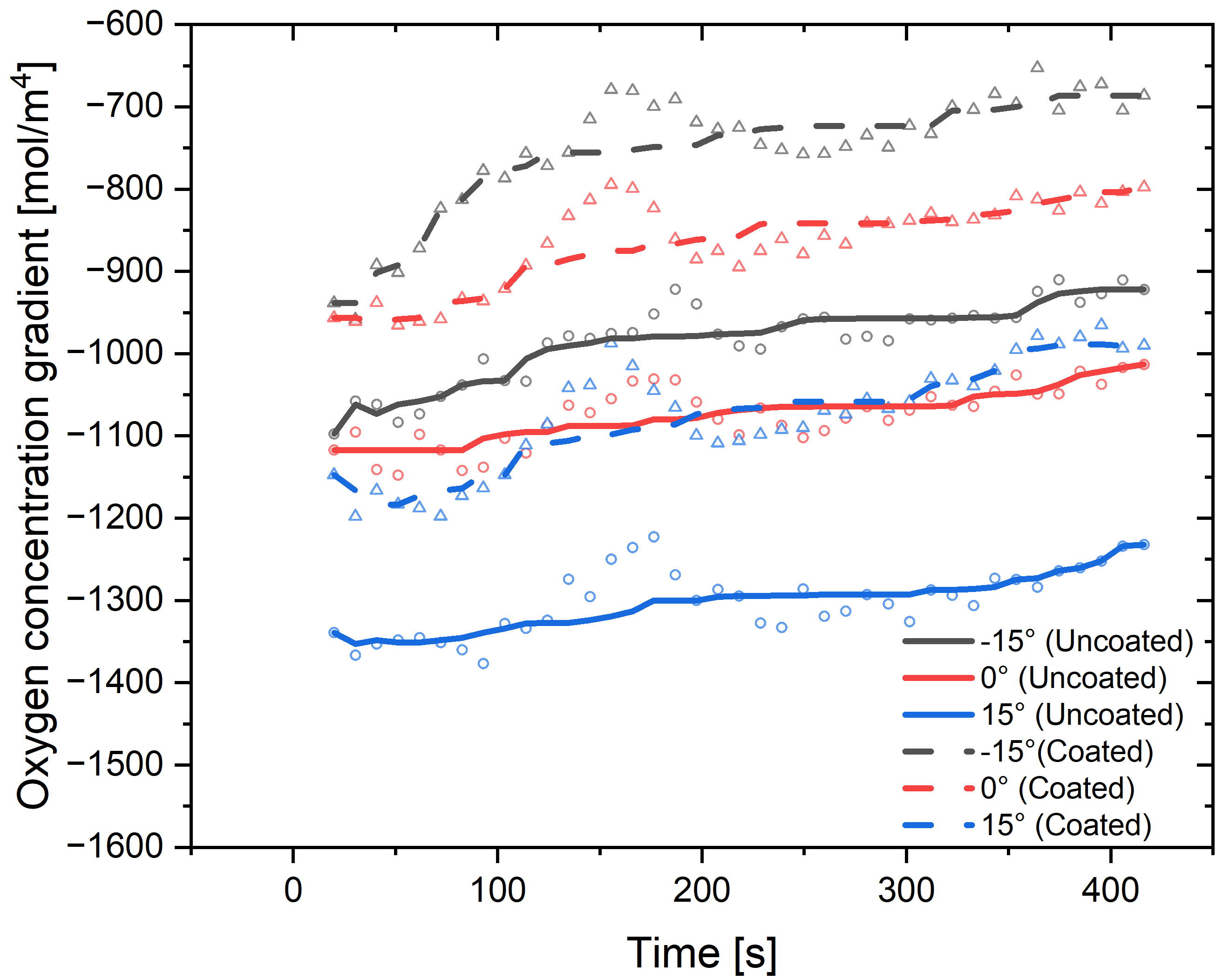}
		\caption{}
		\label{fig:con_tim_ang_pos}
	\end{subfigure}  
	\caption{Dissolution phase II: (a) Radial profiles of the dissolved oxygen concentration for an uncoated (solid lines) and a coated (dashed lines) Ti64 substrate at different oxygen supersaturation at angular position 0\textdegree{} at 228 s; (b) Comparison of the measured radial oxygen concentration profiles (solid lines) and calculated (open squares, \textcolor{black}{ Eq. \ref{eq:cDistri}})   values at angular position 0\textdegree{} at 228 s for the uncoated cases; (c) Concentration gradient vs. angular positions 
	for the uncoated ($\zeta \sim -0.73$) and coated ($\zeta \sim -0.76$) cases; (d) Concentration gradient vs. time at three angular positions of -15\textdegree{}, 0\textdegree{}, and 15\textdegree{} for the uncoated ($\zeta \sim -0.73$) and coated ($\zeta \sim -0.76$) cases. In (d), the thick solid and dashed lines represent the smoothed curve obtained from the original data points.}
	\label{fig:concentration_profile}	
\end{figure}

This is examined in Fig. \ref{fig:concentration_qs} in which the 1D profiles, $c(x)$, of the measured dissolved oxygen concentration  are plotted in the equatorial position of the oxygen bubble,  position 0\textdegree{}, at different oxygen supersaturation for both the uncoated and coated substrates. The time instant is $t =$ 228 s, i.e. between Figs. \ref{fig:Q125.01} and \ref{fig:Q250.02}
, and Figs. \ref{fig:S125.01} and \ref{fig:S250.02} respectively.
The boundary layer of the  dissolved oxygen, characterized by a high radial concentration gradient, is clearly visible in the range $0 < x \lesssim 0.75 \, \text{mm}$. 
By comparing the oxygen concentrations between the uncoated and coated cases at similar $\zeta$,  we find that the concentration profiles for the uncoated substrate always stay below those for the coated case. This implies a larger concentration difference $\Delta c_{O2}$, hence a larger mass transfer in the uncoated case.
As expected,  $\Delta c_{O2}$, scales with the supersaturation, $\zeta$. The larger $|\zeta|$, the larger $\Delta c_{O2}$, because the bulk oxygen concentration (at large distance) is lower in that case.

The oxygen concentration profiles of the experiments (Fig. \ref{fig:concentration_qs}) are compared in  Fig. \ref{fig:fit_concentration} for the uncoated case with the  theoretical profiles for a spherical and non-substrate bound $O_2$ bubble, calculated by Eq. \ref{eq:cDistri}. A good agreement is seen within the boundary layer as well as in the bulk. 
Between boundary layer and bulk, characteristic deviations in between are observed. 
Probably, these changes are a fingerprint of  the remaining weak convective flow shown schematically in Fig. \ref{fig:shear_flow}, which advects  $O_2$-rich liquid from the boundary layer upwards, thereby lifting the profile above the theoretical one. The downflow, forced by continuity, is then responsible for the drop of the experimental values below the theoretical ones.
\textcolor{black}{Buoyancy effects must be considered as a further possible contribution to convection. Since mass transfer of oxygen takes place, the density of the solution will change according to the volume expansion coefficient which characterizes the relationship between gas concentration and liquid density. For oxygen, the density of the aqueous solution increases
with increasing concentration, whereas for nitrogen, it is the opposite \cite{watanabe1985influence}. Thus the liquid density in the oxygen boundary layer exceeds that of the bulk. This would imply a downward flow of this liquid close to the bubble,
which is not compatible with the \textcolor{black}{observed} shape of the concentration profile. This strongly support the hypothesis} \textcolor{black}{that the deviations in the concentration profiles of Fig. \ref{fig:fit_concentration} are caused by the weak decaying convection of the bouncing phase while buoyancy effects can  be neglected.} 


The average radial gradient, $\partial c/\partial r$
, of the dissolved oxygen concentration in the boundary layer ($r< $ 0.5 mm) as a function of the polar angle $\theta$ is plotted in Fig. \ref{fig:con_gra_ang_pos} as open circles. Solid lines correspond to a fit, the parameter of which are given in the supporting information. Note that a polar angle of 0\textdegree{} corresponds to the equator of the bubble. Characteristic for both the uncoated and coated cases is a non-homogenous mass transfer, i.e. a changing gradient along the bubble surface, in contrast to the dissolution of a non-bound spherical bubble. In general, the gradients are weaker close to the substrate, where an accumulation of dissolved oxygen occurs (see Fig. \ref{fig:Q}), and rise towards the lower pole of the bubble opposite to the substrate.
Most notable, the angular profile of $\partial c/\partial r$ 
for the uncoated case (black symbols) stays below the coated one, i.e. it is characterized by higher negative gradients, which lead to a higher oxygen flux out of the bubble (cf. Eq. \ref{eqn:j}).

Finally, in Fig. \ref{fig:con_tim_ang_pos} we plot the temporal evolution of $\partial c/\partial r$
at selected polar angles and fixed supersaturation. 
\textcolor{black}{Throughout the measurement period, we note a smooth decrease in all positions due to continuous mass transfer of oxygen into the bulk.}

\subsection{Phase II - Dissolution: $O_2$ bubble geometry as function of time}
\label{sec:Geometry_bubble}
Next we analyse how the shape of the $O_2$ bubbles changes under the virtue of the mass transfer from the bubble into the bulk
, examined in the previous section.
Table \ref{tab:initial_values} summarizes the $O_2$ bubble geometry at $t= 20$ s, sufficiently after termination of the bouncing phase I, in terms of contact angle,  $\theta_0$, diameter of the three-phase contact line, $w_0$,  and diameter of the bubble, $D_0$,
for the uncoated and coated cases at the  supersaturations, $\zeta$, studied. We consider these values as initial values for the dissolution phase II.
\begin{table}[h!]
	\centering
	\caption{Initial values of oxygen bubble diameter ($D_0$), diameter of three-phase contact line ($w_0$) and contact angle ($\theta_0$) at $t=20 $ s for the uncoated and coated cases at different oxygen supersaturation levels}
		\begin{tabular}{ c c c c c c c } 	\hline 
		 & \multicolumn{3}{c}{Uncoated} & \multicolumn{3}{c}{Coated} \\\hline
        Oxygen supersaturation level & -0.73 & -0.42 & -0.24 & -0.76 & -0.41 & -0.29\\ \hline 
        $\theta_0$(°) & 52.60 & 49.87 & 50.25 & 69.54 & 71.55 & 69.97\\
        $w_0$($\mathrm{mm}$) & 1.39 & 1.34 & 1.35 & 1.75 & 1.78 & 1.74\\ 
        $D_0$($\mathrm{mm}$) & 1.79 & 1.79 & 1.79 & 1.90 & 1.91 & 1.89 \\ 
        \hline
	\end{tabular}
	\label{tab:initial_values}
\end{table}
The hydrophicity of the substrate \textcolor{black}{(coated Ti64 substrate)} forces higher contact angle, $\theta_0$, and also a higher contact line diameter, $w_0$, as reflected by the values in Table \ref{tab:initial_values}. This also leads to a larger bubble diameter, $D_0$, of the bubbles at the  
coated Ti64 substrate. Thus, the wettability of the sample surface dictates the bubble geometry, and hence the bubble surface area available for mass transfer.

For a better comparison between the dissolution dynamics of the $O_2$ bubble at the different substrates, we refer the changes of all three bubble quantities, $\theta$, $w$ and $D$,
to their initial values in Table \ref{tab:initial_values}, i.e.:
\begin{equation}
   D_r=\frac{(D-D_0)}{D_0}, \, \, w_r=\frac{(w-w_0)}{w_0}, \, \,
   \theta_r=\frac{(\theta-\theta_0)}{\theta_0}.
 \label{eq:D_r}
\end{equation}
\begin{figure}[h!]
	\centering
		\begin{subfigure}{0.45\textwidth}
		\centering
		\captionsetup{justification=centering}
		\includegraphics[height=5.25cm]{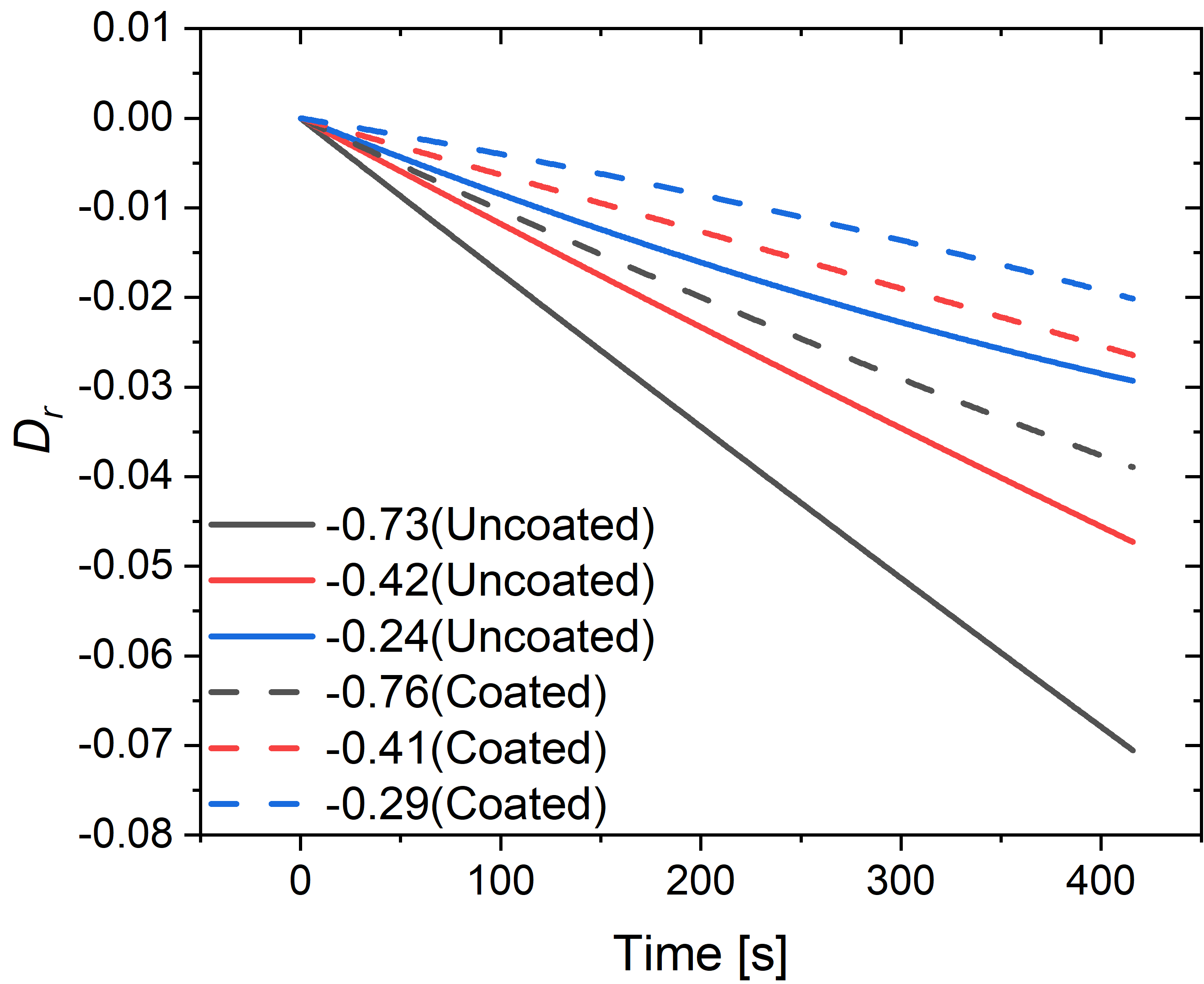}
		\caption{}
		\label{fig:dim_dia}
	\end{subfigure}  \ \ \ \
	\begin{subfigure}{0.45\textwidth}
		\centering
		\captionsetup{justification=centering}
		\includegraphics[height=5.25cm]{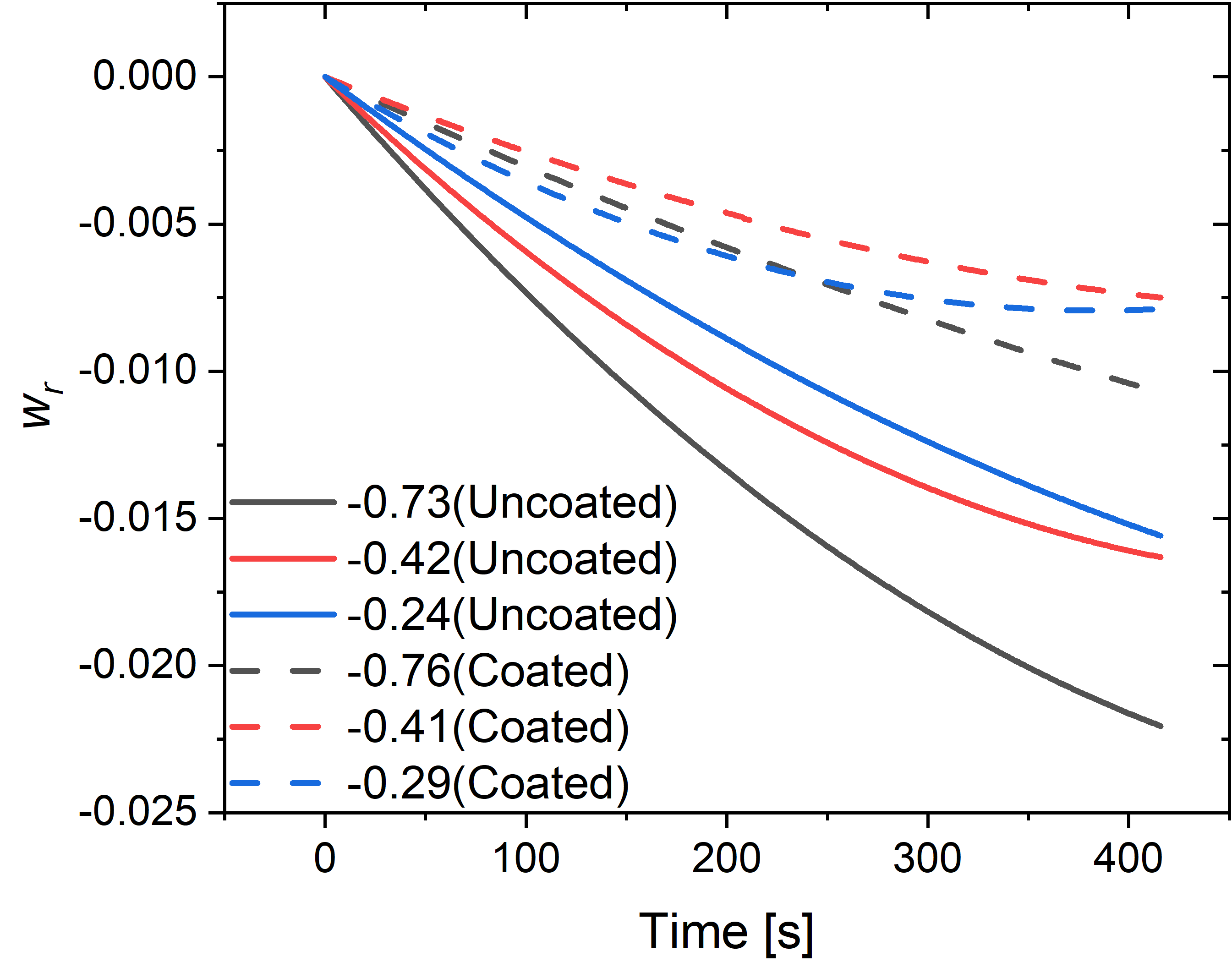}
		\caption{}
		\label{fig:dim_cl}
	\end{subfigure}  \\
	\begin{subfigure}{0.45\textwidth}
		\centering
		\captionsetup{justification=centering}
		\includegraphics[height=5.25cm]{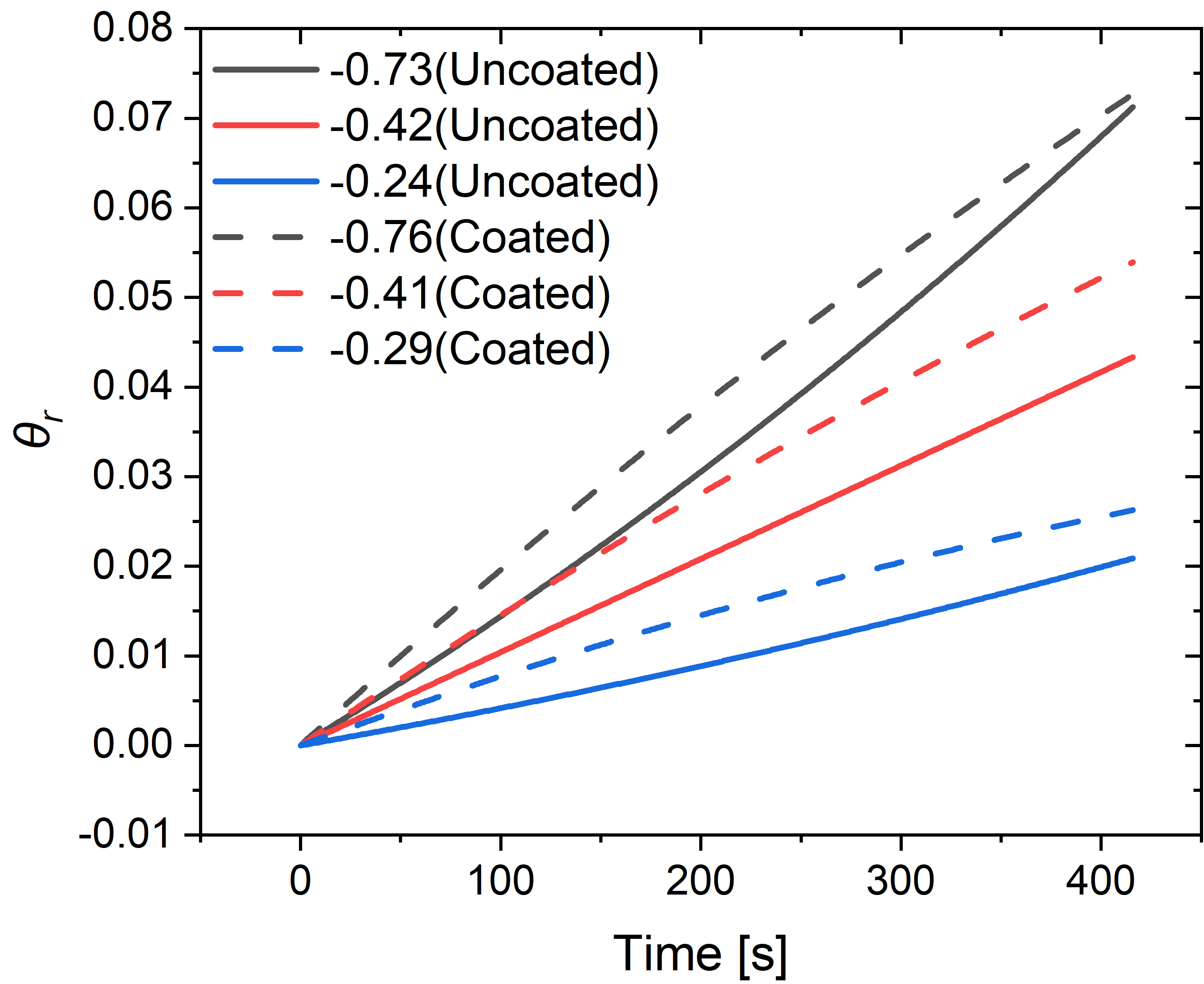}
		\caption{}
		\label{fig:dim_ca}
	\end{subfigure}  \ \ \ \
	\begin{subfigure}{0.45\textwidth}
		\centering
		\captionsetup{justification=centering}
		\includegraphics[height=5.25cm]{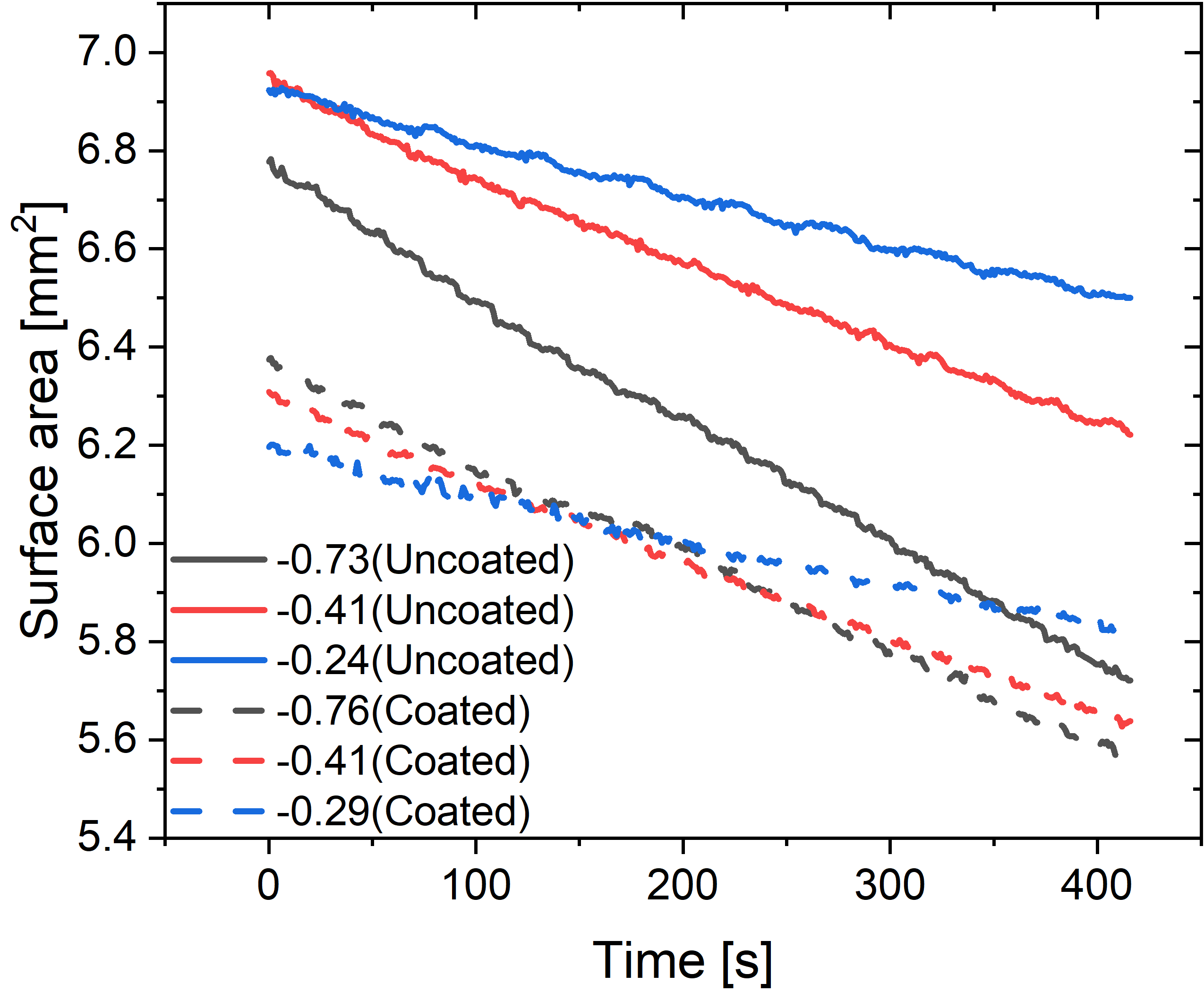}
		\caption{}
		\label{fig:surface_area}
	\end{subfigure}  \\
	\caption{Changes in the geometry of the oxygen bubble during dissolution at different supersaturations (see legend), expressed via relative values (cf. Eq. \ref{eq:D_r}) of diameter $D_r$ (a),  diameter of three-phase contact line $w_r$ (b) and  contact angle $\theta_r$ (c) 
	for the uncoated (solid lines) and the coated Ti64 sample (dashed lines). (d) Change in surface area with time for both cases.}
	\label{fig:dim}		
\end{figure} 
Fig. \ref{fig:dim_dia} shows, for the same supersaturation, that the diameter reduction due to dissolution is roughly twice as high for the uncoated case (up to 7 \% over the observation timespan $t =$ 400 s) compared to the coated one. This holds true for all supersatuations studied. 
Similar trends are also found for the contact line diameter plotted in Fig. \ref{fig:dim_cl}. However, the reduction amount is with less than 2.5 \% smaller compared to the diameter. 
By contrast, the contact angle of the bubbles increases during the dissolutions, irrespectively on whether it takes place on the uncoated or the coated substrate, cf.  Fig. \ref{fig:dim_ca}. However, the trend, namely larger changes with larger supersaturation, is the same as for $D_r$ and $w_r$. The increase of $\theta_r$ of up to 7 \% is comparable to the changes in the diameter. To conclude, the dissolution of the $O_2$ bubbles proceeds via 
contact angle increase and a shrinking contact line diameter, \textcolor{black}{i.e.  along with a moving,  non-pinned contact line}.




To demonstrate that the changes in $D_r, w_r$ and $\theta_r$ are related with the surface area of the bubble, available for the mass transfer, we plot in Fig. \ref{fig:surface_area}
the dimensional surface area of the $O_2$ bubble.
Again the negative slope of all curves scales with the supersaturation $\zeta$. The bigger the $|\zeta|$ is, the larger the decrease. As the bubble surface area is higher in the uncoated case,  the  mass transfer is larger in this case, cf. Eq. \ref{eqn:dmdt1}.

\section{Conclusions}
\label{sec:Conclusions}

The dissolution process of an oxygen bubble, sitting on a Ti64 substrate, is studied for different undersaturations (negative supersaturations) of the surrounding ultrapure water and different wettabilities of the Ti64 sample.
The mass transfer at the oxygen bubble is quantified simultaneously by measuring the bubble geometry via shadowgraphy, and concentration field of dissolved oxygen around the bubble by using planar laser-induced fluorescence (PLIF).

The dissolution process in the present work divides into two phases. First, a short bouncing phase I, $0<t<250 $ ms, takes place which comprises the short rise of the $O_2$ bubble towards the Ti64 sample and decaying bouncing oscillations until its position is fixed there. 
After letting decay the perturbations brought by the bouncing, the second dissolution phase of the $O_2$ bubble is analysed in a time window, $20\,s < t < 400\,s$.

While the $O_2$ bubble performs up to six bouncing periods at the uncoated (hydrophilic) Ti64 sample,
a rupture of the thin water film between bubble and substrate occurs already after the third period at the coated (more hydrophobic) Ti64 sample. As a result, the bubbles remain attached at the sample.
The system of vortices at the bouncing bubble does not only redistribute the boundary layer of dissolved oxygen surrounding the bubble surface but also advects parts into the bulk of ultrapure water. This sets the initial conditions for the later second dissolution phase which starts after the oscillations of the bouncing phase have been damped out.

During dissolution, the $O_2$ mass transfer proceeds non-homogenously across the surface of the bubble with concentration gradients becoming steeper with increasing polar angle. The weakest gradients occur close to the sample interface as the diffusion of dissolved oxygen towards the bulk is hindered there, and an oxygen accumulation takes place. For all polar angles, the magnitude of the concentration gradient, hence the mass transfer, is higher  for the uncoated case compared to the coated one.

This is in line with the measured changes in the bubbles geometry. Here, the dissolution, i.e. the reduction of the bubble diameter, proceeds via contact line motion and contact angle increase. For all three supersaturation levels investigated, the magnitude of changes in the diameter of bubble and contact line 
are bigger for the uncoated than the coated case and increase with the magnitude of supersaturation of the ultrapure water with dissolved oxygen.
Finally, we have shown that the main reason for the notable differences in the $O_2$ bubble dissolution dynamics depending on the substrate wettability are the differences in the surface area of the bubble available for mass transfer.
Due to the smaller contact angle at the uncoated Ti64 sample, the surface area is higher there. Hence, dissolution proceeds faster at uncoated Ti64 sample. Thus, hydrophilic substrates enable a faster dissolution-driven bubble removal.

\section{Declaration of competing interest} The authors declare that they have no known competing financial interests or personal relationships that could have appeared to influence the work reported in this paper.

\section{ACKNOWLEDGMENTS}
The authors would like to acknowledge 
fruitful discussions with Prof. Jens-Uwe Repke, Dr. G. Brösigke, Johann Weigelt (TU Berlin), Dr. Gerd Mutschke, Dr. Mengyuan Huang, Yifan Han and Ming Xu (Helmholtz-Zentrum Dresden-Rossendorf). Financial support by the German Federal Ministry of Education and Research (BMBF) within the framework of the project H2Giga, Sinewave/OxySep (Grant no. 03HY123D) is gratefully acknowledged.

\section{Supplemental Material}
\label{sec:Supplemental Material}

\subsection{Shadowgraphs of bouncing $O_2$ bubbles at Ti64 substrates of different wettability}

\begin{figure}[h!]
	\centering
	    
		\begin{subfigure}{0.15\textwidth}
		\centering
		\captionsetup{justification=centering}
		\includegraphics[height=2cm]{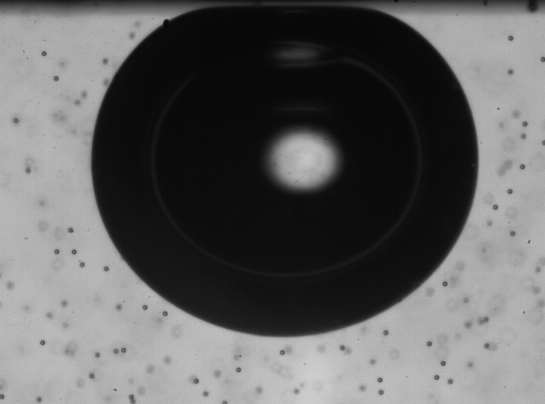}
		\caption{$t=0.0530s$}
		\label{fig:q_145}
	\end{subfigure}  \ \ \ \
	\begin{subfigure}{0.15\textwidth}
		\centering
		\captionsetup{justification=centering}
		\includegraphics[height=2cm]{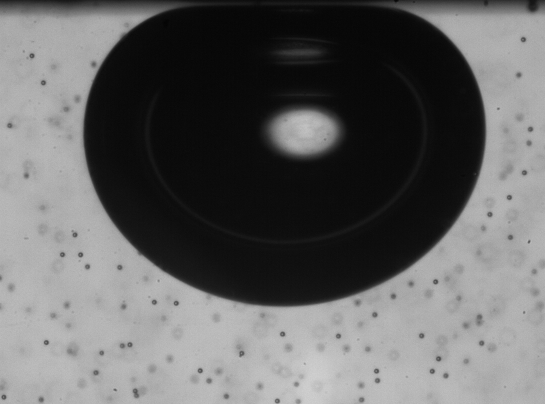}
		\caption{$t=0.0545s$}
		\label{fig:q_148}
	\end{subfigure}    \ \ \ \
	\begin{subfigure}{0.15\textwidth}
		\centering
		\captionsetup{justification=centering}
		\includegraphics[height=2cm]{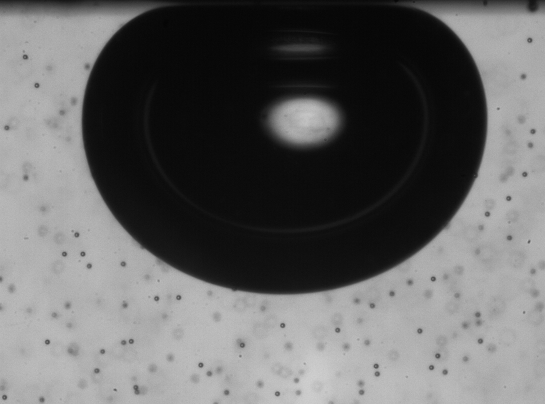}
		\caption{$t=0.0560s$ }
		\label{fig:q_151}
	\end{subfigure}    \ \ \ \
	\begin{subfigure}{0.15\textwidth}
		\centering
		\captionsetup{justification=centering}
		\includegraphics[height=2cm]{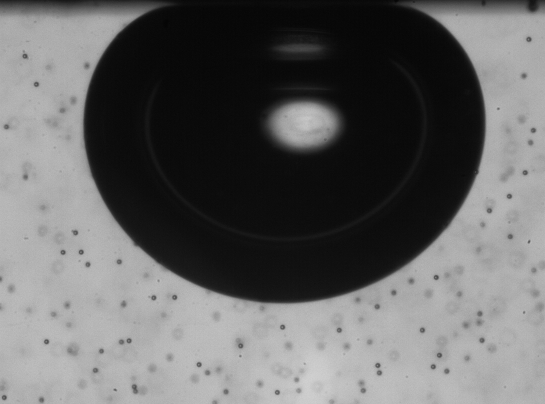}
		\caption{$t=0.0575s$ }
		\label{fig:q_153}
			\end{subfigure}    \ \ \ \
	\begin{subfigure}{0.15\textwidth}
		\centering
		\captionsetup{justification=centering}
		\includegraphics[height=2cm]{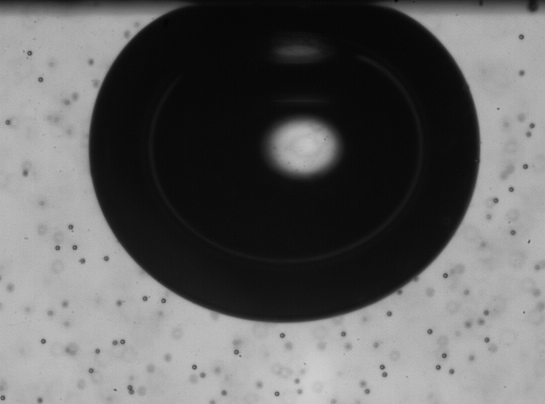}
		\caption{$t=0.0590s$ }
		\label{fig:q_156}
	\end{subfigure}  \\
	\begin{subfigure}{0.15\textwidth}
		\centering
		\captionsetup{justification=centering}
		\includegraphics[height=2cm]{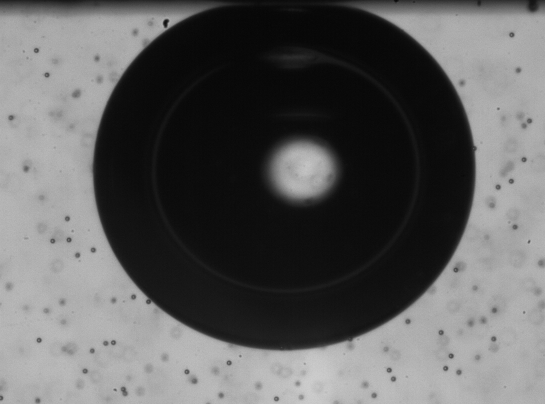}
		\caption{$t=0.0605s$}
		\label{fig:q_159}
	\end{subfigure}  \ \ \ \
	\begin{subfigure}{0.15\textwidth}
		\centering
		\captionsetup{justification=centering}
		\includegraphics[height=2cm]{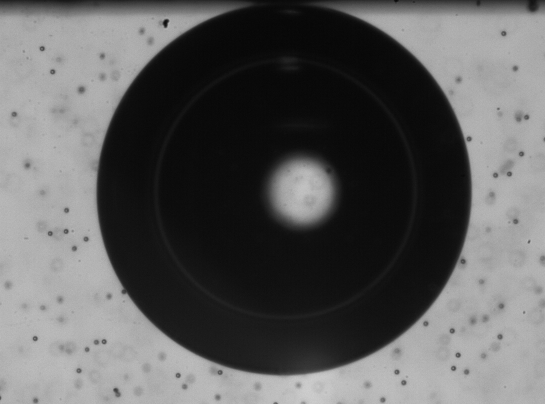}
		\caption{$t=0.0620s$}
		\label{fig:q_162}
	\end{subfigure}    \ \ \ \
	\begin{subfigure}{0.15\textwidth}
		\centering
		\captionsetup{justification=centering}
		\includegraphics[height=2cm]{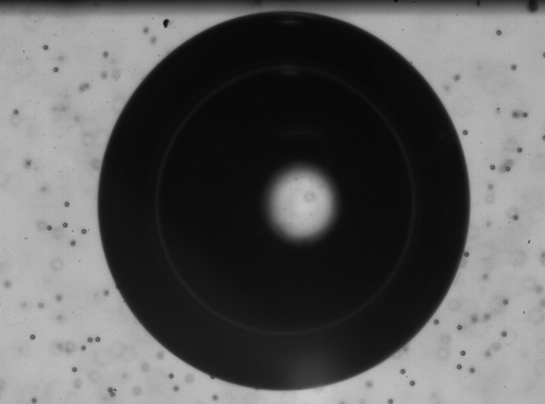}
		\caption{$t=0.0635s$ }
		\label{fig:q_165}
	\end{subfigure}    \ \ \ \
	\begin{subfigure}{0.15\textwidth}
		\centering
		\captionsetup{justification=centering}
		\includegraphics[height=2cm]{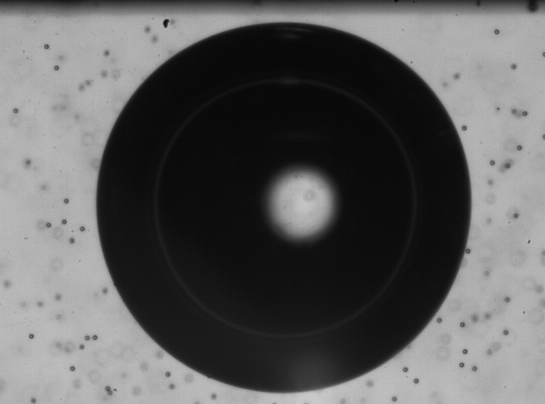}
		\caption{$t=0.0650s$ }
		\label{fig:q_168}
			\end{subfigure}    \ \ \ \
	\begin{subfigure}{0.15\textwidth}
		\centering
		\captionsetup{justification=centering}
		\includegraphics[height=2cm]{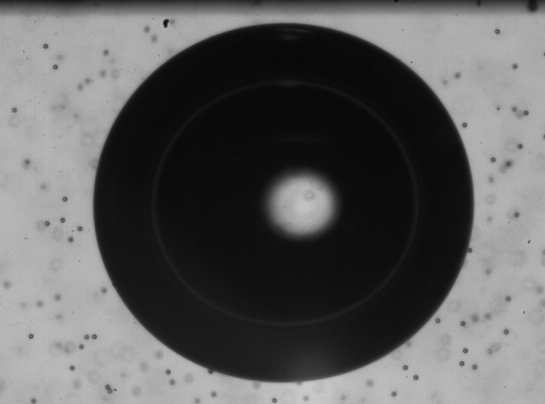}
		\caption{$t=0.0665s$ }
		\label{fig:q_171}
	\end{subfigure}  \\
		\begin{subfigure}{0.15\textwidth}
		\centering
		\captionsetup{justification=centering}
		\includegraphics[height=2cm]{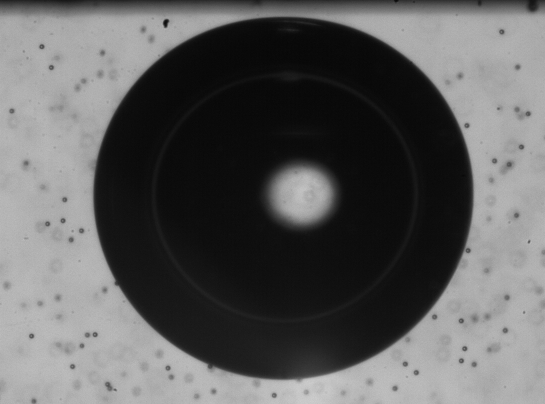}
		\caption{$t=0.0680s$}
		\label{fig:q_174}
	\end{subfigure}  \ \ \ \
	\begin{subfigure}{0.15\textwidth}
		\centering
		\captionsetup{justification=centering}
		\includegraphics[height=2cm]{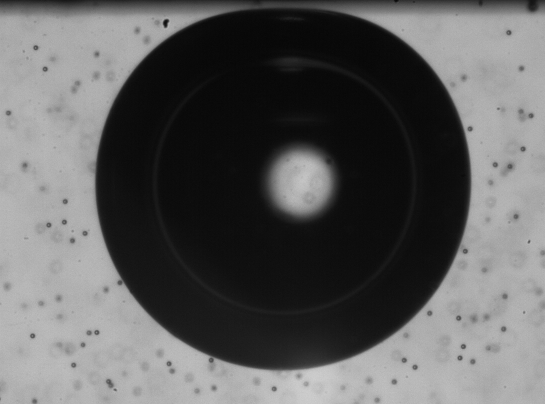}
		\caption{$t=0.0695s$}
		\label{fig:q_177}
	\end{subfigure}    \ \ \ \
	\begin{subfigure}{0.15\textwidth}
		\centering
		\captionsetup{justification=centering}
		\includegraphics[height=2cm]{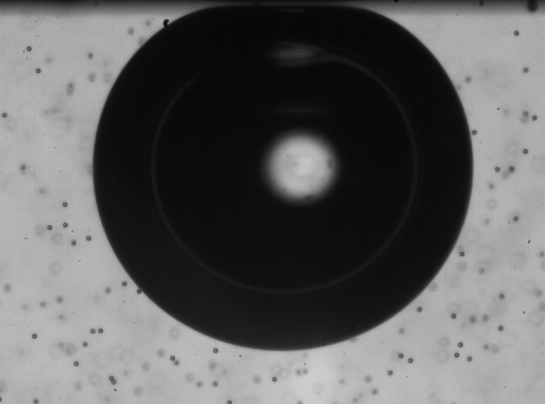}
		\caption{$t=0.0710s$ }
		\label{fig:q_180}
	\end{subfigure}    \ \ \ \
	\begin{subfigure}{0.15\textwidth}
		\centering
		\captionsetup{justification=centering}
		\includegraphics[height=2cm]{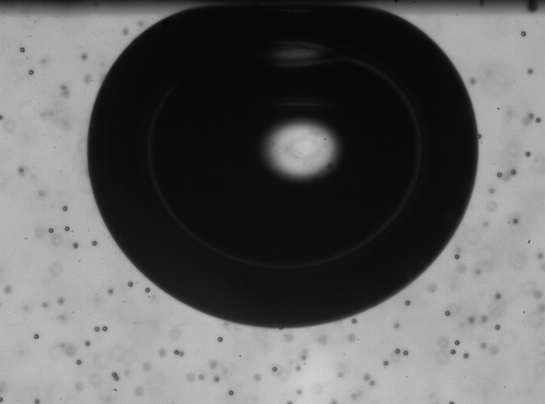}
		\caption{$t=0.0725s$ }
		\label{fig:q_183}
			\end{subfigure}    \ \ \ \
	\begin{subfigure}{0.15\textwidth}
		\centering
		\captionsetup{justification=centering}
		\includegraphics[height=2cm]{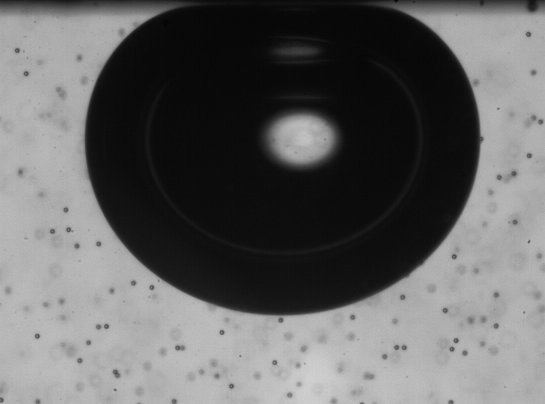}
		\caption{$t=0.0740s$ }
		\label{fig:q_186}
	\end{subfigure}  \\
	
	 \vskip1cm
	
	\begin{subfigure}{0.15\textwidth}
		\centering
		\captionsetup{justification=centering}
		\includegraphics[height=2cm]{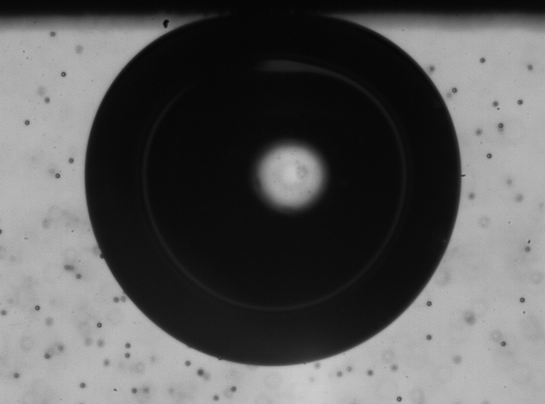}
		\caption{$t=0.0530s$}
		\label{fig:s_142}
	\end{subfigure}  \ \ \ \
	\begin{subfigure}{0.15\textwidth}
		\centering
		\captionsetup{justification=centering}
		\includegraphics[height=2cm]{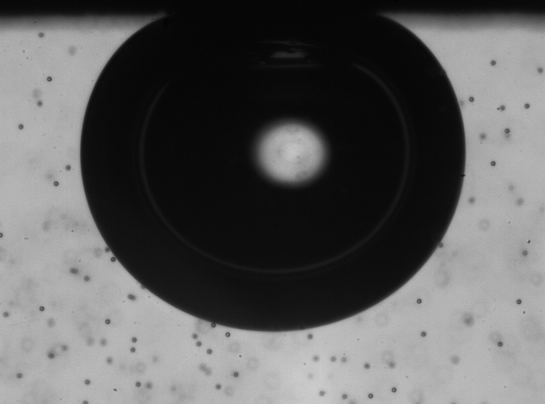}
		\caption{$t=0.0545s$}
		\label{fig:s_145}
	\end{subfigure}    \ \ \ \
	\begin{subfigure}{0.15\textwidth}
		\centering
		\captionsetup{justification=centering}
		\includegraphics[height=2cm]{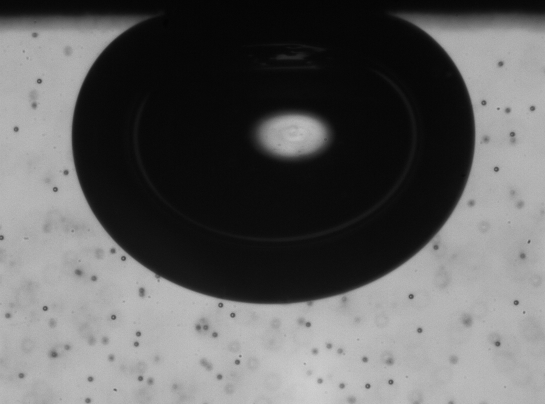}
		\caption{$t=0.0560s$ }
		\label{fig:s_148}
	\end{subfigure}    \ \ \ \
	\begin{subfigure}{0.15\textwidth}
		\centering
		\captionsetup{justification=centering}
		\includegraphics[height=2cm]{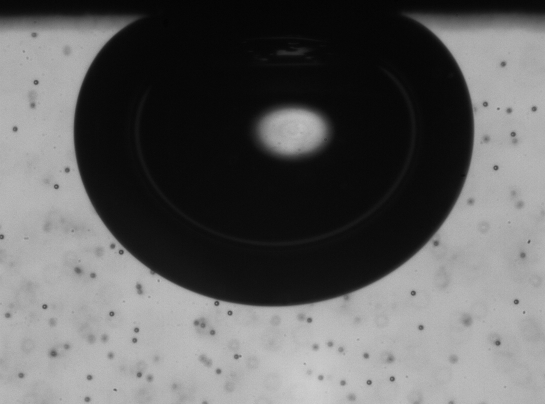}
		\caption{$t=0.0575s$ }
		\label{fig:s_151}
			\end{subfigure}    \ \ \ \
	\begin{subfigure}{0.15\textwidth}
		\centering
		\captionsetup{justification=centering}
		\includegraphics[height=2cm]{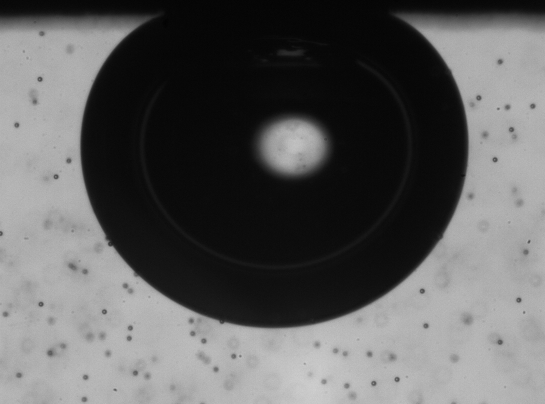}
		\caption{$t=0.0590s$ }
		\label{fig:s_154}
	\end{subfigure}  \\
	\begin{subfigure}{0.15\textwidth}
		\centering
		\captionsetup{justification=centering}
		\includegraphics[height=2cm]{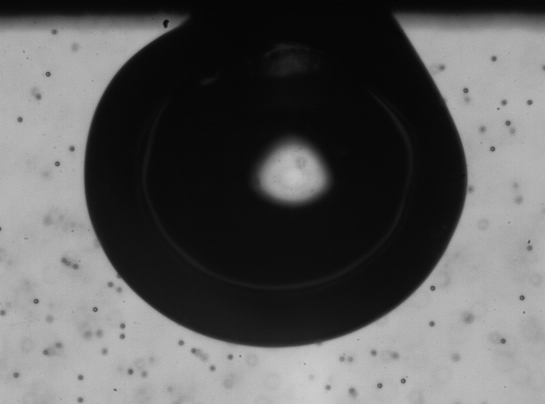}
		\caption{$t=0.0605s$}
		\label{fig:s_157}
	\end{subfigure}  \ \ \ \
	\begin{subfigure}{0.15\textwidth}
		\centering
		\captionsetup{justification=centering}
		\includegraphics[height=2cm]{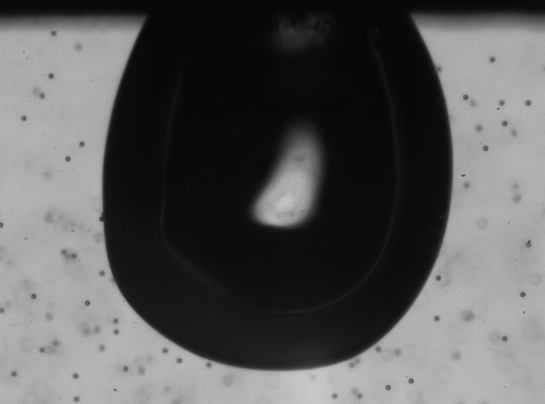}
		\caption{$t=0.0620s$}
		\label{fig:s_160}
	\end{subfigure}    \ \ \ \
	\begin{subfigure}{0.15\textwidth}
		\centering
		\captionsetup{justification=centering}
		\includegraphics[height=2cm]{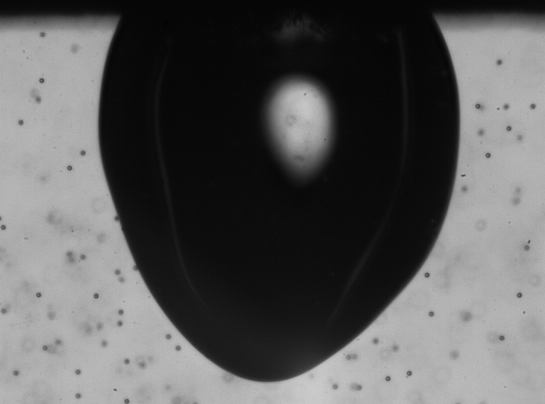}
		\caption{$t=0.0635s$ }
		\label{fig:s_163}
	\end{subfigure}    \ \ \ \
	\begin{subfigure}{0.15\textwidth}
		\centering
		\captionsetup{justification=centering}
		\includegraphics[height=2cm]{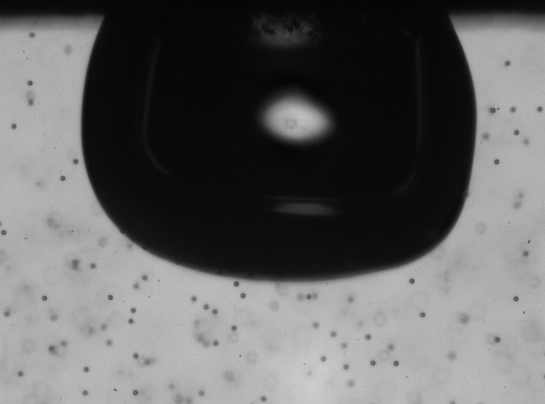}
		\caption{$t=0.0650s$ }
		\label{fig:s_166}
			\end{subfigure}    \ \ \ \
	\begin{subfigure}{0.15\textwidth}
		\centering
		\captionsetup{justification=centering}
		\includegraphics[height=2cm]{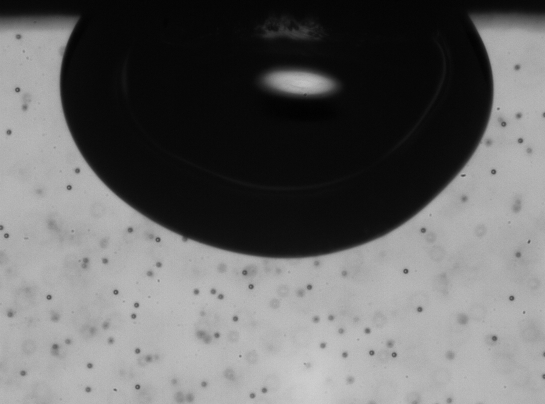}
		\caption{$t=0.0665s$ }
		\label{fig:s_169}
	\end{subfigure}  \\
		\begin{subfigure}{0.15\textwidth}
		\centering
		\captionsetup{justification=centering}
		\includegraphics[height=2cm]{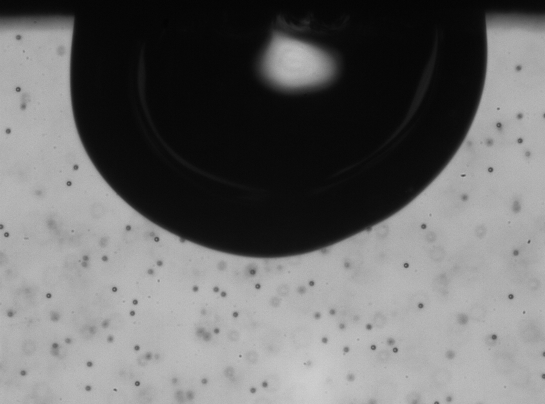}
		\caption{$t=0.0680s$}
		\label{fig:s_172}
	\end{subfigure}  \ \ \ \
	\begin{subfigure}{0.15\textwidth}
		\centering
		\captionsetup{justification=centering}
		\includegraphics[height=2cm]{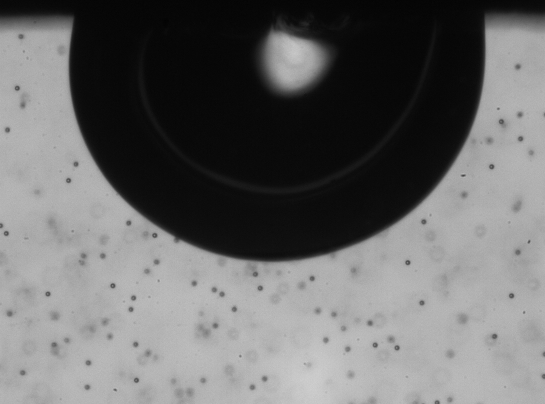}
		\caption{$t=0.0695s$}
		\label{fig:s_175}
	\end{subfigure}    \ \ \ \
	\begin{subfigure}{0.15\textwidth}
		\centering
		\captionsetup{justification=centering}
		\includegraphics[height=2cm]{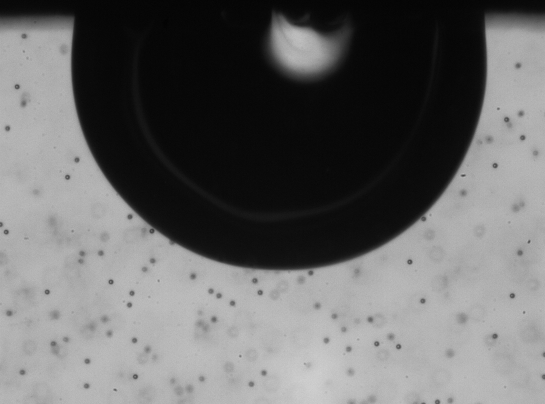}
		\caption{$t=0.0710s$ }
		\label{fig:s_178}
	\end{subfigure}    \ \ \ \
	\begin{subfigure}{0.15\textwidth}
		\centering
		\captionsetup{justification=centering}
		\includegraphics[height=2cm]{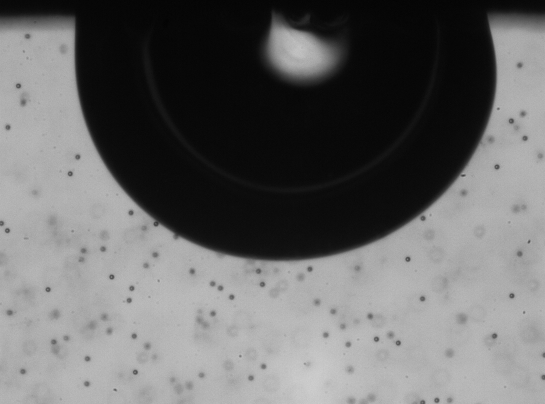}
		\caption{$t=0.0725s$ }
		\label{fig:s_181}
			\end{subfigure}    \ \ \ \
	\begin{subfigure}{0.15\textwidth}
		\centering
		\captionsetup{justification=centering}
		\includegraphics[height=2cm]{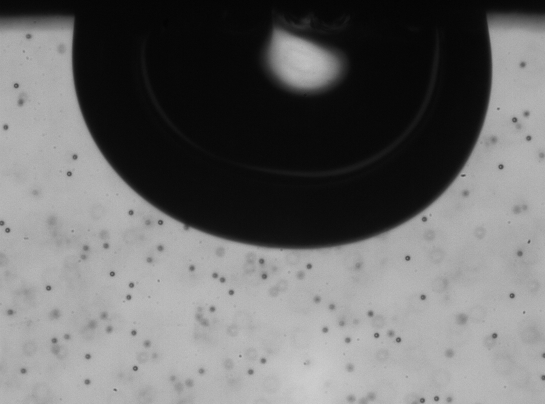}
		\caption{$t=0.0740s$ }
		\label{fig:s_184}
	\end{subfigure}  \\

	\caption{
	\textcolor{black}{The screenshots of an $O_2$ bubble bouncing on an uncoated surface (a - o), and arrested on the coated surface (p – ad) within 0.0530 s - 0.0740 s at similar oxygen supersaturation of $\zeta \sim -0.72 \pm 0.2$.}}
	\label{fig:screenshots_bubble_rupture}
			\end{figure}

\textcolor{black}{The $O_2$ bubble hitting, bouncing off and re-hitting against the uncoated substrate is clearly visible in the Fig (\ref{fig:q_145} - \ref{fig:q_186}) corresponding to 0.0530 s - 0.0740 s. During the bouncing, the bubble shape change on this surface is not as drastic as compared to that on the coated surface.}

Fig. (\ref{fig:s_142} - \ref{fig:s_184}), arranging the images alphabetically from 'a' to 'z' and then from 'aa' to 'ad', visualize the corresponding bouncing process of the $O_2$ bubble on the coated surface. 
Before the $O_2$ bubble is arrested, it exhibits a hitting and bouncing process similar to that against the uncoated substrate, see Fig. (\ref{fig:s_142} – \ref{fig:s_154}). But the three-phase contact is formed \cite{zawala2007influence}, when the upper part of this bubble is arrested by the coated substrate in Fig. \ref{fig:s_157}. 
The inertial force leads to a drastic change in the bubble shape and \textcolor{black}{elongated bubbles},
 see Fig. (\ref{fig:s_160} - \ref{fig:s_163}). Afterwards, the contact line length rapidly increases, causing the bubble to approach the substrate quickly under continuing deformation, see Fig. (\ref{fig:s_169} - \ref{fig:s_181}).



\subsection{Fit of the average concentration gradient as function of the polar angle}
\label{sec:gradient_fit}
We calculate the average radial gradient, $\partial c/\partial r$, of the $O_2$ bubble on uncoated ($\zeta \sim -0.73$) and coated ($\zeta \sim -0.73$) surfaces from the angular position of -20\textdegree{} to 40\textdegree{} at 228 s. After that, these average radial gradients are fitted 
as follows:

\begin{equation}
   \frac{\partial c}{\partial r} = A_2 + \frac{A_1 - A_2}{1 + \exp\left(\frac{\theta - x_0}{d_x}\right)},
   \label{eq:con_gra_ang_pos_fit}
\end{equation}

where $A_1$, $A_2$, $x_0$ and $d_x$ are fitted constants, which are shown in the Table \ref{tab:fitted_constants}, and $\theta$ is the angular position. 
 \textcolor{black}{With this equation, the relationship between all angular positions of the bubble (bound by the coated or uncoated surface) and the concentration gradients can be described.}

\begin{table}[h!]
	\centering
	\caption{The fitted constants of $A_1$, $A_2$, $x_0$ and $d_x$ for the uncoated ($\zeta \sim -0.73$) and coated ($\zeta \sim -0.76$) cases, $R^2$ is the coefficient of determination of the fit, where $R$ is correlation coefficient.}
		\begin{tabular}{ c c c  } 	\hline 
		 & Uncoated case ($\zeta \sim -0.73$) & Coated case ($\zeta \sim -0.76$) \\\hline
        $A_1$ ($\mathrm{mol/m^4}$) & -937.77827 & -530.53824 \\ 
        $A_2$ ($\mathrm{mol/m^4}$) & -1528.17911 & -1323.80901 \\ 
        $x_0$ (\textdegree{}) & 10.31334 & 1.21589\\ 
        $d_x$ (\textdegree{}) & 7.88856 & 12.24128\\\hline
         $R^2$ & 0.99 & 0.98\\\hline
	\end{tabular}
	\label{tab:fitted_constants}
\end{table}

\clearpage
\bibliographystyle{elsarticle-num} 
\bibliography{lib}

\begin{thebibliography}{10}
\expandafter\ifx\csname url\endcsname\relax
  \def\url#1{\texttt{#1}}\fi
\expandafter\ifx\csname urlprefix\endcsname\relax\def\urlprefix{URL }\fi
\expandafter\ifx\csname href\endcsname\relax
  \def\href#1#2{#2} \def\path#1{#1}\fi

\bibitem{angulo2020influence}
A.~Angulo, P.~van~der Linde, H.~Gardeniers, M.~Modestino, D.~F. Rivas, Influence of bubbles on the energy conversion efficiency of electrochemical reactors, Joule 4~(3) (2020) 555--579.

\bibitem{reinecke4704929population}
S.~F. Reinecke, R.~Herrmann-Heber, M.~Oleshova, M.~Meier, S.~Ta{\c{s}}, U.~Hampel, A.~Lerch, Population balance modeling-assisted prediction of oxygen mass transfer coefficients with optical measurements, Available at SSRN 4704929.

\bibitem{herrmann2021experimental}
R.~Herrmann-Heber, F.~Ristau, E.~Mohseni, S.~F. Reinecke, U.~Hampel, Experimental oxygen mass transfer study of micro-perforated diffusers, Energies 14~(21) (2021) 7268.

\bibitem{khani2021novel}
M.~Khani, S.~Habibzadeh, M.~K. Moraveji, H.~A. Ebrahim, J.~Alizadeh, Novel $\alpha$-alumina@\ce{CuO-Fe_2O_3} nanofluid for potential application in pem fuel cell cooling systems: Towards neutralizing the increase of electrical conductivity, Thermochimica Acta 695 (2021) 178818.

\bibitem{epstein1950stability}
P.~S. Epstein, M.~S. Plesset, On the stability of gas bubbles in liquid-gas solutions, The Journal of Chemical Physics 18~(11) (1950) 1505--1509.

\bibitem{scriven1995dynamics}
L.~Scriven, On the dynamics of phase growth, Chemical Engineering Science 50~(24) (1995) 3907--3917.

\bibitem{cable1967spherically}
M.~Cable, D.~Evans, Spherically symmetrical diffusion-controlled growth or dissolution of a sphere, Journal of Applied Physics 38~(7) (1967) 2899--2906.

\bibitem{enriquez2014quasi}
O.~R. Enr{\'\i}quez, C.~Sun, D.~Lohse, A.~Prosperetti, D.~Van Der~Meer, The quasi-static growth of \ce{CO_2} bubbles, Journal of Fluid Mechanics 741 (2014) R1.

\bibitem{li2014growth}
J.~Li, H.~Chen, W.~Zhou, B.~Wu, S.~D. Stoyanov, E.~G. Pelan, Growth of bubbles on a solid surface in response to a pressure reduction, Langmuir 30~(15) (2014) 4223--4228.

\bibitem{kovats2018characterizing}
P.~Kov{\'a}ts, D.~Th{\'e}venin, K.~Z{\"a}hringer, Characterizing fluid dynamics in a bubble column aimed for the determination of reactive mass transfer, Heat and Mass Transfer 54 (2018) 453--461.

\bibitem{rubio2021superhydrophobic}
M.~Rubio-Rubio, R.~Bola{\~n}os-Jim{\'e}nez, C.~Mart{\'\i}nez-Baz{\'a}n, J.~C. Mu{\~n}oz-Herv{\'a}s, A.~Sevilla, Superhydrophobic substrates allow the generation of giant quasi-static bubbles, Journal of Fluid Mechanics 912 (2021) A25.

\bibitem{chen2018contact}
Y.~Chen, W.~Xia, G.~Xie, Contact angle and induction time of air bubble on flat coal surface of different roughness, Fuel 222 (2018) 35--41.

\bibitem{moraila2019wetting}
C.~L. Moraila, F.~J.~M. Ruiz-Cabello, M.~Cabrerizo-V{\'\i}lchez, M.~{\'A}. Rodr{\'\i}guez-Valverde, Wetting transitions on rough surfaces revealed with captive bubble experiments. \uppercase{t}he role of surface energy, Journal of Colloid and Interface Science 539 (2019) 448--456.

\bibitem{xia2021influence}
Y.~Xia, X.~Gao, R.~Li, Influence of surface wettability on bubble formation and motion, Langmuir 37~(49) (2021) 14483--14490.

\bibitem{jo2016single}
H.~Jo, H.~S. Park, M.~H. Kim, Single bubble dynamics on hydrophobic--hydrophilic mixed surfaces, International Journal of Heat and Mass Transfer 93 (2016) 554--565.

\bibitem{shen2018effect}
B.~Shen, J.~Liu, J.~Shiomi, G.~Amberg, M.~Do-Quang, M.~Kohno, K.~Takahashi, Y.~Takata, Effect of dissolved gas on bubble growth on a biphilic surface: A diffuse-interface simulation approach, International Journal of Heat and Mass Transfer 126 (2018) 816--829.

\bibitem{kibar2017bubble}
A.~Kibar, R.~Ozbay, M.~A. Sarshar, Y.~T. Kang, C.-H. Choi, Bubble movement on inclined hydrophobic surfaces, Langmuir 33~(43) (2017) 12016--12027.

\bibitem{huynh2015plastron}
S.~H. Huynh, A.~A.~A. Zahidi, M.~Muradoglu, B.~H.-P. Cheong, T.~W. Ng, Plastron-mediated growth of captive bubbles on superhydrophobic surfaces, Langmuir 31~(24) (2015) 6695--6703.

\bibitem{heinrich2024functionalization}
J.~Heinrich, F.~R{\"a}nke, K.~Schwarzenberger, X.~Yang, R.~Baumann, M.~Marzec, A.~F. Lasagni, K.~Eckert, Functionalization of \uppercase{t}i64 via direct laser interference patterning and its influence on wettability and oxygen bubble nucleation, Langmuir (2024).

\bibitem{Wang2016Investi}
Y.~Wang, X.~Hu, Z.~Cao, L.~Guo, Investigations on bubble growth mechanism during photoelectrochemical and electrochemical conversions, Colloids and Surfaces A: Physicochemical and Engineering Aspects 505 (2016) 86--92.

\bibitem{popov2005evaporative}
Y.~O. Popov, Evaporative deposition patterns: Spatial dimensions of the deposit, Physical Review E 71~(3) (2005) 036313.

\bibitem{paruya2021numerical}
S.~Paruya, J.~Bhati, F.~Akhtar, Numerical model of bubble shape and departure in nucleate pool boiling, International Journal of Heat and Mass Transfer 180 (2021) 121756.

\bibitem{soto2017gas}
{\'A}.~M. Soto, A.~Prosperetti, D.~Lohse, D.~Van Der~Meer, Gas depletion through single gas bubble diffusive growth and its effect on subsequent bubbles, Journal of Fluid Mechanics 831 (2017) 474--490.

\bibitem{vachaparambil2020modeling}
K.~J. Vachaparambil, K.~E. Einarsrud, Modeling interfacial mass transfer driven bubble growth in supersaturated solutions, AIP Advances 10~(10) (2020).

\bibitem{lee2008growth}
S.-L. Lee, W.-B. Tien, Growth and detachment of carbon dioxide bubbles on a horizontal porous surface with a uniform mass injection, in: Proceedings of CHT-08 ICHMT International Symposium on Advances in Computational Heat Transfer, Begel House Inc., 2008.

\bibitem{diddens2021competing}
C.~Diddens, Y.~Li, D.~Lohse, Competing \uppercase{m}arangoni and \uppercase{r}ayleigh convection in evaporating binary droplets, Journal of Fluid Mechanics 914 (2021) A23.

\bibitem{babu2019experimental}
R.~Babu, M.~K. Das, Experimental studies of natural convective mass transfer in a water-splitting system, International Journal of Hydrogen Energy 44~(29) (2019) 14467--14480.

\bibitem{xu2018mass}
F.~Xu, A.~Cockx, G.~H\'ebrard, N.~Dietrich, Mass transfer and diffusion of a single bubble rising in polymer solutions, Industrial \& Engineering Chemistry Research 57~(44) (2018) 15181--15194.

\bibitem{dietrich2018visualisation}
N.~Dietrich, G.~Hebrard, Visualisation of gas-liquid mass transfer around a rising bubble in a quiescent liquid using an oxygen sensitive dye, Heat and Mass Transfer 54 (2018) 2163--2171.

\bibitem{felis2019using}
F.~Felis, F.~Strassl, L.~Laurini, N.~Dietrich, A.-M. Billet, V.~Roig, S.~Herres-Pawlis, K.~Loubi{\`e}re, Using a bio-inspired copper complex to investigate reactive mass transfer around an oxygen bubble rising freely in a thin-gap cell, Chemical Engineering Science 207 (2019) 1256--1269.

\bibitem{kherbeche2020hydrodynamics}
A.~Kherbeche, M.~Mei, M.-J. Thoraval, G.~H{\'e}brard, N.~Dietrich, Hydrodynamics and gas-liquid mass transfer around a confined sliding bubble, Chemical Engineering Journal 386 (2020) 121461.

\bibitem{zhang2020effective}
Z.~Zhang, H.~Zhang, X.~Yuan, K.-T. Yu, Effective \uppercase{uv}-induced fluorescence method for investigating interphase mass transfer of single bubble rising in the hele-shaw cell, Industrial \& Engineering Chemistry Research 59~(14) (2020) 6729--6740.

\bibitem{qi2021modelling}
R.~Qi, C.~Dong, S.~Yu, L.-Z. Zhang, Modelling and experiments of falling film break-up characteristics considering mass transfer for liquid desiccant dehumidification, International Journal of Heat and Mass Transfer 181 (2021) 122027.

\bibitem{chen2013measurement}
J.~Chen, H.~D. Kim, K.~C. Kim, Measurement of dissolved oxygen diffusion coefficient in a microchannel using \uppercase{UV-LED} induced fluorescence method, Microfluidics and Nanofluidics 14 (2013) 541--550.

\bibitem{kim2016quantitative}
J.~Kim, K.~C. Kim, H.~D. Kim, Quantitative visualization study on diffusion of oxygen using \uppercase{UV-LED} induced phosphorescence, Journal of Visualization 19 (2016) 591--601.

\bibitem{ruttinger2018laser}
S.~R{\"u}ttinger, C.~Spille, M.~Hoffmann, M.~Schl{\"u}ter, Laser-induced fluorescence in multiphase systems, ChemBioEng Reviews 5~(4) (2018) 253--269.

\bibitem{xu2020comparison}
F.~Xu, G.~H{\'e}brard, N.~Dietrich, Comparison of three different techniques for gas-liquid mass transfer visualization, International Journal of Heat and Mass Transfer 150 (2020) 119261.

\bibitem{jimenez2013mass}
M.~Jimenez, N.~Dietrich, G.~H{\'e}brard, Mass transfer in the wake of non-spherical air bubbles quantified by quenching of fluorescence, Chemical Engineering Science 100 (2013) 160--171.

\bibitem{von20193d}
A.~von Kameke, F.~Kexel, S.~Ruttinger, R.~Colombi, S.~Kastens, M.~Schl{\"u}ter, 3\uppercase{D}-reconstruction of \ce{O_2} bubble wake concentration fields, in: Proceedings of the 13th International Symposium on Particle Image Velocimetry—ISPIV, 2019.

\bibitem{lebrun2022effect}
G.~Lebrun, S.~Benaissa, C.~Le~Men, V.~Pimienta, G.~H{\'e}brard, N.~Dietrich, Effect of surfactant lengths on gas-liquid oxygen mass transfer from a single rising bubble, Chemical Engineering Science 247 (2022) 117102.

\bibitem{huang2015influence}
J.~Huang, T.~Saito, Influence of bubble-surface contamination on instantaneous mass transfer, Chemical Engineering \& Technology 38~(11) (2015) 1947--1954.

\bibitem{kovats2020influence}
P.~Kov{\'a}ts, D.~Th{\'e}venin, K.~Z{\"a}hringer, Influence of viscosity and surface tension on bubble dynamics and mass transfer in a model bubble column, International Journal of Multiphase Flow 123 (2020) 103174.

\bibitem{kuck2009analyse}
U.~D. K{\"u}ck, M.~Schl{\"u}ter, N.~R{\"a}biger, Analyse des grenzschichtnahen \uppercase{S}tofftransports an frei aufsteigenden \uppercase{G}asblasen, Chemie Ingenieur Technik 81~(10) (2009) 1599--1606.

\bibitem{butler2016modelling}
C.~Butler, E.~Cid, A.-M. Billet, Modelling of mass transfer in \uppercase{T}aylor flow: Investigation with the \uppercase{PLIF-I} technique, Chemical Engineering Research and Design 115 (2016) 292--302.

\bibitem{lebrun2021gas}
G.~Lebrun, F.~Xu, C.~Le~Men, G.~H{\'e}brard, N.~Dietrich, Gas--liquid mass transfer around a rising bubble: Combined effect of rheology and surfactant, Fluids 6~(2) (2021) 84.

\bibitem{gerke2024planar}
S.~J. Gerke, G.~Br{\"o}sigke, J.-U. Repke, Planar light induced fluorescence quenching oxygen diffusivity measurement near the gas--liquid interface in aqueous glycerol and aqueous propylene glycol, Experimental Thermal and Fluid Science 153 (2024) 111131.

\bibitem{zawala2017bubble}
J.~Zawala, A.~Niecikowska, “\text{B}ubble-on-demand” generator with precise adsorption time control, Review of Scientific Instruments 88~(9) (2017).

\bibitem{merzkirch2012flow}
W.~Merzkirch, Flow visualization, Elsevier, 2012.

\bibitem{karnbach2016interplay}
F.~Karnbach, X.~Yang, G.~Mutschke, J.~Fröhlich, J.~Eckert, A.~Gebert, K.~Tschulik, K.~Eckert, M.~Uhlemann, Interplay of the open circuit potential-relaxation and the dissolution behavior of a single \ce{H_2} bubble generated at a \uppercase{P}t microelectrode, The Journal of Physical Chemistry C 120~(28) (2016) 15137--15146.

\bibitem{kursula2022unsteady}
L.~Kursula, F.~Kexel, J.~Fitschen, M.~Hoffmann, M.~Schl{\"u}ter, A.~von Kameke, Unsteady mass transfer in bubble wakes analyzed by lagrangian coherent structures in a flat-bed reactor, Processes 10~(12) (2022) 2686.

\bibitem{babich2023situ}
A.~Babich, A.~Bashkatov, X.~Yang, G.~Mutschke, K.~Eckert, In-situ measurements of temperature field and \uppercase{M}arangoni convection at hydrogen bubbles using schlieren and \uppercase{PTV} techniques, International Journal of Heat and Mass Transfer 215 (2023) 124466.

\bibitem{yang2020analytical}
S.~Yang, S.~Bao, C.~Liu, D.~Yuan, W.~Huang, An analytical model of the growth of invisible bubbles on solid surfaces in a supersaturated solution, Chemical Engineering Science 215 (2020) 114968.

\bibitem{Gro2017diffusion}
T.~F. Gro\ss, P.~F. Pelz, Diffusion-driven nucleation from surface nuclei in hydrodynamic cavitation, Journal of Fluid Mechanism 830 (2017) 138--164.

\bibitem{cussler2009diffusion}
E.~L. Cussler, Diffusion: Mass transfer in fluid systems, Cambridge University Press, 2009.

\bibitem{lohse2015pinning}
D.~Lohse, X.~Zhang, et~al., Pinning and gas oversaturation imply stable single surface nanobubbles, Physical Review E 91~(3) (2015) 031003.

\bibitem{wang2016investigations}
Y.~Wang, X.~Hu, Z.~Cao, L.~Guo, Investigations on bubble growth mechanism during photoelectrochemical and electrochemical conversions, Colloids and Surfaces A: Physicochemical and Engineering Aspects 505 (2016) 86--92.

\bibitem{huysmans2005review}
M.~Huysmans, A.~Dassargues, Review of the use of \uppercase{P}{\'e}clet numbers to determine the relative importance of advection and diffusion in low permeability environments, Hydrogeology Journal 13 (2005) 895--904.

\bibitem{watanabe1985influence}
H.~Watanabe, K.~Iizuka, The influence of dissolved gases on the density of water, Metrologia 21~(1) (1985) 19.

\bibitem{zawala2007influence}
J.~Zawala, M.~Krasowska, T.~Dabros, K.~Malysa, Influence of bubble kinetic energy on its bouncing during collisions with various interfaces, The Canadian Journal of Chemical Engineering 85~(5) (2007) 669--678.

\end{thebibliography}

\end{document}